\documentclass[3p,amsmath,amssymb]{elsarticle}

\usepackage{amsmath}
\usepackage{graphicx}
\usepackage{latexsym}
\usepackage{amsmath,amssymb}
\usepackage{bm} 
\usepackage{xcolor}
\usepackage{stackrel}
\usepackage{accents}

\journal{Annals of Physics}

\numberwithin{equation}{section}

\newcommand{\braket}[1]{\left\langle {#1} \right\rangle}

\newcommand{\kb}{\vex{k}}
\newcommand{\bl}{\vex{l}}
\newcommand{\rb}{\vex{x}}

\newcommand{\lb}{\lambda}
\newcommand{\dd}{\mathcal{D}}
\newcommand{\ww}{\omega}

\newcommand{\rql}{\rho_{\mathsf{q}}}
\newcommand{\rcl}{\rho_{\mathsf{cl}}}
\newcommand{\bql}{{b}_{\mathsf{q}}}
\newcommand{\bd}{\Delta}

\newcommand{\xb}{\vex{x}}
\newcommand{\bv}{\vex{b}}
\newcommand{\blv}{\vex{B}}
\newcommand{\sv}{\vex{s}\,}
\newcommand{\bvql}{\bv_{\mathsf{q}}}
\newcommand{\bvcl}{\bv_{\mathsf{cl}}}
\newcommand{\Bvql}{\blv_{\mathsf{q}}}
\newcommand{\Bvcl}{\blv_{\mathsf{cl}}}

\DeclareMathOperator{\sgn}{sgn}

\newcommand{\intl}[1]{\int\limits_{#1}}
\newcommand{\vex}[1]{\bm{\mathrm{#1}}}

\usepackage{color}

\newcommand{\bsub}{\begin{subequations}}
\newcommand{\esub}{\end{subequations}}

\newcommand{\h}[1]{\hat{#1}}
\newcommand{\tr}{\mathsf{Tr}}
\newcommand{\Nabla}{\bm{\nabla}}

\newcommand{\pup}[1]{{\scriptscriptstyle{({#1})}}}
\newcommand{\ket}[1]{\left| {#1} \right\rangle}
\newcommand{\bra}[1]{\left\langle {#1} \right|}

\newcommand{\msf}[1]{\mathsf{#1}}
\newcommand{\e}{\varepsilon}

\newcommand{\D}{\mathcal{D}}
\newcommand{\G}{\hat{G}}
\newcommand{\T}{\mathsf{T}}
\newcommand{\vexh}[1]{{\hat{\vex{#1}}}}

\newcommand{\sig}{\sigma}

\newcommand{\Sigh}{\hat{\Sigma}}

\newcommand{\sigh}{\hat{\sigma}}

\newcommand{\tauh}{\hat{\tau}}

\newcommand{\tim}{\mathrm{T}}
\newcommand{\atim}{\bar{\mathrm{T}}}
\newcommand{\mf}{\hat{M}_F}

\newcommand{\Geta}{\hat{G}_{\eta}}

\newcommand{\vcl}{V_{\mathsf{cl}}}
\newcommand{\vq}{V_{\mathsf{q}}}
\newcommand{\tvcl}{\tilde{V}_{\mathsf{cl}}}
\newcommand{\tvq}{\tilde{V}_{\mathsf{q}}}
\newcommand{\rhocl}{\rho_{\mathsf{cl}}}
\newcommand{\rhoq}{\rho_{\mathsf{q}}}

\newcommand{\Bq}{B_{\mathsf{q}}}
\newcommand{\Bcl}{B_{\mathsf{cl}}}
\newcommand{\tBq}{\tilde{B}_{\mathsf{q}}}
\newcommand{\tBcl}{\tilde{B}_{\mathsf{cl}}}
\newcommand{\bq}{{b}_{\mathsf{q}}}
\newcommand{\bcl}{{b}_{\mathsf{cl}}}

\newcommand{\dcl}{\Delta_{\mathsf{cl}}}
\newcommand{\dq}{\Delta_{\mathsf{q}}}
\newcommand{\HD}{\hat{\mathfrak{D}}}

\newcommand{\ulo}{\hat{U}_{\scriptscriptstyle{\msf{LO}}}}
\newcommand{\qsp}{\hat{q}_{\scriptscriptstyle{\mathsf{SP}}}}
\newcommand{\qh}{\hat{q}}

\newcommand{\Xh}{\hat{X}}
\newcommand{\Yh}{\hat{Y}}
\newcommand{\Wh}{\hat{W}}

\newcommand{\z}{\zeta}

\newcommand{\suml}[1]{\sum\limits_{#1}}

\begin{document}
\begin{frontmatter}

\title{Response theory of the ergodic many-body delocalized phase:\\
Keldysh Finkel'stein sigma models and the 10-fold way}
\author[Rice]{Yunxiang Liao}
\author[Wisc]{Alex Levchenko}
\author[Rice,RCQM]{Matthew S.\ Foster}
\address[Rice]{Department of Physics and Astronomy, Rice University, Houston, Texas 77005, USA}
\address[Wisc]{Department of Physics, University of Wisconsin-Madison, Madison, Wisconsin 53706, USA}
\address[RCQM]{Rice Center for Quantum Materials, Rice University, Houston, Texas 77005, USA}

\date{\today}
\begin{abstract}

We derive the finite temperature Keldysh response theory for interacting fermions in the presence of quenched short-ranged disorder, as applicable to any of the 10 Altland-Zirnbauer classes in an Anderson delocalized phase with at least a U(1) continuous symmetry. In this formulation of the interacting Finkel'stein nonlinear sigma model, the statistics of one-body wave functions are encoded by the constrained matrix field, while physical correlations follow from the hydrodynamic density or spin response field, which decouples the interactions. Integrating out the matrix field first, we obtain weak (anti)localization and Altshuler-Aronov quantum conductance corrections from the hydrodynamic response function. This procedure automatically incorporates the correct infrared cutoff physics, and in particular gives the Altshuler-Aronov-Khmelnitsky (AAK) equations for dephasing of weak (anti)localization due to electron-electron collisions. We explicate the method by deriving known 
 quantum corrections in two dimensions for the symplectic metal class AII, as well as the spin-SU(2) invariant superconductor classes C and CI. We show that quantum conductance corrections due to the special modes at zero energy in nonstandard classes are automatically cut off by temperature, as previously expected, while the Wigner-Dyson class Cooperon modes that persist to all energies are cut by dephasing. We also show that for short-ranged interactions, the standard self-consistent solution for the dephasing rate is equivalent to a particular summation of diagrams via the self-consistent Born approximation. This should be compared to the corresponding AAK solution for long-ranged Coulomb interactions, which exploits the Markovian noise correlations induced by thermal fluctuations of the electromagnetic field. We discuss prospects for exploring the many-body localization transition as a dephasing catastrophe in short-range interacting models, as encountered by approaching 
 from the ergodic side. 
\end{abstract}


\end{frontmatter}

\tableofcontents


\section{Introduction}\label{sec:intr}

Recently, there has been a renewed surge of interest in disordered interacting quantum systems 
ignited by many-body localization (MBL) \cite{FA,AGKL,BAA,GMP,MBL-Rev}. In 2006, Basko, Aleiner and Altshuler (BAA)~\cite{BAA,BAA2} demonstrated that an isolated electron system, with weak and short-range interaction and strong quenched disorder that localizes all single-particle states, can undergo a finite temperature metal-insulator transition. The insulating state now known as the MBL phase exhibits strictly zero dc conductivity and a number of unique physical properties. Isolated from an external environment, a system in the MBL phase fails to serve as its own heat bath and does not thermalize. Quantum coherence is preserved on all length scales in such systems at energy densities corresponding to nonzero or even infinite temperature~\cite{MBLHighT}. Coherence due to localization can protect some types of topological order \cite{MBLOrder,MBLSPT}, even in the regime where such order is forbidden in thermal equilibrium.

Most of the recent work has focused on the MBL phase \cite{MBL-Rev}, while the MBL-ergodic phase transition in one-dimensional systems has been studied in Refs.~\cite{Imbrie2016,MBL1DT1,MBL1DT2,MBL1DT3}. By contrast, the nature (or even the existence) of the MBL-ergodic transition in dimensions higher than one remains unclear. Another key open question involves the issue of whether rare thermal fluctuations are able to destabilize the MBL phase
in two or more dimensions~\cite{MBLThermalFluct,MBLThermalFluct2}.

\subsection{The ergodic-MBL transition in $2D$ and nonstandard classes}

A strategy to understand MBL in two dimensions is to approach the putative transition 
from the ergodic side. In a system with a many-body mobility edge, the ergodic phase should persist for
temperatures $T > T_{MBL}$. One possibility is to study a system that is completely localized without
interactions, but which can exhibit a zero temperature quantum metal-insulator transition
in the presence of interactions. The latter can occur due to the antilocalizing effect of certain 
Altshuler-Aronov (AA) corrections~\cite{AA,AAG}, which are caused by the elastic scattering of particles off 
of disorder-induced Friedel oscillations in the particle density \cite{Zala}. Since AA corrections
are ineffective at (de)localizing states away from the Fermi energy \cite{AAG}, it implies that such a zero
temperature metal-insulator transition sits at the threshold of MBL with $T_{MBL} = 0$. 
A slight weakening of the interaction strength could then induce a small $T_{MBL} > 0$,
so that in this case the MBL-ergodic transition is a deformation of the zero temperature quantum 
critical point \cite{AAS2010}.

Noninteracting disordered fermion systems are completely classified according to the ``10-fold way'' due to Zirnbauer and Altland \cite{Zirnbauer,AltlandZirnbauer}. The 10-fold way is a random matrix scheme that includes the three standard Wigner-Dyson classes, which describe diffusive metals, and seven ``nonstandard'' classes that describe fermion systems with particle-hole and/or chiral symmetry \cite{AndersonRevC}. 
This additional symmetry in the nonstandard classes gives rise to special characteristics at the center of the one-body spectrum, such as critical scaling of the average density of states \cite{AndersonRevC}. 
The nonstandard classes arise e.g.\ in the description of quasiparticles in superconductors, 
since Pauli exclusion imposes ``automatic'' particle-hole symmetry for Majorana fermions, with
or without additional internal degrees of freedom. The 10-fold way also classifies the strong (fully gapped) topological insulators and superconductors, as well as their edge or 
surface states \cite{TIClass}. 

Key to the physics of noninteracting, nonstandard class models are additional quantum interference corrections \cite{AndersonRevC}, beyond those encountered in the Wigner-Dyson classes that afflict diffusive metals. These modify the statistics of the one-body wave functions near the band center and can lead to anomalous and/or critical behavior of the zero temperature Landauer conductance (at half-filling) \cite{Gade93,Ludwig94}. It is important to note however that the single-particle wave functions away from 
zero energy reside in a standard Wigner-Dyson class, since the particle-hole or chiral symmetry \cite{Zirnbauer,AltlandZirnbauer} responsible for the special properties at the band center is broken by finite frequency or chemical potential. This point is reviewed at length in this paper. 

Despite decades of work, aspects of zero-temperature metal-insulator transitions in $d > 1$ spatial dimensions for interacting Wigner-Dyson class systems remain unsolved or controversial \cite{FNLSM1983,BelitzKirkpatrick94,And50Book,CondInsQPT}. Recent progress includes understanding the interplay of wave function multifractality and interactions \cite{IQHP2,MR-BurmistrovA,MR-BurmistrovD} 
as well as the effects of disorder on interacting surface states of topological insulators \cite{MR-Konig}. Yet interacting versions of the nonstandard classes greatly expand the possibilities for understanding critical delocalization and interaction-driven quantum phase transitions, as shown by Dell'Anna \cite{CLuca,DellAnna16} and others \cite{CMomo,momothesis,FosterLudwig1,FosterLudwig2,FXC14}. In addition, some nonstandard class models in low dimensions can be solved exactly in the absence of interactions \cite{AndersonRevC,Brouwer-PRL00,Bagrets-PRB15}, enabling a nonperturbative starting point (with respect to disorder) for analyzing interaction effects. For example, strong evidence has been provided that AA corrections to the spin or thermal conductivity vanish to all orders at the dirty surface of a bulk topological superconductor \cite{FXC14,XCF15}. As applied to gapless quasiparticles in superconductors, the nonstandard class systems give physical realizations of disordered electronic systems with short-range (vs.\ long-range Coulomb) interactions, mediated by virtual fluctuations of the ``massive'' electromagnetic field \cite{CMomo}. Restriction to short-range interactions is believed to be a necessary ingredient for MBL \cite{Yao2014}.

\subsection{Keldysh response theory and results \label{sec:Keldresults}}

In this paper, we reformulate the problem of disordered interacting fermion systems as a finite-temperature Keldysh response theory. We obtain a version of the Finkel'stein nonlinear sigma model (FNL$\sigma$M)~\cite{FNLSM1983}, applicable to any Altland-Zirnbauer class with at least a U(1) continuous symmetry. 
Our approach is similar to the Keldysh formulation for the Wigner-Dyson classes previously exploited in Ref.~\cite{AlexAlex}; see also \cite{CLN-PRB99,KA-PRB99,Schwiete14}. The FNL$\sigma$M provides a systematic framework to study the combined effects of interactions and disorder, wherein the inverse dimensionless conductance is usually treated as a perturbation parameter (but see, e.g., \cite{FXC14}). In our version, the FNL$\sigma$M is a disorder-averaged theory containing two types of interacting fields: a dynamic matrix field subject to nonlinear constraints, and Hubbard-Stratonovich (H.-S.) field(s) introduced to decouple the interactions. The matrix field encodes the statistics of the one-body wave functions in the presence of disorder and describes the diffusive motion of electrons in a delocalized phase. By contrast, the H.-S.\ field corresponds to a quantity conserved in every realization of disorder potential, i.e., a hydrodynamic response mode associated to a continuous symmetry. The theory of the ergodic phase can be formulated as the hydrodynamic response theory at finite temperature using the Keldysh technique. 

The advantage of this dual-field Keldysh framework is that one is able to describe and clearly distinguish 
virtual and real scattering processes. In an isolated system, the latter arise entirely due to inelastic collisions between electrons, responsible for dephasing weak (anti)localization conductance corrections at finite temperature
\cite{AA,AAG,AAK,ChakravartySchmid,LoganWolynes87}.
So long as quantum interference corrections to dc transport are cut off in the infrared, 
the system behaves as a nonintegrable classical system on the largest scales and is guaranteed to equilibrate deformations away from thermal equilibrium. 

In the present paper, we set up and calculate explicitly the linear response function of 2D disordered systems, 
and obtain the quantum corrections to the conductivity which consist of weak localization (WL) 
[or weak anti-localization (WAL) in case of spin-orbit interaction] and Altshuler-Aronov (AA) corrections~\cite{AA,AAG,Zala}. Our framework has two key advantages. First, it automatically integrates ``tricky'' field-theoretic effects such as wave function renormalization in a natural way; these are pervasive in nonstandard class calculations. Second, it incorporates the correct infrared cutoffs to all quantum corrections. 
In particular, Wigner-Dyson class quantum conductance corrections that arise at all one-body energies are cut by dephasing. We show how to derive the Altshuler-Aronov-Khmelnitsky (AAK) \cite{AAK} equations for the dephasing of the weak (anti)localization correction. We expect that higher-loop calculations would give the corresponding generalization for the dephasing of higher-order quantum conductance corrections. 
By contrast, we show that additional nonstandard class WL/WAL corrections that arise due to the special modes at zero energy are automatically cut by temperature \cite{BTDLC05}, as are the AA corrections \cite{AAG}.


The specific models we consider here are 2D disordered conductors in the Wigner-Dyson symmetry class AII 
(also known as symplectic or spin-orbit metal class) and in the nonstandard class C, both with short-range interaction. The symplectic metal has been thoroughly studied (for a review, see~\cite{BelitzKirkpatrick94}) and serves as a benchmark. On the other hand, class C is a nonstandard class with particle-hole symmetry. 
It can be viewed as a superconductor quasiparticle system with broken time-reversal symmetry, and yet possessing spin-rotational invariance in every disorder realization \cite{CRealization1}. Class C could be realized experimentally in a type II superconductor, in which gapless quasiparticles hop between randomly-pinned vortex cores~\cite{CRealization1,CRealization2,AndersonRevC}. For quasiparticles in a superconductor, electric charge is not a hydrodynamic mode because an electron can be Andreev reflected as a hole. In class C, spin SU(2) symmetry implies that spin is a hydrodynamic mode. 
We consider the spin-spin (exchange) interaction and the spin conductance in class C.

An important exceptional aspect of class C is that, contrary to most other 2D systems, the spin-spin 
interaction strength is not renormalized to one-loop order \cite{CMomo,CLuca}, and possibly not to three loops \cite{momothesis,Hikami1992}. This should be contrasted against the original Finkel'stein model calculation in the orthogonal metal class AI, which features a notorious one-loop divergence in the spin triplet interaction channel \cite{FNLSM1983,BelitzKirkpatrick94,FP2005} that may signal a magnetic instability. 
We emphasize that the only small parameter in the FNL$\sigma$M loop expansion is the inverse dimensionless conductance; the interaction strength is treated to all orders. Formally, the sigma model sums interaction corrections as in a large-$N$ expansion \cite{FXC14,XCF15}. Because the interaction is not renormalized, by balancing the contribution from WL and AA, class C can undergo a controlled zero-temperature metal-insulator transition in the spin conductance~\cite{CMomo,CLuca}. This property makes it a promising candidate for investigating the MBL-ergodic transition in two dimensions by deforming the zero-temperature metal-insulator transition. 

We derive the one-loop results for these two models and one more (class CI) as follows,
\bsub\label{eq:correction}
\begin{align}
	\label{eqAII}
	\delta \sigma
	= \, &
	\frac{1}{4 \pi^2}  \ln \left(\frac{\Lambda}{\tau_{\phi}^{-1}}\right)
	-
	\frac{1}{2 \pi^2}
	\left[ 1-\left(1-\frac{1}{\gamma}\right)\ln (1-\gamma) \right] 
	\ln \left(\frac{\Lambda}{T}\right),	
	\quad \text{class AII},
	\\	
	\delta \sigma
	= \, &
	-\frac{1}{4 \pi^2}  \ln \left(\frac{\Lambda}{T}\right)
	-
	\frac{3}{2 \pi^2}
	\left[ 1-\left(1-\frac{1}{\gamma}\right)\ln (1-\gamma) \right] 
	\ln \left(\frac{\Lambda}{T}\right),
	\quad \text{class C},
	\label{eqC}
	\\	
	\delta \sigma
	= \, &
	-\frac{1}{4 \pi^2}  \ln \left(\frac{\Lambda}{\tau_{\phi}^{-1}}\right)
	-\frac{1}{4 \pi^2}  \ln \left(\frac{\Lambda}{T}\right)
	-
	\frac{3}{2 \pi^2}
	\left[ 1-\left(1-\frac{1}{\gamma}\right)\ln (1-\gamma) \right] 
	\ln \left(\frac{\Lambda}{T}\right),
	\quad \text{class CI}.
	\label{eqCI}
\end{align}
\esub
Eq.~(\ref{eqAII}) [(\ref{eqC})] gives the quantum correction to the electric (spin) conductivity 
for class AII (C). Here and throughout this paper, we work in units such that $\hbar$, the Boltzmann constant $k_B$, and the electric charge $e$ or spin charge $s = \hbar/2$ are set equal to one. 
The ultraviolet cutoff appearing in all corrections is the inverse of the elastic scattering time $\Lambda=\tau_{\mathsf{el}}^{-1}$. Eq.~(\ref{eqCI}) provides the one-loop corrections for class CI.
This is the same as class C, but with time-reversal preserved instead of broken \cite{CRealization1}. 
Eqs.~(\ref{eq:correction}) are valid to all orders of interaction strength $\gamma$.
In classes AII and CI, for simplicity we ignore the BCS interaction channel \cite{BelitzKirkpatrick94} in this work (see Refs. \cite{MR-Feigelman,AL-PRB07,AL-PRB10,MR-BurmistrovB,Konig-BKT,MR-BurmistrovC} where various effects stemming from the Cooper channel renormalizations were scrutinized). 

Although the results in Eq.~(\ref{eq:correction})
were obtained previously in the form of renormalization group (RG) equations~\cite{BelitzKirkpatrick94,kotliarAII} (AII) \cite{momothesis,CMomo,CLuca} (C,CI), 
here we rederive them in the response framework 
since the purpose of this paper is to present a method applicable to disordered systems at finite temperature in any symmetry class. 
In Eq.~(\ref{eqAII}) [Eq.~(\ref{eqC})], the first term corresponds to the WAL (WL) correction for class AII (C),
whereas the second terms in these equations are AA corrections due to the charge (spin) interaction channel in class AII (C).

Different from the RG method, our calculations directly give the correct infrared cutoffs to all conductance corrections. 
The WAL correction to class AII is cut off in the infrared by the dephasing rate $\tau_{\phi}^{-1}$.
The dephasing time $\tau_\phi$ is a function of the diffusion constant $D$ and the interaction strength $\gamma$ such that
$\tau_\phi(D,\gamma) \rightarrow \infty$ for $\gamma \rightarrow 0$ or $D \rightarrow \infty$.  
The rate $\tau_{\phi}^{-1}$ must be determined by solving the appropriate AAK equations \cite{AAK}, 
as we review in the next subsection and in Sec.~\ref{sec:dephasing}. 
By contrast, the WL correction to class C is directly cut by the temperature $T$. This is because
this correction arises due to the special nonstandard class diffusion modes present only 
at zero energy \cite{BTDLC05}. Since this is a set of measure zero for the energy integration,
it is regularized automatically for any $T > 0$, as are the AA corrections \cite{AAG}. 

Except for the first term in Eq.~(\ref{eqCI}), the result for class CI is identical to Eq.~(\ref{eqC}) for class C.
The first term is the WL correction due to the usual orthogonal Wigner-Dyson class AI Cooperon,
as we show here in the noninteracting model.  
The Cooperon is enabled in class CI by time-reversal symmetry, which is absent in class C.
Since this mode persists to all one-body energies, it must be cut by the dephasing rate $\tau_\phi^{-1}(D,\gamma)$ 
\cite{BTDLC05,KYG01}.

\begin{figure}[t]
\centering
\includegraphics[width=0.5\linewidth]{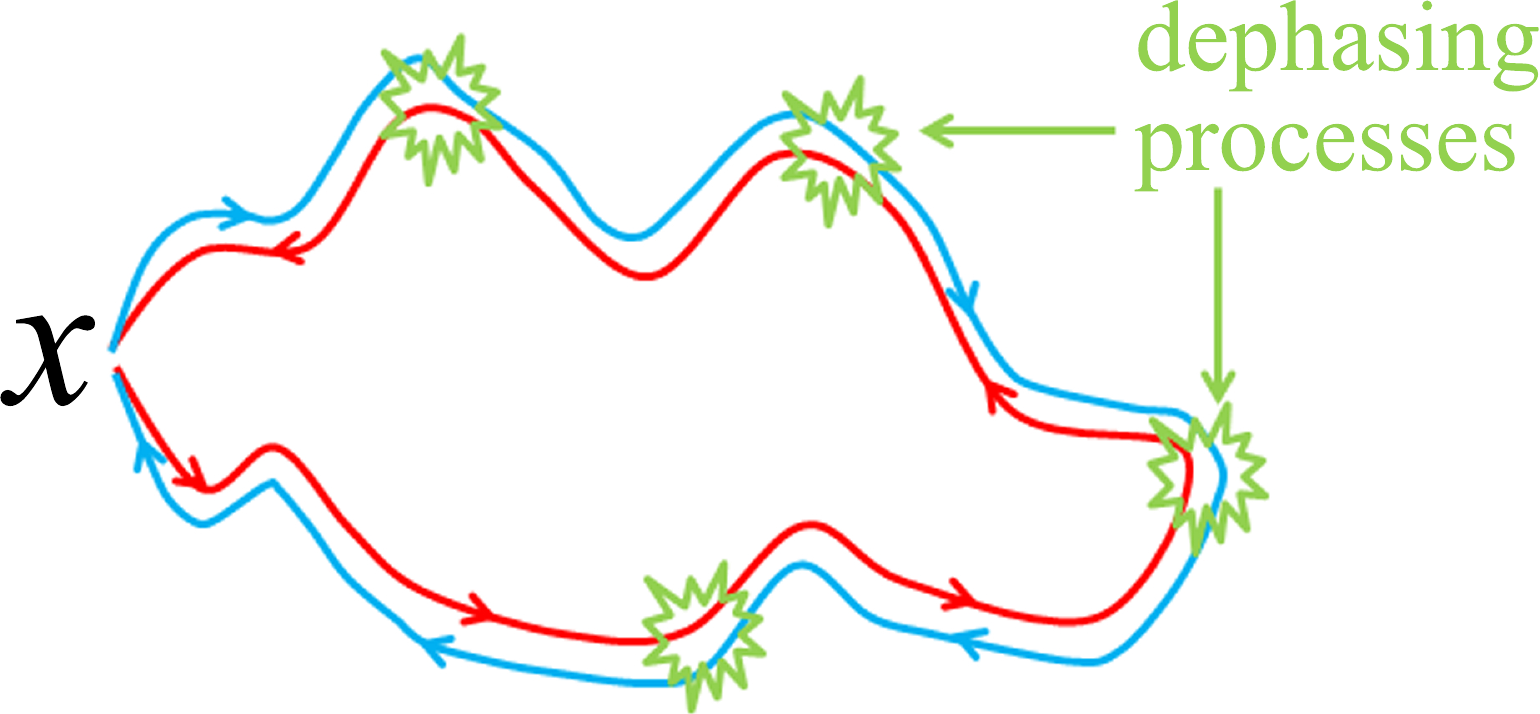}
\caption{Dephasing of quantum interference between time-reversed paths [Cooperon weak (anti)localization correction]. 
The dephasing ``events'' occur due to the interaction between the virtually diffusing quantum particle
and the stochastic, diffusive fluctuations of the density at temperature $T$. 
Dephasing suppresses the contribution of paths longer than the length 
$L_\phi = \sqrt{D \tau_\phi}$, where $1/\tau_\phi$ is the dephasing rate. 
For a system with short-range interactions, both the rate of virtual diffusion 
and 
of the thermal density fluctuations 
are controlled by the same diffusion constant $D$. 
So long as $1/\tau_\phi > 0$, the system serves as its own heat bath (many-body delocalized, ergodic phase).
By contrast, for weak localization in two spatial dimensions, $1/\tau_\phi = 0$ would signal localization,
since the Cooperon correction is logarithmically divergent in the infrared. 
Different from the case of dynamically screened long-range Coulomb interactions \cite{AAK}, 
for short-range interactions 
the thermal fluctuations of the density are a \emph{non-Markovian} dephasing mechanism for the virtual diffusion.}
\label{fig:Dephasing}
\end{figure}

\subsection{Self-dephasing of weak (anti)localization by diffusive density fluctuations \label{sec:self-dephasing}}

One of the main goals of this paper is to show how the problem of dephasing
quantum interference corrections can be precisely derived from the Keldysh sigma model.
This should allow a systematic investigation (order by order in the inverse dimensionless conductance) 
of \emph{self-dephasing} in a closed,
two-dimensional disordered many-body quantum system with short-range interactions.  
In Sec.~\ref{sec:dephasing_C}, we return to this problem and explain how class C may admit a perturbatively
controlled investigation of a many-body delocalization transition in 2D in the form
of a failure of self-dephasing. The class C scenario however requires a two-loop calculation,
which we leave to future work. 

We now summarize the technical statement of the dephasing problem
for the lowest order Cooperon correction, as formulated by AAK \cite{AAK}. 
As is well known, the one-loop WL or WAL Cooperon correction arises due to interference
between pairs of time-reversed paths \cite{ChakravartySchmid,AAG}. 
In section \ref{sec:dephasing}, we derive the AAK problem for the symplectic metal class
using our Keldysh formalism. The WAL correction [first term in Eq.~(\ref{eqAII})] 
obtains from return probability, equal to the integral of the Cooperon,  
\begin{align}\label{eq:dccorrection--INTRO}
	\delta \sigma_{\msf{WAL}} 
	= 
	&
	\frac{2}{\pi} 
	\intl{\eta}  
	\braket{C_{\eta,-\eta}^{t}(\rb,\rb)}_{\rho},	
\end{align}
where the Cooperon $C_{\eta,-\eta}^{t}(\rb,\rb)$ is the solution of
\begin{align}\label{eq:Schrodinger--INTRO}
\begin{aligned}
	\left\lbrace 
	\partial_{\eta} 
	-
	\frac{D}{2}
	\nabla ^2
	+
	 \frac{i}{2}
	\left[  \rcl\left(\rb,t+\frac{\eta}{2}\right) -\rcl\left(\rb,t-\frac{\eta}{2}\right) \right] 
	\right\rbrace 
	C_{\eta,\eta'}^{t}(\rb,\rb')
	=\,&
	\frac{D}{2}\delta(\eta-\eta')\delta(\rb-\rb').
\end{aligned}
\end{align}	
Here $D$ denotes the diffusion constant, $t$ is the average time on the forward and backward
(time-reversed) paths, and $\eta$ is the relative time. The field $\rcl(\rb,t)$
is the ``classical'' component of the hydrodynamic electric charge density, a bosonic
mode in the Keldysh formalism. The Cooperon interacts with (emits particle-hole pairs \cite{AleinerBlanter} via)
$\rcl(\rb,t)$ along the forward and backward paths at times $(t + \eta/2)$ and $(t - \eta/2)$, respectively. 

To obtain the WAL correction to conductivity, one needs to perform the average 
$\braket{C_{\eta,-\eta}^{t}(\rb,\rb)}_{\rho}$ in Eq.~(\ref{eq:dccorrection--INTRO})
over the thermal fluctuations of the density field $\rhocl$. 
The latter is Gaussian with the Keldysh (classical-classical) correlation function 
given by 
\begin{align}\label{eq:Gcl--INTRO}
\begin{aligned}
	i \Delta_{\rho}^{(K)} (\kb,\ww)
	\approx \,& 
	T  \frac{\gamma^2}{\kappa}
	\left(
	\frac{1}{D_c k^2 + i \ww}
	+
	\frac{1}{D_c k^2 - i \ww}		
	\right), 	
\end{aligned}
\end{align}
with the corresponding space-time expression 
\begin{align}\label{eq:Gclxt--INTRO}
\begin{aligned}
	i \Delta_{\rho}^{(K)} (\rb,t)
	\approx\, &
	T  \frac{\gamma^2}{\kappa}
	\left( \frac{1}{4 \pi D_c |t|} \right) 
	\exp \left( {-\frac{\rb^2}{4 D_c |t| }} \right). 
\end{aligned}
\end{align}
In these equations, we assume that the frequency $|\ww| \lesssim k_B T$, 
corresponding to real collision processes between thermally activated carriers 
responsible for dephasing. 
Here 
\begin{align}
	D_c \equiv \frac{D}{1 - \gamma}, \qquad
	\kappa \equiv (1 - \gamma) 2 \nu_0,
\end{align}
where $D_c$ is the charge diffusion constant,
$\gamma$ is the short-range interaction strength,
$\kappa$ is the charge compressibility, 
and 
$\nu_0$ is the bare density of states per spin 
\cite{BelitzKirkpatrick94}. 
The physics of Eqs.~(\ref{eq:dccorrection--INTRO})--(\ref{eq:Gclxt--INTRO}) is sketched in Fig.~\ref{fig:Dephasing}. 

Eqs.~(\ref{eq:Schrodinger--INTRO}) and (\ref{eq:Gclxt--INTRO}) show that for short-range interactions,
both the 
rate of virtual diffusion 
and 
the rate of thermal density fluctuations 
are controlled by the same diffusion constant $D$.
Different from the usual case of dynamically screened long-range Coulomb interactions,
the ``noise kernel'' in Eq.~(\ref{eq:Gclxt--INTRO}) is \emph{non-Markovian}; 
equivalently, the ``heat bath'' kernel in Eq.~(\ref{eq:Gcl--INTRO}) is non-Ohmic. 
The Markovian (memoryless) character of density fluctuations due to 
screened Coulomb interactions allows an exact solution to Eq.~(\ref{eq:dccorrection--INTRO})
\cite{AAK}. In that case the solution is equivalent to a self-consistent condition
imposed on the lowest order cumulant expansion for the averaged Cooperon \cite{ChakravartySchmid,AAG}. 
For short-range interactions (non-Markovian noise), the same self-consistent condition gives the result \cite{Zala-Deph}
\begin{align}\label{DephasingRate--INTRO}
	\tau_\phi^{-1}
	= 
	\frac{1}{4 \pi D \nu_0}
	\frac{\gamma^2}{\left(2-\gamma\right)}
	\,
	T 
	\ln \left( \frac{T}{\tau_\phi^{-1}}\right), 
\end{align}
as we derive in Sec.~\ref{sec:CWL}.
For $\gamma < 0$, $\tau_\phi^{-1}$ is nonzero except in the zero temperature limit $T \rightarrow 0$. 
Since the noise mediated by the heat bath is slow due to diffusion, there could
be corrections to Eq.~(\ref{DephasingRate--INTRO}) that are not captured by the self-consistent condition. 
This is another interesting direction for future work.

\subsection{Outline}

The rest of paper is organized as follows. 
In Sec.~\ref{sec:FNLSM}, we review and derive the FNL$\sigma$M in the Keldysh formalism applicable to a finite-temperature system 
in any symmetry class. 
Sec.~\ref{sec:AII} presents the detailed derivation of the response function for the symplectic metal, 
while Sec.~\ref{sec:C} is devoted to an analogous calculation for 
quasiparticle spin transport in a		
class C superconductor. 
The additional WL correction due to the Cooperon mode in class CI is extracted in 
Sec.~\ref{sec:CI}. 
We close the technical part of this work with Sec.~\ref{sec:dephasing}, wherein we derive the AAK equations 
(\ref{eq:dccorrection--INTRO})--(\ref{eq:Gclxt--INTRO})
for dephasing of  
the WAL correction. 
We show that the usual self-consistent solution \cite{ChakravartySchmid,AAG,Zala-Deph} 
is equivalent to a particular diagrammatic summation. 

In the final section~\ref{sec:dephasing_C}, we sketch a ``dephasing catastrophe'' scenario
for class C
that may allow perturbatively controlled access to a many-body delocalization transition in 
two dimensions.


\section{Derivation of the Nonlinear $\sigma$ Model in the Keldysh Formalism}\label{sec:FNLSM}

\subsection{Keldysh path integral}
In this section, we give the derivation of the Finkel'stein nonlinear sigma model (FNL$\sigma$M) in the Keldysh formalism for various universality classes.
We consider a system of spin-1/2 fermions subject to a disorder potential. 
We also include repulsive short-range density-density interactions with interaction strength $U$ and attractive spin singlet BCS interaction with coupling $W$. 
The starting point is the generating function for the closed Keldysh contour going from $t=-\infty$ to $t=+\infty$ and then back to $t=-\infty$:
\begin{align}\label{eq:ZKsymp}
	Z[V]
	\equiv
	\int 
	\D \bar{\psi}
	\D \psi
	\,
	\exp
	\left\{
	\begin{aligned}
	&\,
	 i
	\intl{\omega,\vex{x},\vex{x'}}
	\bar{\psi}(\omega,\vex{x})
	\;
	\G^{-1}(\omega;\vex{x},\vex{x'})	
	\;
	\psi(\omega,\vex{x'})
	\\&\,
	-
	{\frac{i}{2}}
	\,
	U
	\sum_{a = 1,2}
	\tauh^3_{a,a}
	\intl{t,\vex{x}}
	\left(\bar{\psi}_a \psi_a\right)^2
	+
	i
	\,
	\frac{W}{4}
	\sum_{a = 1,2}
	\tauh^3_{a,a}
	\intl{t,\vex{x}}
	\left( - i \bar{\psi}_a \hat{s}^2 \bar{\psi}_a^\T \right)
	\,
	\left( i \psi^\T_a \hat{s}^2 \psi_a \right)
	\\&\,
	-
	i
	\intl{\omega,\omega',\vex{x}}	
	\left[
	\vcl(\omega - \omega') 
	\,
	\bar{\psi}(\omega) \tauh^3 \psi(\omega') 
	+
	\vq(\omega - \omega') 
	\,
	\bar{\psi}(\omega) \, \psi(\omega')
	\right]
	\end{aligned}
	\right\}.
\end{align}
Here $\psi \rightarrow \psi_{a,s}(t,\vex{x})$ carries Keldysh $a \in \{1,2\}$ and spin $s \in \{\uparrow,\downarrow\}$
labels.
The index $a=1$ ($a=2$) corresponds to the forward (backward) part of time contour.
$\tauh,\hat{s}$ stand for Pauli matrices acting on the Keldysh and spin spaces, respectively. 
$\G$ is the 
noninteracting 	
Green's function defined on the Keldysh contour. 
In the space-time basis, it is given by:
\begin{align}
\begin{aligned}
	i
	\G(t,t';\xb,\xb')
	\equiv
	\begin{bmatrix}
	i \G_{\tim}
	&
	i \G_<
	\\
	i \G_>
	&
	i \G_{\atim}
	\end{bmatrix}
	=
	\begin{bmatrix}
	\,
	\left\langle \tim \, \psi(\xb,t) \, \bar{\psi}(\xb',t')\right\rangle_0
	&
	\,
	-
	\left\langle \bar{\psi} (\xb',t') \, \psi(\xb,t)   \right\rangle_0
	\,
	\\
	\,
	\left\langle \psi(\xb,t) \,  \bar{\psi} (\xb',t')\right\rangle_0
	&
	\,
	\left\langle \atim \, \psi(\xb,t) \, \bar{\psi}(\xb',t') \right\rangle_0
	\,
	\end{bmatrix},
\end{aligned}	
\end{align}
where $\tim$ and $\atim$ are time-ordering and anti-time-ordering operators, respectively.
A detailed review on the Keldysh formalism can be found in \cite{AlexAlex,Kamenev}.
$V$ is a scalar potential that is incorporated so we can compute the density response to an external field. 
Its classical component $\vcl$ is the external electric potential, while the quantum one $\vq$ couples to the density operator. 
The net potential field on the forward (backward) part of time contour $V_1$ ($V_2$) is given by
\begin{align}\label{eq:V1V2}
	V_1
	=\,
	\vcl + \vq,
	\qquad
	V_2
	=\,
	\vcl - \vq.
\end{align}

We further decouple the interactions with Hubbard-Stratonovich (H.-S.) fields $\rho$ and $\Delta$, and obtain
\begin{align}\label{eq:ZKsympDC}
\begin{aligned}
	Z[V]
	=
	\int 
	&\,
	\D \bar{\psi}
	\D \psi
	\D \rho
	|\D \Delta|^2
	\,
	\exp
	\left[
	i \frac{2 }{U}
	\intl{t,\vex{x}}
	\rhoq \rhocl
	+
	i \frac{2 }{W}
	\intl{t,\vex{x}}
	\left(\dq^* \dcl + \dq \dcl^*\right) 
	\right]
	\\
	&\,	
	\times
	\exp
	\left\{
	\begin{aligned}
	&\,
	 i
	\intl{\omega,\vex{x},\vex{x'}}
	\bar{\psi}(\omega,\vex{x})
	\;
	\G^{-1}(\omega;\vex{x},\vex{x'})	
	\;
	\psi(\omega,\vex{x'})
	\\&\,
	-
	i
	\intl{\omega,\omega',\vex{x}}	
	\left[
	\tvcl(\omega - \omega') 
	\,
	\bar{\psi}(\omega) \tauh^3 \psi(\omega') 
	+
	\tvq(\omega - \omega') 
	\,
	\bar{\psi}(\omega) \, \psi(\omega')
	\right]
	\\&\,
	-
	\frac{i}{2}
	\intl{\omega,\omega',\vex{x}}
	\left[
	\begin{aligned}
	&\,
	\phantom{-}\,\,
	\dcl(\omega + \omega') \,
	\bar{\psi}(\omega) \hat{s}^2 \tauh^3 \bar{\psi}^\T(\omega')
	+	
	\dq(\omega + \omega') \,
	\bar{\psi}(\omega) \hat{s}^2 \bar{\psi}^\T(\omega')
	\\&\,
	-
	\dcl^*(\omega + \omega') \, 
	\psi^\T(\omega) \hat{s}^2 \tauh^3 \psi(\omega')
	-
	\dq^*(\omega + \omega') \,
	\psi^\T(\omega) \hat{s}^2 \psi(\omega')
	\end{aligned}
	\right]
	\end{aligned}
	\right\},
\end{aligned}
\end{align}
where $\tilde{V}$ denotes the sum of source field $V$ and H.-S.\ field $\rho$:
\begin{align}
	\tilde{V}_{\mathsf{cl},\mathsf{q}}=\,
	V_{\mathsf{cl},\mathsf{q}}+\rho_{\mathsf{cl},\mathsf{q}}.
\end{align}

\subsection{Keldysh and ``thermal'' rotations}

The inverse of the Green's function can be expressed as
\begin{align}\label{eq:Ginverse}
	\G^{-1}(\omega;\vex{x},\vex{x'})	
	=\, &
	\ulo
	\mf(\omega)
	\;
	\Geta^{-1}(\omega;\vex{x},\vex{x'})	
	\;
	\mf(\omega) 
	\ulo^\dagger
	\tauh^3,
\end{align} 	
where
\begin{align}
	\Geta(\omega)
	\equiv
	\left[\omega + i \eta \tauh^3 - \hat{h}\right]^{-1}
	\!\!\!\!\!\!\!\!,
	\qquad
	\ulo = 
	{\textstyle{\frac{1}{\sqrt{2}}}}
	(\hat{1} + i \tauh^2),
	\qquad
	\mf(\omega)
	=
	\begin{bmatrix}
	1  	& F(\omega) 	\\
	0 	& -1  
	\end{bmatrix}.	
\end{align}
Here $\hat{h}$ refers to the static single particle Hamiltonian. 
$F(\omega)$ is the generalized Fermi distribution function at temperature $T$,
\begin{align}
	F(\omega) \equiv \tanh \left(\frac{ \omega}{2 T}\right).
\end{align}
Notice that $\G_{\eta}$ depends only on the spectrum but not the occupation number; 
its diagonal components are the retarded and advanced noninteracting Green's functions.	
We implement the nonunitary change of variables,
\begin{align}\label{CoV}
\begin{aligned}
	\psi(\omega,\vex{x}) 
	\rightarrow \, 
	\tauh^3 \ulo \mf(\omega) \, \psi(\omega,\vex{x}),
	\qquad
	\bar{\psi}(\omega,\vex{x}) 
	\rightarrow \,
	\bar{\psi}(\omega,\vex{x}) \, \mf(\omega) \ulo^\dagger,
\end{aligned}
\end{align} 
to eliminate the distribution function in the noninteracting part of the fermion action, i.e.,
$-i \int \bar{\psi} \G^{-1} \psi$.
Under this transformation, the generating function in Eq.~(\ref{eq:ZKsympDC}) becomes
\begin{align}\label{eq:ZTrans-symp}
\begin{aligned}
	Z[V]
	=\,
	\int 
	\D \bar{\psi}
	\D \psi
	\D \rho
	|\D \Delta|^2
	&\,
	\exp
	\left[
	i \frac{2 }{U}
	\intl{t,\vex{x}}
	\rhoq \rhocl
	+
	i \frac{2 }{W}
	\intl{t,\vex{x}}
	\left(\dq^* \dcl + \dq \dcl^*\right) 
	\right]
	\\
	\times
	&\,
	\exp
	\left\{
	\begin{aligned}
	&\,
	i
	\intl{\omega,\vex{x},\vex{x'}}
	\bar{\psi}(\omega,\vex{x})
	\Geta^{-1}(\omega;\vex{x},\vex{x'})	
	\psi(\omega,\vex{x'})
	\\&\,
	-
	i
	\intl{\omega,\omega',\vex{x}}	
	\left[
	\begin{aligned}
	&\,
	\tvcl(\omega - \omega') 
	\,
	\bar{\psi}(\omega) 
	\,
	\mf(\omega)	 \mf(\omega') 
	\,
	\psi(\omega')
	\\
	&\,
	+
	\tvq(\omega - \omega') 
	\,
	\bar{\psi}(\omega) 
	\,
	\mf(\omega) \tauh^1 \mf(\omega')
	\,
	\psi(\omega')
	\end{aligned}	
	\right]
	\\&\,
	-
	\frac{i}{2}
	\intl{\omega,\omega',\vex{x}}
	\left[
	\begin{aligned}
	&\,
	\phantom{-}\,\,
	\dcl(\omega + \omega') \,
	\bar{\psi}(\omega) \hat{s}^2		
	\mf(\omega) \tauh^1 \mf^\T(\omega') \, 
	\bar{\psi}^\T(\omega')
	\\&\,
	+	
	\dq(\omega + \omega') \,
	\bar{\psi}(\omega) \hat{s}^2 		
	\mf(\omega) \mf^\T(\omega') 
	\, \bar{\psi}^\T(\omega')
	\\&\,
	-
	\dcl^*(\omega + \omega') \, 
	\psi^\T(\omega) \hat{s}^2 		
	\mf^\T(\omega) \tauh^1  \mf(\omega')
	\, \psi(\omega')
	\\&\,
	-
	\dq^*(\omega + \omega') \,
	\psi^\T(\omega) \hat{s}^2 	
	\mf^\T(\omega) \mf(\omega')
	\, \psi(\omega')
	\end{aligned}
	\right]
	\end{aligned}
	\right\}.
\end{aligned}
\end{align}
The distribution function now appears only in the external and H.-S.\ potentials. 
This is physically and mathematically desirable, since the noninteracting, unperturbed
theory encodes only the problem of single-particle wave function localization, which is independent 
of mode occupation numbers or temperature.

\subsection{Keldysh action for a static Bogoliubov-de Gennes Hamiltonian}

Eq.~(\ref{eq:ZTrans-symp}) can also be used as the starting point for the study of 
unpaired quasiparticles in a BCS superconductor.
At the level of static mean field theory, we have
\begin{align}\label{eq:BCS-MFT}
\begin{aligned}
	\dcl(\omega + \omega') =\, i \dcl^{\pup{0}} \delta_{\omega + \omega',0},
	\qquad	
	\dcl^*(\omega + \omega') =\, i (\dcl^{\pup{0}})^* \delta_{\omega + \omega',0},
	\qquad
	\dq(\omega + \omega') =\, 0,
	\qquad	
	\dq^*(\omega + \omega') =\, 0. 
\end{aligned}
\end{align}
and the generating function $Z[V]$ [see Eq.~(\ref{eq:ZTrans-symp})] acquires the form
\begin{align}\label{eq:ZTrans-BdG}
\begin{aligned}
	Z[V]
	=\,
	\int 
	\D \bar{\psi}
	\D \psi
	\D \rho
	&\,
	\exp
	\left\{
	\begin{aligned}
	\,&
	i
	\intl{\omega,\vex{x},\vex{x'}}
	\bar{\psi}(\omega,\vex{x})
	\Geta^{-1}(\omega;\vex{x},\vex{x'})	
	\psi(\omega,\vex{x'})
	+
	i \frac{2 }{U}
	\intl{t,\vex{x}}
	\rhoq \rhocl
	\\
	\,&
	+
	\frac{i}{2}
	\intl{\omega,\vex{x}}
	\left[
	\dcl^{\pup{0}}
	\bar{\psi}(\omega) i \hat{s}^2 \tauh^1 \, \bar{\psi}^\T(-\omega)
	+
	\left(\dcl^{\pup{0}}\right)^*
	\psi^\T(-\omega) (-i) \hat{s}^2 \tauh^1 \, \psi(\omega)
	\right] 
	\\
	\,&
	-
	i
	\intl{\omega,\omega',\vex{x}}	
	\left[ 
	\begin{aligned}
	\,&
	\tvcl(\omega - \omega') 
	\,
	\bar{\psi}(\omega) 
	\,
	\mf(\omega)	 \mf(\omega') 
	\,
	\psi(\omega')
	\\
	\, + &	
	\tvq(\omega - \omega') 
	\,
	\bar{\psi}(\omega) 
	\,
	\mf(\omega) \tauh^1 \mf(\omega')
	\,
	\psi(\omega')
	\end{aligned}
	\right] 
	\end{aligned}
	\right\}.
\end{aligned}
\end{align}
Here we have exploited the following identity
\begin{align}\label{eq:mfid}
	\tauh^1 \mf^\T(-\omega)  \tauh^1 
	=\,
	 - \mf(\omega),
	\qquad
	\mf^{-1}(\omega)
	 =\,
	\mf(\omega).
\end{align}

\subsection{Majorana spinor reformulation}

It is useful to introduce the Majorana spinors 
\begin{align}\label{eq:chiDef}
	\chi
	\equiv
	\begin{bmatrix}
	\psi
	\\
	\hat{s}^2 \tauh^1 \Sigh^1
	\bar{\psi}^\T
	\end{bmatrix},
	\qquad
	\bar{\chi} 
	=
	\begin{bmatrix}
	\bar{\psi}
	&
	- 
	\psi^\T \hat{s}^2 \tauh^1 \Sigh^1
	\end{bmatrix}.	
\end{align}
which carry indices in 
particle-hole ($\sigma$), 
Keldysh ($\tau$), 
and 
spin ($s$) 
spaces. 
In addition, we view $\chi$ and $\bar{\chi}$ as having a continuous index $|\omega|$ 
that ranges over the positive real axis, and a discrete sign index 
$ \Sigma \equiv \sgn(\omega) \in \pm$. The Pauli matrix $\Sigh^1$ is an inversion operator on frequency space:
\begin{align}
	\bra{\omega} \Sigh^1 \ket{\omega'} = 2 \pi \delta(\omega + \omega').
\end{align}
$\chi$ and $\bar{\chi}$ are not independent of each other but are related by
\begin{align}\label{eq:chiMajorana}
	\bar{\chi} 
	=\,
	- \chi^\T 	
	\hat{s}^2 \sigh^1 \tauh^1 \Sigh^1,	
\end{align}
where $\sigh$ indicates a Pauli matrix in the particle-hole space. 

Using Eq.~(\ref{eq:mfid}), the generating function $Z[V]$ [Eq.~(\ref{eq:ZTrans-symp})] can be rewritten as
\begin{align}\label{eq:ZTrans-sympchi}
\begin{aligned}
	Z[V]
	=
	\int 
	\D \chi
	\D \rho
	|\D \Delta|^2
	&\,
	\exp
	\left[
	i \frac{2 }{U}
	\intl{t,\vex{x}}
	\rhoq \rhocl
	+
	i \frac{2 }{W}
	\intl{t,\vex{x}}
	\left(\dq^* \dcl + \dq \dcl^*\right) 
	\right]
	\\
	\times & \,
	\exp
	\left\{
	\frac{i}{2}
	\bar{\chi}
	\left[
	\sigh^3
	\,
	\hat{\omega} 
	+
	i \eta \tauh^3 \sigh^3
	-
	\sigh^3\hat{h}_{\mathsf{BdG}}
	-
	\hat{\mathcal{V}}
	-
	\HD 
	\right]
	\chi
	\right\}
	\!\!,\!\!\!\!\!\!\!\!
	\end{aligned}
	\end{align}
where  $\hat{h}_{\mathsf{BdG}}$ takes the form
\begin{align}\label{eq:hBdG}
	\hat{h}_{\mathsf{BdG}}
	=\, &
	\begin{bmatrix}
	\hat{h}   &    -i \dcl^{\pup{0}}
	\\
	i \left(\dcl^{\pup{0}}\right)^*    & -\hat{s}^2 \hat{h}^T \hat{s}^2
	\end{bmatrix}_{\sig},
\end{align}
and the kernels $\hat{\mathcal{V}}$ and $\HD$ are defined as
\begin{align}\label{eq:kernelVD}
\begin{aligned}
	\hat{\mathcal{V}}_{\omega,\omega'}(\vex{x})
	=\, &
	\tvcl(\omega - \omega',\vexh{x}) 
	\,
	\mf(\omega) \mf(\omega')
	+
	\tvq(\omega - \omega',\vexh{x}) 
	\,
	\mf(\omega) \tauh^1 \mf(\omega'),
	\\
	\HD_{\omega,\omega'}(\vex{x})
	=\,&
	-
	\left[
	\sigh^+ \dcl(\omega - \omega',\vex{x})  
	+
	\sigh^- \dcl^*(-\omega + \omega',\vex{x}) 
	\right]
	\,
	\mf(\omega)  \mf(\omega')
	\\
	&- 
	\left[
	\sigh^+ \dq(\omega - \omega',\vex{x})  
	+
	\sigh^- \dq^*(-\omega + \omega',\vex{x}) 
	\right]
	\,
	\mf(\omega) \tauh^1 \mf(\omega').
\end{aligned}
\end{align}
Here $\sigh^{\pm}$ denotes $(\sigh^1 \pm i \sigh^2)/2$.

\subsection{Target manifold}\label{sec:manifold}

Next, we follow the standard route to derive the Finkel'stein nonlinear sigma model starting from Eq.~(\ref{eq:ZTrans-sympchi}). 
To begin with, we want to identify the nonlinear sigma model target manifold for various symmetry classes.

\subsubsection{Class AI}

As an example, we first consider the time-reversal and spin-rotational invariant orthogonal metal (AI) class. For this class, $\hat{h}_{\mathsf{BdG}}$ satisfies the following conditions:
\bsub
\label{eq:hsym}
\begin{align}
	- \hat{s}^2 \sigh^2 \hat{h}_{\mathsf{BdG}}^\T \hat{s}^2 \sigh^2
	=\,&
	\hat{h}_{\mathsf{BdG}},
	&&\text{``Majorana'' condition (automatic particle-hole symmetry)},
	\\
	\hat{s}^2 \sigh^3 \hat{h}_{\mathsf{BdG}}^\T \hat{s}^2 \sigh^3
	=\,&
	\hat{h}_{\mathsf{BdG}},
	&&\text{time-reversal invariance},
	\\
	\hat{s}^i \hat{h}_{\mathsf{BdG}} \hat{s}^i 
	=\,&
	\hat{h}_{\mathsf{BdG}},		
	&&\text{spin SU(2) invariance},
	\\
	\sigh^3 \hat{h}_{\mathsf{BdG}}  \sigh^3
	=\,&
	\hat{h}_{\mathsf{BdG}},
	&&\text{electric charge U(1) invariance}.
\end{align}
\esub
Here Eq.~(\ref{eq:hsym}a) is true in all cases
[due to Eq.~(\ref{eq:chiMajorana})], whereas Eqs.~(\ref{eq:hsym}b) and (\ref{eq:hsym}c) arise from the time-reversal and spin-rotational invariance, respectively.  
Moreover, Eq.~(\ref{eq:hsym}d) corresponds to the electric charge conservation, i.e.\ $\dcl^{\pup{0}} = 0$ in Eq.~(\ref{eq:hBdG}).
Since the particle-hole condition in Eq.~(\ref{eq:hsym}a) is ``automatic'' (i.e.\ merely a consequence of Pauli exclusion), 
we can combine it with Eq.~(\ref{eq:hsym}b) to obtain an equivalent, alternative ``chiral'' version of time-reversal
symmetry: 
\begin{align}
	-\sigh^1 \hat{h}_{\mathsf{BdG}}  \sigh^1 =  \hat{h}_{\mathsf{BdG}},
	\qquad
	\text{``chiral'' form of time-reversal invariance}.
\end{align}		

We want to find the unitary transformation $\chi \rightarrow \hat{U} \chi$ under which Hamiltonian part of action
\begin{align}\label{eq:ShDef}
	S_{h} \equiv \frac{i}{2} \bar{\chi} \, \sigh^3 \, \hat{h}_{\mathsf{BdG}} \, \chi
\end{align}
remains invariant. This requires
\begin{align}
	\hat{U}^\T
	\hat{s}^2 
	\sigh^2
	\tauh^1
	\Sigh^1
	\hat{h}_{\mathsf{BdG}}
	\hat{U}
	=\,
	\hat{s}^2 
	\sigh^2
	\tauh^1
	\Sigh^1
	\hat{h}_{\mathsf{BdG}}.
\end{align}
Taking into account the conditions imposed on $\hat{h}_{\mathsf{BdG}}$ [Eq.~(\ref{eq:hsym})], we find
\begin{align}\label{eq:AI-U1}
	\hat{U}^\T
	\hat{s}^2
	\tauh^1
	\sigh^1
	\Sigh^1
	\hat{U}
	=\,&
	\hat{s}^2
	\tauh^1
	\sigh^1
	\Sigh^1.
\end{align}
This implies that $\hat{U} \in $ Sp(16$N$), where $N$ is the total number of absolute frequencies. Only a subgroup of transformations leaves the infinitesimal part of the action
$S_{\eta}= \frac{\eta}{2} \bar{\chi} \tauh^3 \sigh^3 \chi$ 
invariant. Besides Eq.~(\ref{eq:AI-U1}), they are subject to 
\begin{align}\label{eq:AI-U2}
	\hat{U}^\dagger
	\tauh^3 \sigh^3
	\hat{U}
	= \,&
	\tauh^3 \sigh^3,
\end{align}
and as a result belong to the group Sp(8$N$) $\times$ Sp(8$N$). 
The target manifold for the orthogonal class sigma model is therefore Sp(16$N$)/$\left[\text{Sp}(8N) \times \text{Sp}(8N)\right]$. 
See Ref.~\cite{AndersonRevC} for an enumeration of (noninteracting) sigma model target manifolds in the 10-fold way.

\subsubsection{Class AII}

If we introduce the spin-orbit scattering, the spin-rotational invariance is broken but the time-reversal symmetry is preserved, and we arrive at the symplectic metal (AII) class. 
In this case, $\hat{h}_{\mathsf{BdG}}$ no longer obeys the condition in Eq.~(\ref{eq:hsym}c).  Eq.~(\ref{eq:AI-U1}) which gives the symmetry of $S_h$ now becomes
\begin{align}\label{eq:AII-U1}
	\hat{U}^\T
	\tauh^1
	\sigh^1
	\Sigh^1
	\hat{U}
	=\,
	\tauh^1
	\sigh^1
	\Sigh^1,
	\qquad
	\hat{s}^i
	\hat{U}
	\hat{s}^i
	=\,
	\hat{U},
\end{align}
while Eq.~(\ref{eq:AI-U2}) defining the symmetry-breaking subgroup remains the same. 
Unlike the orthogonal class, here the transformation matrix $\hat{U}$ does not act on the spin space. One can then easily deduce that target manifold of the AII class is 
O(8$N$)/$\left[\text{O}(4N) \times \text{O}(4N)\right]$.

\subsubsection{Class A with spin SU(2) invariance}

Now we turn to the unitary metal (A) class with spin SU(2) invariance. 
The time-reversal symmetry is broken, and the associated condition in Eq.~(\ref{eq:hsym}b) is no longer imposed on $\hat{h}_{\mathsf{BdG}}$. The set of transformations that preserves the action $S_h$
satisfies
\begin{align}\label{eq:A-U1}
	\hat{U}^\T
	\hat{s}^2
	\tauh^1
	\sigh^1
	\Sigh^1
	\hat{U}
	=\,
	\hat{s}^2                
	\tauh^1
	\sigh^1
	\Sigh^1,
	\qquad
	\sigh^3
	\hat{U}
	\sigh^3
	=\,
	\hat{U},
\end{align}
while imposing invariance of $S_\eta$, in addition, gives Eq.~(\ref{eq:AI-U2}).	
The two independent conditions in Eq.~(\ref{eq:A-U1}) can be solved via the particle-hole space decomposition
\begin{align}\label{eq:A-U}
	\hat{U}
	=\,
	\begin{bmatrix}
	\hat{U}_1 & 0 \\
	0 & \hat{s}^2 \tauh^1 \Sigh^1 \hat{U}_1^* \hat{s}^2 \tauh^1 \Sigh^1
	\end{bmatrix}_{\sig},
	\qquad
	\hat{U}_1 \in \text{ U(8$N$)}.
\end{align}
Therefore, the unitary metal with spin SU(2) invariance possesses the sigma model target manifold \linebreak 
U(8$N$)/$\left[\mathrm{U}(4N) \times \mathrm{U}(4N)\right]$.

\subsubsection{Class C}

Our final example 
consists of gapless quasiparticles in 
the class C superconductor \cite{CMomo,CLuca}, which has broken time reversal symmetry and preserved spin-rotational invariance. It can be considered as a descendant of class A with spin SU(2) symmetry, after relinquishing charge U(1) symmetry. Now $\hat{h}_{\mathsf{BdG}}$ only follows conditions in Eqs.~(\ref{eq:hsym}a) and (\ref{eq:hsym}c). The invariance of the action $S_h$ [Eq.~(\ref{eq:ShDef})]  requires
\begin{align}\label{eq:C-U1}
	\hat{U}^\T
	\hat{s}^2
	\tauh^1
	\Sigh^1
	\hat{U}
	=\,
	\hat{s}^2
	\tauh^1
	\Sigh^1,
	\qquad
	\sigh^i
	\hat{U}
	\sigh^i
	=\,
	\hat{U}.
\end{align}
Here the second equation means the solution does not act on the particle-hole space.
The invariance of $S_{\eta}$ further restricts
\begin{align}\label{eq:C-U2}
	\hat{U}^\dagger
	\tauh^3
	\hat{U}
	= \,&
	\tauh^3,
\end{align}
which can be solved by the decomposition in the Keldysh space:
\begin{align}
\begin{aligned}
	\hat{U}
	=
	\begin{bmatrix}
	\hat{U}_1 & 0 \\
	0 &  \hat{s}^2 \Sigh^1  \hat{U}_1^* \hat{s}^2 \Sigh^1 
	\end{bmatrix}_{\tau},
	\qquad
	\hat{U}_1 \in \text{ U(4$N$)}.
\end{aligned}
\end{align}
The target manifold is therefore Sp(8$N$)/U(4$N$) (c.f.\ \cite{AndersonRevC}).

\subsection{Hamiltonian description for non-standard classes}

Below we will obtain the Keldysh FNL$\sigma$M for the non-standard class C
as a formal ``descendant'' of the orthogonal metal class AI model. 
This is possible because class AI has more symmetry than class C, namely time-reversal
invariance and electric charge conservation. Suppressing these symmetries
makes massive some of the quantum diffusion modes in the parent class,
immediately determining the structure of the lower symmetry sigma model \cite{CLuca}.
In Sec.~\ref{sec:CI}, we analyze the class CI model that restores time-reversal symmetry;
the FNL$\sigma$M is obtained from class AI in the same way. 
It is however instructive to provide ``microscopic'' Hamiltonians
for these non-standard class systems, in order to ground the interpretation
of the interaction channels. 

A class C system can be realized in principle in a type II
$s$-wave superconductor, driven into the quasi-2D Abrikosov vortex lattice
phase via a perpendicular magnetic field $B$ with $H_{c1} < B < H_{c2}$
\cite{CRealization2}. Here $H_{c\{1,2\}}$ denote the lower and upper critical field strengths. 
The idea is that for $H_{c1} \lesssim B \ll H_{c2}$, the density of vortices
is very low and the system is a spin and thermal insulator, with localized bound state
quasiparticles residing in the vortex cores. Note that to obtain class C,
it is necessary to neglect the Zeeman coupling to spin.
By increasing the orbital field strength, the vortex density becomes higher,
enabling hopping between isolated vortices. In the presence of nonmagnetic disorder,
the vortex positions will deviate from a perfect lattice, forming a pinned ``vortex glass.'' 
This system can be gapless, i.e.\ possess quasiparticle states at the Fermi energy \cite{CRealization2}.
These gapless quasiparticles could undergo an Anderson insulator-metal transition 
as a function of increasing $B < H_{c2}$. Because class C localizes without interactions in two dimensions,
the metallic phase is in fact only possible in 2D due to the delocalizing effect of the Altshuler-Aronov (AA)
correction, see Eq.~(\ref{eqC}). The AA correction arises due to residual quasiparticle interactions
mediated by spin exchange scattering \cite{CMomo,CLuca}.

The Hamiltonian incorporating disorder, mean-field superconductivity, an external magnetic field, and 
electron-electron interactions is given by \cite{CRealization2,momothesis}
\bsub
\begin{align}
	H^\pup{C}
	\equiv&\,
	H^\pup{C}_D
	+
	H^\pup{C}_I,
\\
	H^\pup{C}_D
	=&\,
	\intl{\vex{x}}
	\left[
	\begin{aligned}
	&\,
		\psi^\dagger_{s}(\vex{x})
		\left\{
			\frac{1}{2m}
			\left[ - i \Nabla - \frac{e}{c} \vex{A}(\vex{x}) \right]^2
			-
			E_F
			+
			u(\vex{x})
		\right\}
		\psi_s(\vex{x})
	\\&\,
	+
	\Delta(\vex{x})
	\,
	\psi^\dagger_\uparrow(\vex{x})
	\,
	\psi^\dagger_\downarrow(\vex{x})
	+
	\Delta^*(\vex{x}) 
	\,
	\psi_\downarrow(\vex{x})
	\,
	\psi_\uparrow(\vex{x})
	\end{aligned}
	\right],
\label{HCD}
\\
	H^\pup{C}_I
	=&\,
	\intl{\vex{x}}
	\left[
	U_\rho
	\,
	\rho^2(\vex{x}) 
	+
	U_S
	\,
	\vex{S}(\vex{x}) \, \cdot \vex{S}(\vex{x})
	+
	U_\Delta
	\,
	\big(
	\psi^\dagger_\uparrow
	\,
	\psi^\dagger_\downarrow
	\big)(\vex{x})
	\big(
	\psi_\downarrow
	\psi_\uparrow
	\big)(\vex{x})
	\right].
\label{HCI}
\end{align}	
\esub
In Eq.~(\ref{HCD}), $\psi_{s}(\vex{x})$ annihilates an electron with spin $s \in \{\uparrow,\downarrow\}$
(and the repeated index is summed). This term incorporates 
the static magnetic field via $B = \Nabla \times A(\vex{x})$,
quenched disorder via the potential $u(\vex{x})$, 
and the inhomogeneous mean-field pairing potential $\Delta(\vex{x})$.
The latter must be self-consistently determined in the presence of $B$ and $u(\vex{x})$. 

The interactions in Eq.~(\ref{HCI}) are the three channels that generically arise for a finite density 
spin-1/2 electron system. All are four-fermion interactions, where the electric charge density $\rho$
and spin density $\vex{S}$ are defined via
\begin{align}
	\rho = \psi^\dagger_s \psi_s, 
\quad 
	\vex{S} = \psi^\dagger_{s_1} \hat{\vex{s}}_{s_1,s_2} \psi_{s_2}, 
\end{align}
and $\hat{\vex{s}}$ is the vector of Pauli matrices acting on the physical spin. 
The interactions are charge-charge ($U_\rho$), spin exchange ($U_S$), and 
residual pairing ($U_\Delta$). Long-range Coulomb interactions are assumed
to be screened by the condensate, so that $U_\rho$ incorporates only the
short-range component. 

As written, all three interaction terms in Eq.~(\ref{HCI}) 
are in fact equivalent due to the Pauli principle, i.e.\ there is only one local product of four independent fermion 
fields. However, equation (\ref{HCI}) should be interpreted differently: it is a short-hand notation for interactions that should be 
defined along the Fermi surface in the unpaired system, and then projected into the low-energy effective theory for
the gapless quasiparticle states that arise in the disordered Abrikosov vortex lattice. 
To derive the \emph{form} of the sigma model, it is not necessary to provide this level of detail.
Symmetry dictates the structure of the allowed interaction terms in the FNL$\sigma$M. 
A microscopic description is necessary only to derive the bare values of the coupling strengths $U_{\rho,S,\Delta}$. 

For a system in class C which possesses only spin SU(2) symmetry in every realization of disorder,
it is straightforward to show that both the charge-charge $U_\rho$ and residual pairing $U_\Delta$ interactions
drop out of the sigma model. This is because charge is not conserved, and time-reversal symmetry is broken. 
Only the spin-spin interaction survives \cite{CMomo,CLuca}. 
The dimensionless interaction parameter $\gamma$ appearing in the AA correction in Eq.~(\ref{eqC}) 
is proportional to $U_S$, and incorporates in addition a Fermi liquid renormalization. 
See Eq.~(\ref{eq:hlbgamDef}).  

By contrast, class CI describes gapless quasiparticles in a superconductor with time-reversal and spin SU(2) 
symmetries. In this case, both $U_S$ and $U_\Delta$ would enter the full Keldysh FNL$\sigma$M \cite{CLuca}, although
we neglect the residual pairing channel to obtain Eq.~(\ref{eqCI}). 
The kinetic term in class CI can also take the form shown in Eq.~(\ref{HCD}), but with $\vex{A}(\vex{x}) = 0$. 
Class CI can describe gapless 2D Dirac quasiparticles in the $d$-wave cuprates, subject to nonmagnetic disorder \cite{AndersonRevC}.
We note however that the derivation of the sigma model from a gapless, disordered Dirac model in two spatial dimensions requires special care,
as the standard self-consistent Born approximation used to obtain the saddle-point configuration for the matrix field
$\hat{q}$ (see below) is known to be invalid \cite{AleinerEfetov06}. A better method exploits the nonabelian bosonization of the clean 
Dirac quasiparticles, and incorporates the disorder into this \cite{ZirnbauerNAB}. The nonabelian bosonization method becomes ``exact''
for surface states of a class CI topological superconductor, where it directly gives the class CI FNL$\sigma$M,
but augmented with a Wess-Zumino-Novikov-Witten term. 
For topological superconductor surface states, the residual pairing interaction $U_\Delta$ can induce spontaneous time-reversal symmetry breaking and surface spin or thermal quantum Hall order. 
See Refs.~\cite{FosterYuzbashyan2012,FXC14,XCF15} for details.

\subsection{Effective $\qh$-matrix field theory}

$\hat{h}_{\msf{BdG}}$ in Eq.~(\ref{eq:ZTrans-sympchi}) can be written as a summation of two terms:
\begin{align}
	\hat{h}_{\msf{BdG}}
	=\,
	\hat{h}_0+u(\vex{x})\hat{\sigma}^3
\end{align}
where $\hat{h}_0$ is the corresponding $\hat{h}_{\msf{BdG}}$ of the clean system. $u(\vex{x})$ indicates the static impurity potential and is assumed to be 
Gaussian 
white-noise
distributed
\begin{align}\label{eq:Vdis}
	P[u]
	=
	\exp 
	\left[ 
	-\pi \nu_0 \tau_{\mathsf{el}}
	\intl{\vex{x}} 
	u^2(\vex{x})
	\right] .
\end{align}
Here $\tau_{\mathsf{el}}$ denotes the elastic scattering time and $\nu_0$ is the density of states per spin species. Although we only consider potential disorder, the results are independent of this assumption.

The disorder-dependent part of the action takes the form 
\begin{align}
	S_{\msf{dis}}
	=\,
	\frac{i}{2}
	\intl{\vex{x}}
	\bar{\chi}(\vex{x})
	u(\vex{x})
	\chi(\vex{x}).
\end{align}
Averaging the disorder part of the generating function $Z[V]$ over the distribution in Eq.~(\ref{eq:Vdis}), we obtain
\begin{align}\label{eq:disavg}
\begin{aligned}
	\braket{e^{-S_{\msf{dis}}}}
	=\,
	\exp
	\left\{
	\frac{1}{16 \pi \nu_0 \tau_{\mathsf{el}}}
	\intl{\vex{x}}
	\tr
	\left[
	\left(
	\chi
	\bar{\chi}
	\right)
	\left(
	\chi
	\bar{\chi}
	\right)
	\right]
	\right\}.
\end{aligned}
\end{align}	
Then the quartic action induced by disorder average is decoupled by the H.-S.\ matrix field $\hat{q}$, 
\begin{align}\label{eq:HSQ}
\begin{aligned}
	\braket{e^{-S_{\msf{dis}}}}
	=&\,
	\int \mathcal{D}\hat{q}
	\exp
	\left\{	
	-\frac{\pi \nu_0 }{8 \tau_{\mathsf{el}} }
	\intl{\vex{x}}
	\tr\left(
	\hat{q}^2
	\right)
	-	
	\frac{1}{4 \tau_{\mathsf{el}} }
	\bar{\chi}
	\,
	\hat{q}
	\,
	\chi
	\right\}.
\end{aligned}
\end{align}
$\hat{q}$ is a Hermitian matrix with indices in particle-hole, spin, Keldysh, and frequency spaces.

After the H.-S.\ transformation, we integrate the disorder-averaged partition function $Z[V]$ over the fermion field $\chi$, and obtain an effective $\qh$-matrix field theory:
\begin{align}\label{eq:ZSymp-Start}
\begin{aligned}
	Z[V]
	=&\,
	\int
	\D \hat{q}
	\D \rho
	|\D \Delta|^2
	\exp(-S),
	\\
	S 
	=&\,
	- i 
	\frac{2}{U}
	\intl{t,\vex{x}}
	\rhoq \rhocl
	- i
	\frac{2}{W}
	\intl{t,\vex{x}}
	\left(\dq^* \dcl + \dq \dcl^*\right) 
	+
	\frac{\pi \nu_0 }{8 \tau_{\mathsf{el}} }
	\intl{\vex{x}}
	\tr\left(
	\hat{q}^2
	\right)
	\\
	&\,
	-
	\frac{1}{2}
	\tr
	\log
	\left[
	\sigh^3
	\,
	\hat{\omega} 
	+
	i \eta \tauh^3 \sigh^3
	-
	\sigh^3 \hat{h}_0
	-
	\hat{\mathcal{V}}
	-
	\HD 
	+
	i 
	\frac{1}{2\tau_{\mathsf{el}}}
	\,
	\hat{q}
	\right].
\end{aligned}
\end{align}
Neglecting the interactions, and varying the action with respect to the matrix $\hat{q}$ yields the saddle-point equation 
\begin{align}
\begin{aligned}
	- 
	i \pi \nu_0
	\,
	\hat{q}
	=
	\intl{\vex{k}}
	\left[
	\sigh^3
	\,
	\hat{\omega} 
	+
	i \eta \tauh^3 \sigh^3
	-
	\sigh^3 \hat{h}_0(\vex{k})
	+
	i 
	\frac{1}{2\tau_{\mathsf{el}}}
	\,
	\hat{q}
	\right]^{-1},
\end{aligned}
\end{align}
whose solution is $\qsp = \tauh^3 \sigh^3 \hat{1}_{s} \hat{1}_{\omega}$, 
determined by the symmetry-breaking $i \eta$ term.	

We then expand the action in terms of the fluctuations around the saddle point. 
The fluctuations of the massive modes are ignored, while the massless mode can be parameterized as
\begin{align}\label{eq:qfluct}
	\hat{q}=\hat{U}^{\dagger} \qsp \hat{U}.
\end{align}
Here $\hat{U}$ belongs to the set of transformations that preserve the symmetry of $S_h$ [Eq.~(\ref{eq:ShDef})], 
and as a result its explicit form depends upon the universality class of the system [see Sec.~\ref{sec:manifold}].

\subsubsection{Class AI}

For class AI, the transformation matrix $\hat{U}$ in Eq.~(\ref{eq:qfluct}) satisfies condition Eq.~(\ref{eq:AI-U1}). 
Using this, and performing the gradient expansion (see e.g.\ \cite{AlexAlex}), 
we arrive at the FNL$\sigma$M of the orthogonal class:
\begin{align}\label{eq:FNLSM_AI}
\begin{aligned}
	Z[V]
	=&\,
	\int
	\D \hat{q}
	\D \rho
	|\D \Delta|^2
	\exp(-S),
	\\
	S
	=&\,
	\frac{1}{8 \lambda}
	\intl{\vex{x}}
	\tr\left[\Nabla \qh \cdot \Nabla \qh\right]
	+
	i
	\frac{h}{2}	
	\intl{\vex{x}}
	\tr\left[\qh(\sigh^3 \hat{\omega} + i \eta \sigh^3 \tauh^3)\right]
	\\&\,
	-\frac{i h}{2}
	\intl{\vex{x}}
	\tr
	\left[
	\left( \tvcl + \tvq \tauh^1 \right) 
	\mf(\hat{\omega}) \hat{q}(\vex{x}) \mf(\hat{\omega})
	\right]
	\\&\,
	+
	\frac{ih}{2}
	\intl{\vex{x}}
	\tr\left[
	\left(
	\dcl \,\sigh^+ 
	+
	\dcl^* \,\sigh^-
	+
	\dq \, 	\sigh^+ \tauh^1 
	+
	\dq^* \,  \sigh^- \tauh^1 
	\right)
	\mf(\hat{\omega}) 
	\hat{q}(\vex{x})
	\mf(\hat{\omega}) 
	\right]	
	\\&\,	
	-i \frac{4 }{\pi} h
	\intl{t,\vex{x}}	
	\tvcl
	\,
	\tvq
    \,
	-i \frac{4}{\pi} h
	\frac{(1-\gamma)}{\gamma}
	\intl{t,\vex{x}}
	\rhoq \rhocl
	-i 
	\frac{2}{W}
	\intl{t,\vex{x}}
	\left(\dq^* \dcl + \dq \dcl^*\right),	
\end{aligned}	
\end{align}
where $\hat{q}$ is subject to the following constraints
\begin{align}\label{eq:qsym_AI}
	\hat{q}^2=\,1,
	\qquad\
	\tr \, \hat{q}=\,0,
	\qquad
	\hat{s}^2 \sigh^1 \tauh^1 \Sigh^1 \qh^\T \hat{s}^2 \sigh^1 \tauh^1 \Sigh^1 = \qh,
\end{align}
deduced from Eqs.~(\ref{eq:qfluct}) and (\ref{eq:AI-U1}).
The sigma model coupling constants $h$, $\lb$ and $\gamma$ are defined 
in terms of bare parameters as		
\begin{align}\label{eq:hlbgamDef}
	h 
	\equiv 
	\frac{\pi (2 \nu_0)}{2},
	\qquad
	\frac{1}{\lambda} 
	\equiv
	D h,
	\qquad
	\gamma
	\equiv 
	\frac{\frac{2}{\pi} h U}{1  + \frac{2}{\pi} h U}.
\end{align}
Here $D$ is the diffusion constant and takes the value $D = v_F^2 \tau_{\mathsf{el}}/2$.
The parameter $\gamma$ is the interaction strength that takes into account Fermi liquid renormalization \cite{FNLSM1983}.	

The first term on the last line of Eq.~(\ref{eq:FNLSM_AI}) obtains from the diagonal (retarded-retarded, advanced-advanced) 
piece of the second-order gradient expansion \cite{AlexAlex}. 
It supplies the charge compressibility to the density polarization function in the static $\omega \rightarrow 0$ limit.	

The FNL$\sigma$M for the other classes mentioned in subsection~(\ref{sec:manifold}) can be derived similarly. However, they can also be deduced directly from Eq.~(\ref{eq:FNLSM_AI}) by restricting the $\qh$-matrix fluctuations relative to the orthogonal case.

\subsubsection{Class AII}

With respect to class AII, the associated rotation matrix $\hat{U}$ does not act on the spin space 
(since the latter is no longer hydrodynamic, due to spin-orbit coupling), 
and is subject to the constraints in Eq.~(\ref{eq:AII-U1}). As a result, one can simplify the problem by parameterizing $\qh$ as $\qh=\qh_1 \otimes \hat{1}_s$ and eliminating the spin space. 
The partition function of the nonlinear sigma model reduces to
\begin{align}\label{eq:FNLSM_AII}
\begin{aligned}
	Z[V]
	=&\,
	\int
	\D \hat{q}_1
	\D \rho
	|\D \Delta|^2
	\exp(-S),
	\\
	S 
	=&\,
	\frac{1}{4 \lambda}
	\intl{\vex{x}}
	\tr\left[\Nabla \qh_1 \cdot \Nabla \qh_1 \right]
	+
	i h
	\intl{\vex{x}}
	\tr\left[\qh_1(\sigh^3 \hat{\omega} + i \eta \sigh^3 \tauh^3)\right]
	\\&\,
	-i h
	\intl{\vex{x}}
	\tr
	\left[
	\left( \tvcl + \tvq \tauh^1 \right) 
	\mf(\hat{\omega}) \hat{q}_1 (\vex{x}) \mf(\hat{\omega})
	\right]
	\\&\,
	+
	i h
	\intl{\vex{x}}
	\tr\left[
	\left(
	\dcl \,\sigh^+ 
	+
	\dcl^* \,\sigh^-
	+
	\dq \, 	\sigh^+ \tauh^1 
	+
	\dq^* \,  \sigh^- \tauh^1 
	\right)
	\mf(\hat{\omega}) 
	\hat{q}_1(\vex{x})
	\mf(\hat{\omega}) 
	\right]	
	\\&\,	
	-i \frac{4 }{\pi} h
	\intl{t,\vex{x}}	
	\tvcl
	\,
	\tvq
	\,
	-i \frac{4}{\pi} h
	\frac{(1-\gamma)}{\gamma}
	\intl{t,\vex{x}}
	\rhoq \rhocl
	+
	\frac{2}{W i}
	\intl{t,\vex{x}}
	\left(\dq^* \dcl + \dq \dcl^*\right).	
\end{aligned}
\end{align}
$\hat{q}_1$ carries indices in particle-hole, Keldysh and frequency spaces, and obeys
	\begin{align}\label{eq:qsym_AII}
	\hat{q}_1^2=\,1,
	\qquad\
	\tr \, \hat{q}_1=\,0,
	\qquad
	\sigh^1 \tauh^1 \Sigh^1 \qh_1^\T \sigh^1 \tauh^1 \Sigh^1 = \qh_1.
\end{align}
The saddle point is $\qsp = \tauh^3 \sigh^3$.

\subsubsection{Class A with spin SU(2) invariance}

The FNL$\sigma$M of the unitary metal with spin SU(2) invariance can be derived in a similar fashion. Given the particular form matrix $\hat{U}$ takes [see Eq.~(\ref{eq:A-U})], $\qh$ is parameterized in the particle-hole space as
\begin{align}
\begin{aligned}
	\qh
	=\,
	\begin{bmatrix}
	\qh_1 & 0 \\
	0 & \hat{s}^2 \tauh^1 \Sigh^1 \qh_1^\T \hat{s}^2 \tauh^1 \Sigh^1
	\end{bmatrix}_{\sig}, 
\end{aligned}
\end{align}
where 
$	\qh_1 \equiv \, \hat{U}_1^{\dagger} \tauh^3 \hat{U}_1 $
is a matrix in spin, Keldysh and frequency spaces.

Using the identity $\Sigh^1 \hat{\omega} \Sigh^1 = - \hat{\omega}$, we arrive at the sigma model action
\begin{align}\label{eq:FNLSM_A}
	S
	=
	\frac{1}{4 \lambda}
	\intl{\vex{x}}
	\tr\left[\Nabla \qh_1 \cdot \Nabla \qh_1 \right]
	+
	i
	h
	\intl{\vex{x}}
	\tr\left[\qh_1(\hat{\omega} + i \eta \tauh^3)\right],
\end{align}
where $\hat{q}_1$ is restricted by
\begin{align}\label{eq:qsym_A}
	\hat{q}_1^2=\,1,
	\qquad\
	\tr \, \hat{q}_1=\,0.
\end{align}
Here for simplicity we have dropped interacting part of the action, which is given by the same expression as that in Eq.~(\ref{eq:FNLSM_AII}) (except for the BCS channel interaction, which vanishes in this case due to broken time reversal symmetry).
The saddle point of this sigma model is $\qsp=\tau^3$.

\subsubsection{Class C}

For the class C superconductor, we discard the dynamical charge density and BCS channel interactions,
but incorporate the spin triplet interactions. Eq.~(\ref{eq:ZTrans-symp}) now becomes
\begin{align}\label{ZTrans-sympchi-C}
\begin{aligned}
	Z[\vex{B}]
	=
	\int 
	\D \chi
	\D b
	&\,
	\exp
	\left\{
	\frac{i}{2}
	\bar{\chi}
	\left[
	\sigh^3
	\,
	\hat{\omega} 
	+
	i \eta \tauh^3 \sigh^3
	-
	\sigh^3\hat{h}_{\mathsf{BdG}}
	-
	\hat{\mathcal{B}}
	\right]
	\chi
	\right\}
	\exp
	\left[
	\frac{2 i}{U}
	\intl{t,\vex{x}}
	\bq^i \bcl^i
	\right]\!\!,\!\!\!\!\!\!\!\!
\end{aligned}
\end{align}
where
\begin{align}\label{kernelB}
\begin{aligned}
	\hat{\mathcal{B}}_{\omega,\omega'}(\vex{x})
	=\, &
	\tBcl^i(\omega - \omega',\vexh{x}) \hat{s}^i \hat{\sigma}^3
	\,
	\mf(\omega) \mf(\omega')
	+
	\tBq^i(\omega - \omega',\vexh{x})  \hat{s}^i \hat{\sigma}^3
	\,
	\mf(\omega) \tauh^1 \mf(\omega').
\end{aligned}
\end{align}
$U$ now stands for the coupling strength of the spin triplet interaction.
We denote the source and H.-S.\ magnetic fields as $B$ and $b$, respectively, and call the combined field $\tilde{B}$. Classical and quantum components of the magnetic field $B$ are defined similarly as the scalar potential $V$ [see Eq.~(\ref{eq:V1V2})]: The classical component $\Bcl$ is an external Zeeman field, and the quantum component $\Bq$ couples to the physical spin density operator. 

Notice that the rotation matrix $\hat{U}$ for this class does not act on particle-hole space [see Eq.~(\ref{eq:C-U1})]. Therefore, we parameterize $\qh$ as $\qh=\qh_1 \otimes \sigh^3$,
where $\qh_1$ is a matrix in spin, Keldysh, and frequency spaces. 
The nonlinear sigma model for class C acquires the form
\begin{align}\label{eq:FNLSM_C}
\begin{aligned}
	& Z[\vex{B}] \,=
	\int \dd \bv \dd \qh_1  \exp (-S),
	\\
	S =\,& 
	\frac{1}{4 \lambda}
	\intl{\vex{x}}
	\tr\left[\Nabla \qh_1 \cdot \Nabla \qh_1 \right]
	+
	i h
	\intl{\vex{x}}
	\tr\left[\qh_1(\hat{\omega} + i \eta \tauh^3)\right]
	\\ 
	& 
	-i h \intl{\vex{x}} \tr 
	\left[
	\left( \tilde{\blv}_{\mathsf{cl}} +\tilde{\blv}_{\mathsf{q}} \hat{\tau}^1\right) 
	\cdot \h{\sv} \mf(\hat{\omega}) \hat{q}_1 (\vex{x}) \mf(\hat{\omega})
	\right] 
	\\
	& 
	-i \frac{4}{\pi} h \intl{t,\vex{x}} \tilde{\blv}_{\mathsf{cl}} \cdot \tilde{\blv}_{\mathsf{q}} 
	-i \frac{4}{\pi} h \frac{(1-\gamma)}{\gamma} \intl{t,\vex{x}} \bv_{\mathsf{cl}} \cdot \bv_{\mathsf{q}}. 
\end{aligned}
\end{align}
Here the reduced matrix $\qh_1$ possesses the saddle point $\qsp=\tauh^3$, and satisfies the conditions 
\begin{align}\label{eq:qsym_C}
	\hat{q}_1^2=\,1,
	\qquad\
	\tr \, \hat{q}_1=\,0,
	\qquad
	-\h{s}^2 \h{\tau}^1 \h{\Sigma}^1 \qh_1^{\T} \h{s}^2 \h{\tau}^1 \h{\Sigma}^1 =\, \qh_1.
\end{align}
We have used the same definitions for $h$, $\lb$ and $\gamma$ as in Eq.~(\ref{eq:hlbgamDef}), although now $U$ is the coupling constant of the spin triplet interaction,
and $D$ denotes the bare spin diffusion constant (in the absence of interactions). 
	
In the next few sections, we work with the FNL$\sigma$Ms derived here and compute the (spin) density response function and conductivity in the disordered class AII metal and class C superconductor.

\section{Class AII}\label{sec:AII}

\subsection{Density linear response function}

The density linear response function is defined as
\begin{align}\label{eq:LinRes}
\begin{aligned}
	\Pi 
	\left( 
	\vex{k} , \ww 
	\right) 
	=
	\frac{\delta n(\kb,\ww) }{\delta \vcl (\kb, \ww)}\bigg|_{\vcl=0}
	=
	 \frac{i}{2} \frac{\delta ^2 Z[V]}{\delta \vcl(\kb,\ww) \delta \vq (-\kb, -\ww)}
	\bigg|_{\vcl=\vq=0},
\end{aligned}
\end{align}
where $n$ is the density averaged over the forward and backward contour copies, and the generating function $Z[V]$ for the symplectic class is given by Eq.~(\ref{eq:FNLSM_AII}). In what follows, we drop the BCS pairing channel interaction as we are only interested in the density linear response and the conductivity.

It is convenient to apply the transformation: $\rho_{\mathsf{cl,q}} \rightarrow \rho_{\mathsf{cl,q}} - V_{\mathsf{cl,q}}$ after which the problem reduces to performing the functional integration
\begin{align}\label{eq:AII_Pi}
\begin{aligned}
	& \Pi \left( \vex{k} , \ww \right) 
	= \, 
	- 
	\frac{2}{\pi} h \frac{(1-\gamma)}{\gamma}
	\int \dd \rho \dd \qh  \, e^{-S_q-S_{c}-S_{\rho} }
	\left[
	1+
	i\frac{4}{\pi} h \frac{(1-\gamma)}{\gamma}
	\rcl (\kb, \ww)  \rql (-\kb, -\ww)
	\right],
\end{aligned}
\end{align}
where 
\bsub
\label{eq:AII_S}
\begin{align}
	S_{q} 
	=&\,
	\frac{1}{4 \lambda}
	\intl{\vex{x}}
	\tr\left[\Nabla \qh \cdot \Nabla \qh \right]
	+
	i h
	\intl{\vex{x}}
	\tr\left[\qh (\sigh^3 \hat{\omega} + i \eta \sigh^3 \tauh^3) \right],
	\\
	S_{c}
	=&\,
	-i h
	\intl{\vex{x}}
	\tr
	\left[
	\left( \rcl + \rql \tauh^1 \right) 
	\mf(\hat{\omega}) \hat{q} (\vex{x}) \mf(\hat{\omega})
	\right],
	\\
	S_{\rho}
	=&\,
	-i \frac{4}{\pi} h
	\frac{1}{\gamma}
	\intl{t,\vex{x}}
	\rhoq \rhocl.
\end{align}
\esub
The physical response function obtains from the classical-quantum (retarded) correlation function of
the hydrodynamic charge density field $\rho$.


\subsection{Parameterization}

To simplify the parameterization of $\hat{q}$ around the saddle point $\h{q}_{\msf{sp}}=\h{\tau}^3 \h{\sigma}^3$, we perform a rotation 
such that $\h{q}_{\msf{sp}} \rightarrow \tauh^3$. 	
This can be achieved by the similarity transformation for $\hat{q}$:
\begin{align}\label{eq:AII_qRot}
	\h{q} \rightarrow \h{R} \h{q} \h{R}^{\dagger},
\end{align}
where 
\begin{align}\label{eq:AII_R}
	 \h{R} 
	 \equiv \, & 
	  \frac{\h{1} + \h{\sigma}^3}{2} +\frac{\h{1} - \h{\sigma}^3}{2} \h{\tau}^1.
\end{align}	
Under this transformation, $S_q$ remains invariant, while $S_{c}$ acquires the form
\begin{align}\label{eq:AII_Scrot}
\begin{aligned} 
	S_{c} 
	= \, &
	-2 ih \int \tr 
	\left[
	\left( \rcl +\rql \hat{\tau}^1 \right) 
	\h{M}_{F}(\h{\ww}) \left( \frac{\h{1} + \h{\sigma}^3}{2} \h{q} \right) \h{M}_{F}(\h{\ww})
	\right].
\end{aligned}
\end{align}
$\h{q}$ is still subject to the first two constraints in Eq.~(\ref{eq:qsym_AII}), but the similarity transformation changes the last condition to
\begin{align}\label{eq:AII_qsym2}
\begin{aligned} 
	\h{\sigma}^1 \h{\Sigma}^1 \h{q}^{\T} \h{\sigma}^1 \h{\Sigma}^1 
	=\, &
	 \qh.
\end{aligned}
\end{align}

We then parameterize $\h{q}$ in the Keldysh space as
\begin{align}\label{eq:AII_qKeldysh}
	\hat{q}
	= \, &
	\begin{bmatrix}
		\sqrt{1-\Wh^\dagger \Wh} & \Wh^\dagger
		\\
		\Wh                    & -\sqrt{1-\Wh \Wh^\dagger }
	\end{bmatrix}
	_{\tau}.
\end{align}
This parameterization resolves the nonlinear constraint $\hat{q}^2=1$. $\Wh$ is a matrix in both the particle-hole and frequency spaces, satisfying the constraint
\begin{align}\label{eq:AII_Wsym1}
\begin{aligned}
	\Wh
	= \, &
	\h{\sigma}^1 \h{\Sigma}^1 (\Wh^{\dagger})^{\T} \h{\sigma}^1 \h{\Sigma}^1. 
\end{aligned}
\end{align}

We introduce unconstrained matrix fields $\Xh$ and $\Yh$ defined as
\begin{align}\label{eq:AII_XY}
\begin{aligned}
	\Xh_{1,2}(\kb) 
	\equiv \, 
	\Wh^{1,1}_{1,2} (\kb), 
	  \qquad
	\Yh_{1,2}(\kb) 
	\equiv \, 
	\Wh^{1,2}_{1,2} (\kb).
\end{aligned}	
\end{align}		
Here superscripts index the particle-hole space, while
the subscripts $\left\lbrace 1,2\right\rbrace $ represent frequencies $\left\lbrace  \ww_1,\ww_2\right\rbrace $. In what follows, we also adopt the notation:
\begin{align}
	F_1 \equiv F(\omega_1) = \tanh\left(\frac{\omega_1}{2 T}\right), 
	\qquad
	\delta_{1,2} \equiv \delta_{\ww_1,\ww_2}.
\end{align}
I.e., the numeric subscripts appearing in these formulas index the frequency. 
Moreover, we use subscript $-1$ to indicate $-\ww_1$.
Using Eq.~(\ref{eq:AII_Wsym1}) and Eq.~(\ref{eq:AII_XY}), $\Wh$ can be parameterized as
\begin{align}
\begin{aligned}
	W_{1,2}
	= \,
	\begin{bmatrix}
		\Xh_{1,2} & \Yh_{1,2}
		\\
		\Yh^{\dagger}_{-2,-1} & \Xh^{\dagger}_{-2,-1}
	\end{bmatrix}_\sigma,
\end{aligned}
\end{align}
in the particle-hole space.

Next, we expand the action $S_q+S_c$ in powers of $\Xh$ and $\Yh$ which are then rescaled by
\begin{align}\label{eq:AII_scale}
\begin{aligned}
	& \Xh \rightarrow \sqrt{\lb} \Xh, 
	\qquad\,\,
	\Yh \rightarrow \sqrt{\lb} \Yh,  	
\end{aligned}
\end{align}
in order to simplify the power counting of the perturbation-theory parameter $\lb$.
Up to quadratic order in $\Xh$ and $\Yh$, the action $S_q+S_c$ contains two parts: $S_{X}^{(2)}$ and $S_{Y}^{(2)}$, depending on the matrix fields $\Xh$ and $\Yh$, respectively,
\bsub
\label{eq:AII_SQ2}
\begin{align}
	S_{X}^{(2)}[\Xh^\dagger,\Xh] 
	= \, &
	 \int 
	\left[ 
	\Xh^{\dagger}_{1,2}(\kb_1)
	M_{2,1;4,3}(\kb_1,\kb_2)
	\Xh_{3,4}(\kb_2)
	+
	J^{\dagger}_{2,1}(\kb) \Xh_{1,2}(\kb)
	+
	J_{2,1}(\kb) \Xh^{\dagger}_{1,2}(\kb)
	\right] ,
	\\
	S_{Y}^{(2)}[\Yh^\dagger,\Yh] 
	= \, &
	\int 
	\Yh^{\dagger}_{1,2}(\kb_1)
	N_{2,1;4,3}(\kb_1,\kb_2)
	\Yh_{3,4}(\kb_2).	
\end{align}
\esub
Here $M$, $N$, $J^{\dagger}$ and $J$ are defined by the following equations,
\bsub \label{eq:AII_MNJ}
\begin{align}
	& \begin{aligned}
		M_{2,1;4,3}(\kb_1,\kb_2) 
		\equiv \,
		&
		\left[ k_1^2 - i h \lb (\ww_1-\ww_2) \right] 
		\delta_{1,4} \delta_{2,3} \delta_{\kb_1,\kb_2}
		\\
		& + i h \lb 
		\left[
		\rcl(\kb_1-\kb_2,\ww_4-\ww_1)
		+
		F_{4} \rql(\kb_1-\kb_2,\ww_4-\ww_1)
		\right]
		\delta_{2,3}
		\\
		& + i  h \lb
		\left[
		- \rcl(\kb_1-\kb_2,\ww_2-\ww_3)
		+
		F_{3} \rql(\kb_1-\kb_2,\ww_2-\ww_3)
		\right]
		\delta_{1,4},
	\end{aligned}	
		\\
	& \begin{aligned}	
		N_{2,1;4,3}(\kb_1,\kb_2) 
		\equiv \, 
		&
		 \left[ k_1^2 + i h \lb  (\ww_1+\ww_2) \right] 
		\delta_{1,4} \delta_{2,3} \delta_{\kb_1,\kb_2}
		\\
		& + i h \lb
		\left[
		\rcl(\kb_1-\kb_2,\ww_4-\ww_1)
		-
		F_{1} \rql(\kb_1-\kb_2,\ww_4-\ww_1)
		\right]
		\delta_{2,3}
		\\
		& + i h \lb 
		\left[
		- \rcl(\kb_1-\kb_2,\ww_2-\ww_3)
		+
		F_{3} \rql(\kb_1-\kb_2,\ww_2-\ww_3)
		\right]
		\delta_{1,4},
	\end{aligned}	
		\\	 
	& \begin{aligned}		
		J^{\dagger}_{2,1}(\kb) 
		\equiv \, &
		2 i h \sqrt{\lb}
		\left[
		(F_{2}-F_{1})\rcl(-\kb,\ww_2-\ww_1)
		+(1-F_{1}F_{2})\rql(-\kb,\ww_2-\ww_1)
		\right] ,
	\end{aligned}		
		\\
	& \begin{aligned}		
		J_{2,1}(\kb) 
		\equiv \, &
		2 i h \sqrt{\lb} \rql(\kb,\ww_2-\ww_1).	
	\end{aligned}				
\end{align}
\esub
($J^{\dagger}$ is actually independent of $J$.) 
We also keep the higher-order terms in the $S_q$ expansion. The cubic term vanishes, whereas the quartic term takes the form 
\begin{align}\label{eq:AII_SQ4}
\begin{aligned}
	 S_q^{(4)}[\Xh^\dagger,\Xh,\Yh^\dagger,\Yh]
	= 
	\int \delta_{\kb_1+\kb_3,\kb_2+\kb_4}
	 \frac{\lb}{4} 
	\,& 
	\left[ 
		\begin{aligned}
		&\,
	 	-(\kb_1 \cdot \kb_3 +\kb_2 \cdot \kb_4) 
	 	+\frac{1}{2}(\kb_1 + \kb_3)\cdot (\kb_2 + \kb_4) 
	 	\\&\,
		+i \frac{h}{2} \lb (\ww_1-\ww_2+\ww_3-\ww_4) 
		\end{aligned}
	 \right] 	
	 \\
	\times& 
	\left[ 
	\begin{aligned}	
	& \Xh_{1,2} (\kb_1) \Xh ^{\dagger}_{2,3}(\kb_2)
	\Xh_{3,4} (\kb_3) \Xh ^{\dagger}_{4,1}(\kb_4)
	\\
	+ &
	\Yh_{1,-2} (\kb_1) \Yh ^{\dagger}_{-2,3}(\kb_2)
	\Yh_{3,-4} (\kb_3) \Yh ^{\dagger}_{-4,1}(\kb_4)
    \\
	+ & 
	2 \Xh_{1,2} (\kb_1) \Xh ^{\dagger}_{2,3}(\kb_2)
	\Yh_{3,-4} (\kb_3) \Yh ^{\dagger}_{-4,1}(\kb_4)
	\\
	+ &
	2 \Xh_{1,2} (\kb_1) \Yh_{3,-2}(-\kb_2)
	\Yh ^{\dagger}_{-4,3} (-\kb_3) \Xh^{\dagger}_{4, 1}(\kb_4)
	\\			
	+ &
	2 \Xh_{1,2} (\kb_1) \Yh_{3,-2}(-\kb_2)
	\Xh ^{\dagger}_{4,3} (-\kb_3) \Yh^{\dagger}_{-4, 1}(\kb_4)	
	\end{aligned}	
	\right] .
\end{aligned}
\end{align}


\subsection{Feynman rules}

In this subsection, we present the Feynman rules for the matrix fields $\Xh$ and $\Yh$. 
Before continuing, note that the rotation matrix $\hat{R}$ [see Eq.~(\ref{eq:AII_R})] is diagonal in the particle-hole space, and thus the transformation in Eq.~(\ref{eq:AII_qRot}) does not mix the diagonal and off-diagonal components of $\Wh$ in this space. Therefore, the diagonal elements of the transformed matrix field $\Wh^{\sigma,\gamma}$, i.e. $\Xh$ and $\Xh^{\dagger}$, represent the 
``diffuson'' 
mode, while the off-diagonal ones $\Yh$ and $\Yh^{\dagger}$
correspond to the 
``Cooperon'' mode \cite{AlexAlex}.

\begin{figure}
\centering
\includegraphics[width=0.7\linewidth]{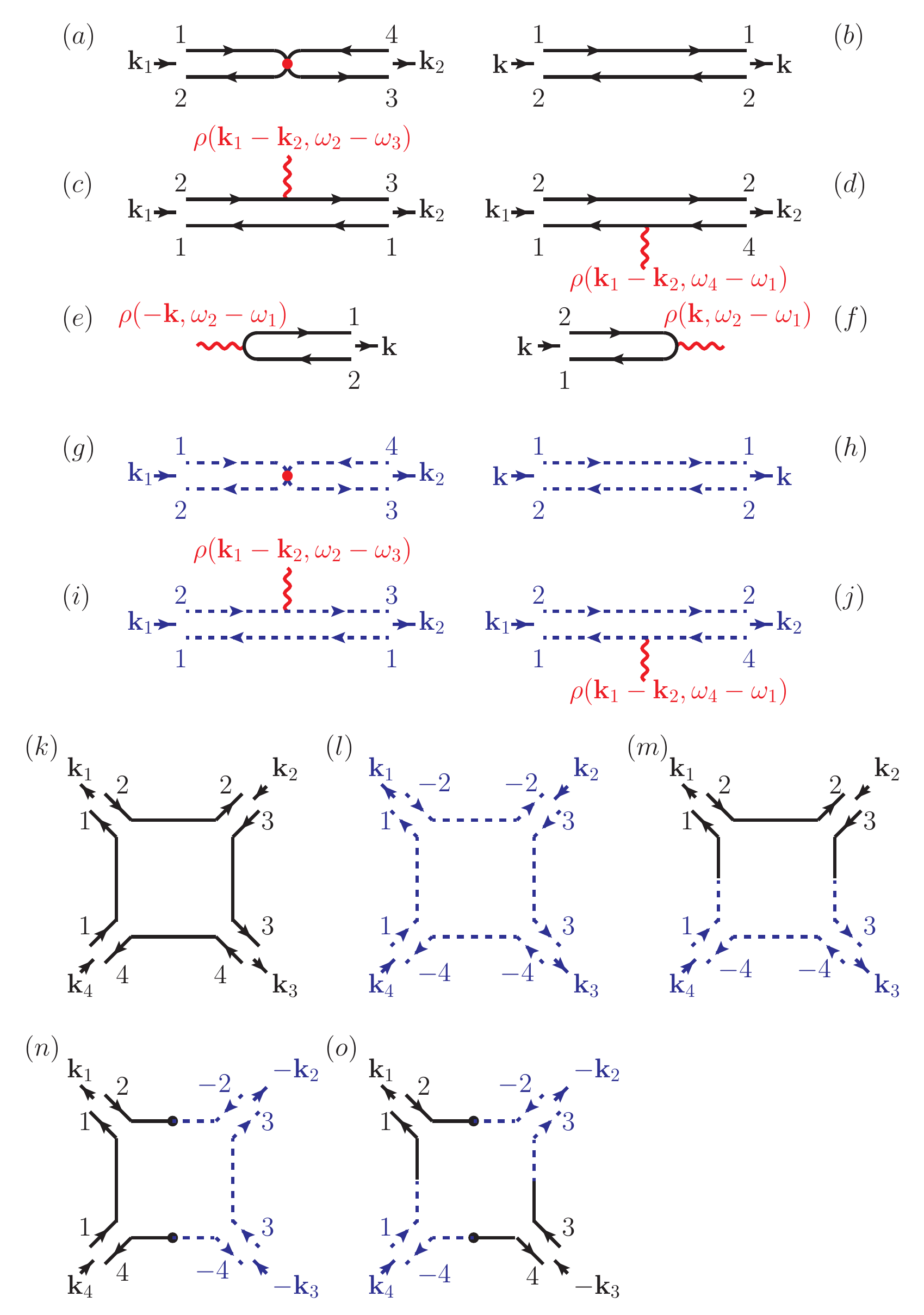}
\caption{(Color online) Feynman rules for class AII: 
	(b) and (h) show the bare propagators for the diffuson [Eq.~(\ref{eq:AII_diffuson})] and the Cooperon [Eq.~(\ref{eq:AII_cooperon})], respectively. 
	Their full propagators, whose expressions are stated in Eq.~(\ref{eq:AII_fullpropagator}), are illustrated in (a) and (g).
	(c)--(f) depict interaction vertices coupling between the matrix field $\Xh$ and the H.-S.\ field $\rho$, while those coupling together $\Yh$ and $\rho$ are pictured in (i) and (j). 
	(k)--(o) show the 4-point diffusion vertices with amplitudes represented by Eq.~(\ref{eq:AII_4PI}).
	In this figure and all other ones in Sec.~\ref{sec:AII},
	the black solid line represents the diffuson $\Xh$, while the blue dashed one corresponds to the Cooperon $\Yh$. 
	The H.-S.\ field $\rho$ is indicated by the red wavy line.
	}
\label{fig:AII_1}
\end{figure}

\subsubsection{Bare propagators}

Using Eqs.~(\ref{eq:AII_SQ2}) and (\ref{eq:AII_MNJ}) and neglecting the interaction terms, we obtain the bare propagators for the diffuson  
\begin{equation}\label{eq:AII_diffuson}
	\braket{\Xh_{1,2}(\kb)\Xh^{\dagger}_{2,1}(\kb)}_0
	=\,
	\Delta_0(k,\ww_1-\ww_2),
\end{equation}
and Cooperon
\begin{equation}\label{eq:AII_cooperon}
	\braket{\Yh_{1,2}(\kb)\Yh^{\dagger}_{2,1}(\kb)}_0
	=\, 
	\Delta_0(k,\ww_1+\ww_2).
\end{equation}
Here we have defined
\begin{equation}\label{eq:Delta0}
	\Delta_0(k,\ww) 
	\equiv \, 
	 \frac{1}{k^2 + i h \lb \ww}.
\end{equation}

In Fig.~\ref{fig:AII_1}(b), the diffuson propagator in Eq.~(\ref{eq:AII_diffuson}) is represented diagrammatically by two black solid lines with arrows pointing in the opposite directions. The numeric labels on the top and the bottom of these lines denote the frequency indices of matrices $\Xh$ and $\Xh^{\dagger}$. The Cooperon propagator in Eq.~(\ref{eq:AII_cooperon}) is depicted in the same manner with blue dashed lines, see Fig.~\ref{fig:AII_1}(h).
The short arrows indicate momentum flow and differentiate $\Xh$ from $\Xh^{\dagger}$:
flow into (out of) the propagator indicates $\Xh$ ($\Xh^{\dagger}$).

\subsubsection{Interaction vertices}

Figs.~\ref{fig:AII_1}(c), \ref{fig:AII_1}(d), \ref{fig:AII_1}(i) and \ref{fig:AII_1}(j) illustrate vertices arising from the interaction terms quadratic in $\Xh$ or $\Yh$ in $S_{X}^{(2)}+S_{Y}^{(2)}$ [see Eqs.~(\ref{eq:AII_SQ2}) and (\ref{eq:AII_MNJ})]. Their amplitudes are given by the following expressions, in respective order,
\begin{align}\label{eq:AII_vertex2}
\begin{aligned}
	(c)=\,&
	-i h \lb
	\left[
	- \rcl(\kb_1-\kb_2,\ww_2-\ww_3)
	+
	F_{3} \rql(\kb_1-\kb_2,\ww_2-\ww_3)
	\right]
	,
	\\
	(d)=\,&
	 -i h \lb
	 \left[
	 \rcl(\kb_1-\kb_2,\ww_4-\ww_1)
	 +
	 F_{4} \rql(\kb_1-\kb_2,\ww_4-\ww_1)
	 \right]
	 ,
	 \\
	 (i)=\,& 
	- i h \lb
	 \left[
	 - \rcl(\kb_1-\kb_2,\ww_2-\ww_3)
	 +
	 F_{3} \rql(\kb_1-\kb_2,\ww_2-\ww_3)
	 \right]
	,
	 \\
	 (j)=\,&
	- i h \lb
	 \left[
	 \rcl(\kb_1-\kb_2,\ww_4-\ww_1)
	 -
	 F_{1} \rql(\kb_1-\kb_2,\ww_4-\ww_1)
	 \right] 
	.
\end{aligned}
\end{align}
Here the H.-S.\ density field $\rho$ is represented by a red wavy line.
In action $S_{X}^{(2)}$, there are also interaction terms linear in the 
diffuson
$\Xh$ field. Figs.~\ref{fig:AII_1}(e) and \ref{fig:AII_1}(f) show the associated vertices whose amplitudes are
\begin{align}\label{eq:AII_vertex1}
\begin{aligned}
	(e)=\,&
	-2 i h \sqrt{\lb}
	\left[
	(F_{2}-F_{1})\rcl(-\kb,\ww_2-\ww_1)
	+(1-F_{1}F_{2})\rql(-\kb,\ww_2-\ww_1)
	\right],
	\\
	(f)=\,&
	-2 i h \sqrt{\lb} \rql(\kb,\ww_2-\ww_1) .
\end{aligned}
\end{align}
It is easy to check that, in all these diagrams, the conservation of momentum and energy holds at each intersection point.

\subsubsection{Full propagators}

If we do not consider the quadratic interaction terms perturbatively, but group them with the diffusion part, we arrive at the (formal) full propagators for the diffuson and Cooperon:
\begin{align}\label{eq:AII_fullpropagator}
\begin{aligned}
	 \braket{\Xh_{1,2}(\kb_1)\Xh^{\dagger}_{3,4}(\kb_2)}
	 =\,& M^{-1}_{1,2;3,4}(\kb_1,\kb_2),
	 \\
	 \braket{\Yh_{1,2}(\kb_1)\Yh^{\dagger}_{3,4}(\kb_2)}
	 =\,& N^{-1}_{1,2;3,4}(\kb_1,\kb_2).
\end{aligned} 
\end{align}
Diagrammatic representation of the full diffuson (Cooperon) propagator is shown in Fig. \ref{fig:AII_1}(a) [Fig. \ref{fig:AII_1}(g)].

\subsubsection{4-point diffusion vertices}

Diagrams in Figs.~\ref{fig:AII_1}(k)--\ref{fig:AII_1}(o) show the 4-point diffusion vertices arising from $S_q^{(4)}$ [see Eq.~(\ref{eq:AII_SQ4})], and each of them gives the identical contribution
\begin{align}\label{eq:AII_4PI}
	-\frac{\lb}{2}
	\left[ 
	-(\kb_1 \cdot \kb_3 +\kb_2 \cdot \kb_4) 
	+\frac{1}{2}(\kb_1 + \kb_3)\cdot (\kb_2 + \kb_4) 
	+i \frac{h}{2} \lb (\ww_1-\ww_2+ \ww_3-\ww_4) 
	\right] \delta_{\kb_1+\kb_3,\kb_2+\kb_4}.
\end{align}
Here the amplitudes of diagrams in Figs.~\ref{fig:AII_1}(k) and \ref{fig:AII_1}(l) have been multiplied by a symmetry factor of 2.


\subsection{Effective response theory for the H.-S.\ field}

\subsubsection{Effective action}

Since the density response function $\Pi (\kb,\ww)$ depends only on the correlator of the H.-S.\ field $\rho$ [see Eq.~(\ref{eq:AII_Pi})], one can integrate out the matrix field $\qh$ to reduce the degrees of freedom. We introduce the effective action $E_\rho$ defined as
\begin{align}\label{eq:AII_E}
\begin{aligned}
	E_{\rho} 
	\equiv \, & 
	S_{\rho}-\ln \left( \int \dd \qh  \, e^{-S_q-S_{c}} \right), 			
\end{aligned}
\end{align}  
and rewrite the partition function as
\begin{align}\label{eq:AII_Z}
\begin{aligned}
	Z
	= \, & 
	\int \dd \rho \,  e^{-E_{\rho}}.
\end{aligned}
\end{align} 

As elaborated in the previous section, after expanding the action in powers of $\Xh$ and $\Yh$, we keep the quadratic terms in both $S_q$ and $S_c$ [Eq.~(\ref{eq:AII_SQ2})], together with the quartic term in $S_q$ [Eq.~(\ref{eq:AII_SQ4})], i.e.,

\begin{align}
\begin{aligned}
	S_q+S_c
	= \, &
		\int 
		\left(  
		\Xh^\dagger M \Xh + J^\dagger \Xh + \Xh^\dagger J
		+\Yh^\dagger N \Yh 
		\right) 
		+ S_{q}^{(4)} [\Xh^\dagger,\Xh,\Yh^\dagger,\Yh],	
\end{aligned}
\end{align}	
where $J^{\dagger}$, $J$, $M$ and $N$ are defined in Eq.~(\ref{eq:AII_MNJ}).
Integrating out $\Xh$ and $\Yh$ matrix fields, we obtain
\begin{align}
		E_{\rho} \approx
		S_{\rho}
		-\int J^\dagger M^{-1} J
		+ \tr \ln M
		+ \tr \ln N
		+
		\braket{ S_{D4}},
\end{align}
where $\braket{ S_{D4} }$ stands for
\begin{align}
\begin{aligned}
	\braket{ S_{D4}}
	\equiv \, &
	\braket{ S_{q}^{(4)} [\Xh^\dagger-J^\dagger M^{-1},\Xh-M^{-1} J,\Yh^\dagger,\Yh] }_{X,Y}
	\\
	\equiv \, & 
	\frac{
			\int \dd \Xh^\dagger \dd \Xh  \dd \Yh^\dagger \dd \Yh
			\,
			\exp \left[ -\int \Xh^\dagger M \Xh -\int \Yh^\dagger N \Yh    \right] 
			\,
			S_{q}^{(4)} [\Xh^\dagger-J^\dagger M^{-1},\Xh-M^{-1} J,\Yh^\dagger,\Yh]	
		}
		{
			 \int \dd \Xh^\dagger \dd \Xh  \dd \Yh^\dagger \dd \Yh
			 \,
			\exp \left[ -\int \Xh^\dagger M \Xh -\int \Yh^\dagger N \Yh \right] 
		}.		
\end{aligned}		
\end{align}
We approximate here 
$\braket{ \exp \left[ {-S_{q}^{(4)}}\right] }_{X,Y} $
with 
$\exp \left[ {-\braket{S_{q}^{(4)}}_{X,Y}}\right] $. 
This is a valid assumption since only the first-order term $\braket{S_{q}^{(4)}}_{X,Y}$ in the cumulant expansion is needed.

Next, we expand the effective action $E_{\rho}$ in terms of the small parameter $\lb$, and find the zeroth-order term $E_0$ acquires the form
\begin{align}\label{eq:AII_S00}
\begin{aligned}
	E_0 
	= &
	 S_\rho  - \int J^\dagger M^{-1}|_{\rho=0} J + \tr \ln M|_{\rho=0} +\tr \ln N|_{\rho=0}.
\end{aligned}
\end{align}
Here $\tr \ln M|_{\rho=0}$ and $\tr \ln N|_{\rho=0}$ are two $\rho$-independent constants whose exact values are unimportant, and as a result are neglected. The 2nd term $-\int J^\dagger M^{-1}|_{\rho=0} J$ is depicted diagrammatically in Fig.~\ref{fig:AII_2}(a). 
Substituting Eqs.~(\ref{eq:AII_S}c) and (\ref{eq:AII_MNJ}) into Eq.~(\ref{eq:AII_S00}),
we find the explicit form of $E_0$:
\begin{align}\label{eq:AII_S0}
\begin{aligned}
	E_0 
	= &
	-i \frac{4}{\pi} h \frac{1}{\gamma} 
	\intl{\kb,\omega}
	 \rql(-\kb,-\ww)\rcl(\kb,\ww)
	\frac{\bd_0 (k, -\ww)}{\bd_u (k, -\ww)}
	\\ &
	-
	(2 i h)^2 \lb
	\intl{\kb,\omega}
	\rql(-\kb, -\ww)\rql(\kb,\ww)
	\frac{\ww}{\pi}
	\coth{\left( \frac{\ww}{2 T}\right) }
	\bd_0 (k, -\ww),	 	
\end{aligned}
\end{align}
where $\bd_u$ is defined as
\begin{align}\label{eq:Deltau}
	\bd_u(k,\ww) \equiv \frac{1}{k^2 + i h (1-\gamma) \lb \ww}.
\end{align}

\subsubsection{Bare propagator}

Using Eq.~(\ref{eq:AII_S0}), we find the bare Green's function of the H.-S.\ field $\rho$ arising from action $E_0$, 
\begin{align}\label{eq:AII_G0}
\begin{aligned}
	\braket{\rho_{a}(\kb,\ww)\rho_{b} (-\kb,-\ww)}_0 = i \bd_{\rho} (\kb,\ww)
	= \, &
	i	
	\begin{bmatrix}
		\bd_{\rho}^{(K)}(\kb,\ww)  & \bd_{\rho}^{(R)} (\kb,\ww)
		\\ 
		\bd_{\rho}^{(A)} (\kb,\ww)  & 0
	\end{bmatrix}.
\end{aligned}
\end{align}
Here $a,b \in \left\lbrace  \msf{cl}, \msf{q} \right\rbrace $ indicate the classical or quantum component. 
The retarded, advanced, Keldysh components are given by,
\begin{align}\label{eq:Deltarho}
\begin{aligned}
	\bd_{\rho}^{(R)} (\kb,\ww)
	=&
	 \frac{\pi \gamma }{4 h} 
	 \frac{\bd_u (k, -\ww)}{\bd_0 (k, -\ww)}, 
	\\
	\bd_{\rho}^{(A)} (\kb,\ww)
	=&
	\frac{\pi \gamma }{4 h} 
	\frac{\bd_u (k, \ww)}{\bd_0 (k, \ww)} 
	=\left[  \bd_{\rho}^{(R)}(\kb,\ww)\right]^*=\bd_{\rho}^{(R)}(\kb,-\ww),
	\\
	\bd_{\rho}^{(K)} (\kb,\ww)	
	= & 
	\left[ \bd_{\rho}^{(R)} (\kb,\ww) -\bd_{\rho}^{(A)} (\kb,\ww) \right]
	\coth{\left( \frac{\ww}{2 T}\right) }.
\end{aligned}
\end{align}
The bare propagator of the H.-S.\ field $\rho$ has the typical form of a bosonic Green's function in the Keldysh formalism, and is represented diagrammatically in the following by a red wavy line with a dot in the middle, see Fig.~\ref{fig:HS}.

\subsubsection{Interaction vertices}

\begin{figure}
	\centering
	\includegraphics[width=0.7\linewidth]{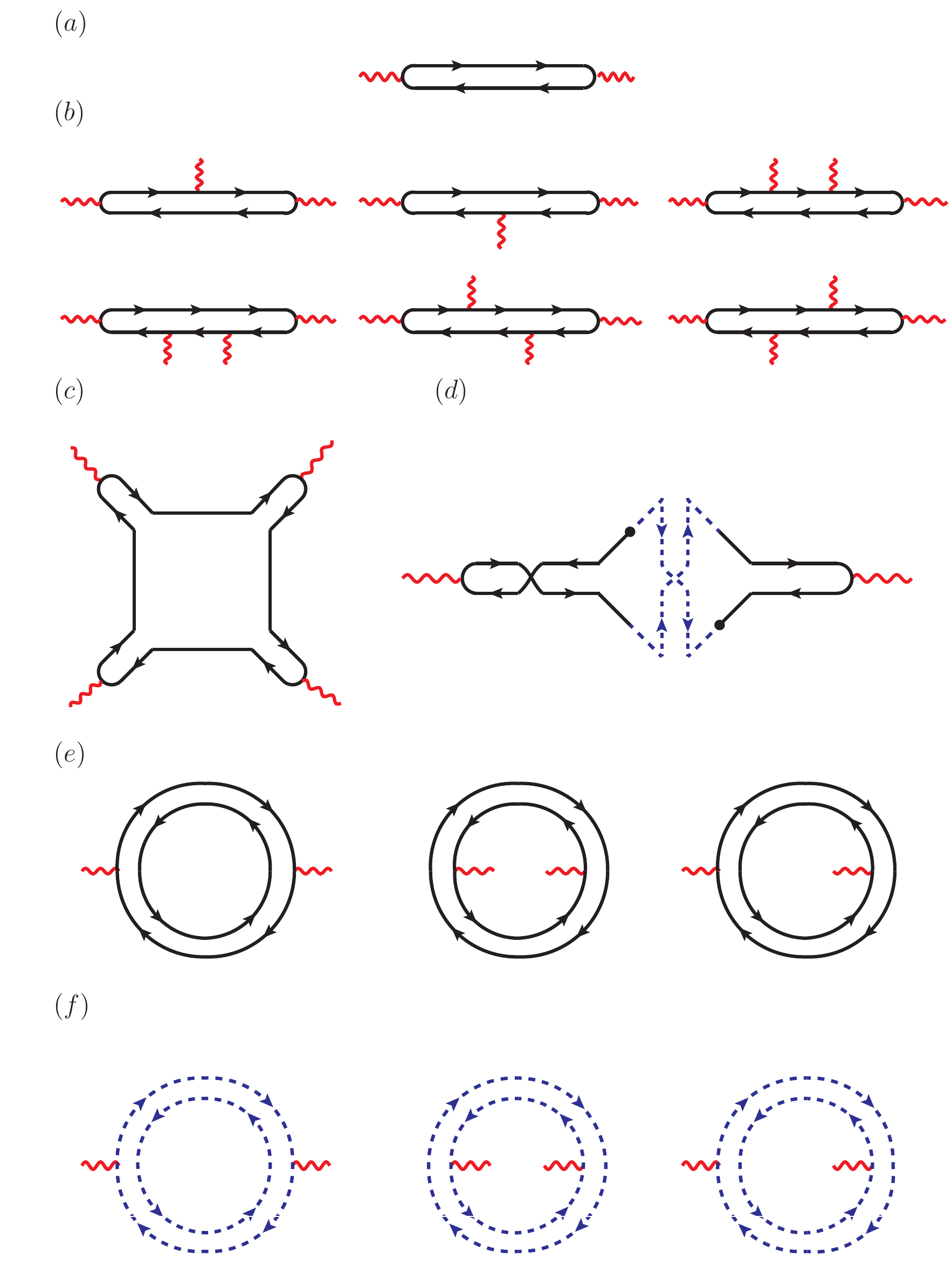}
	\caption{(Color online) Vertices of the H.-S.\ charge density field for class AII.}
	\label{fig:AII_2}
\end{figure}

The remaining part of the effective action $E_{\rho}$ can be considered as interactions and 
encodes quantum corrections to 
the density response function. Vertices from the leading-order interaction terms are shown in Figs.~\ref{fig:AII_2}(b)--\ref{fig:AII_2}(f): Vertices in Figs.~\ref{fig:AII_2}(b), \ref{fig:AII_2}(e) and \ref{fig:AII_2}(f) arise from $-J^{\dagger}M^{-1}J$, $\tr \ln M$ and $\tr \ln N$, respectively; those in Figs.~\ref{fig:AII_2}(c) and \ref{fig:AII_2}(d) are from $\braket{S_{D4}}$. 
Diagrams with a closed Keldysh loop vanish, and are not shown in Fig.~\ref{fig:AII_2}.

\subsubsection{Causality structure of the dressed propagator and self energy}

Before proceeding, we review the general structure of the Green's function and self energy in the Keldysh formalism for the bosonic field $\rho$ \cite{AlexAlex,Kamenev}. The dressed Green's function should have the same structure as the bare one, i.e.,
\begin{align}\label{eq:AII_G}
\begin{aligned}
	 & \braket{\rho_{a}(\kb,\ww)\rho_{b} (-\kb,-\ww)}
	 =\, iG_{\rho} (\kb,\ww)
	 =\,
	i	
	\begin{bmatrix}
	G^{(K)}_{\rho}(\kb,\ww)  & G^{(R)}_{\rho} (\kb,\ww)
	\\ 
	G^{(A)}_{\rho} (\kb,\ww)  & 0
	\end{bmatrix},
\end{aligned}
\end{align}
and also satisfies the condition 
(fluctuation-dissipation theorem)	
\begin{align}\label{eq:FDT}
	G^{(K)}_{\rho}(\kb,\ww)
	= \, &
	\left[  G^{(R)}_{\rho}(\kb,\ww) - G^{(A)}_{\rho}(\kb,\ww) \right] 
	\coth\left(\frac{\ww}{2T}\right).
\end{align}
Here, ``R'', ``A'' and ``K'' denote, respectively, the retarded, advanced and Keldysh components.
The dressed Green's function $G_{\rho}(\kb,\ww)$ can be calculated using 
\begin{align}\label{eq:Dyson}
\begin{aligned}
	G_{\rho}(\kb,\ww)
	=\, &
	\left[ \bd_{\rho}^{-1}(\kb,\ww) - \Sigma_{\rho}(\kb,\ww) \right] ^{-1}.
\end{aligned}
\end{align}
where the self energy $\Sigma_{\rho}(\kb,\ww)$ acquires the following structure
\begin{align}\label{eq:AII_Sigma}
\begin{aligned}
	\Sigma_{\rho} (\kb,\ww)
	=\, &	
	\begin{bmatrix}
		0  & \Sigma_{\rho}^{(A)} (\kb,\ww)
		\\ 
		\Sigma_{\rho}^{(R)} (\kb,\ww)  & \Sigma_{\rho}^{(K)} (\kb,\ww)
	\end{bmatrix}.
\end{aligned}
\end{align}
The Keldysh component of the self energy is related to its retarded and advanced counterparts in the same way as the Green's function, see Eq.~(\ref{eq:FDT}) 
(detailed balance).

\begin{figure}
	\centering
	\includegraphics[width=0.2\linewidth]{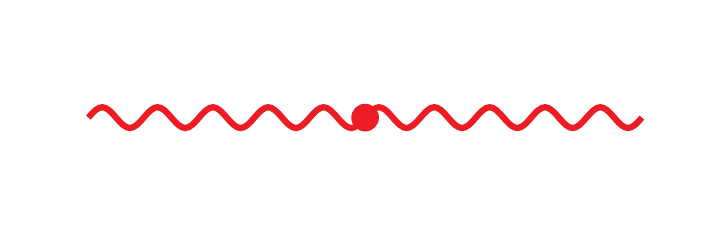}
	\caption{ 
		Propagator of the H.-S.\ field. For class AII, the red wavy line stands for the charge density field $\rho$, while for class C it indicates the spin magnetization density field $\vex{b}$.
		}
	\label{fig:HS}
\end{figure}

\subsubsection{Density response and Green's function}

Following Eq.~(\ref{eq:AII_G}), we see that only the retarded Green's function $G^{(R)}_{\rho} (\kb, \ww)$ enters the calculation of density response function $\Pi \left( \vex{k} , \ww \right)$, and Eq.~(\ref{eq:AII_Pi}) can be rewritten as
\begin{align}\label{eq:AII_Pi0}
\begin{aligned}
	 \Pi \left( \vex{k} , \ww \right) 
	= \, & 
	-\frac{2}{\pi} h \frac{(1-\gamma)}{\gamma}
	\left[
	1-\frac{4}{\pi} h \frac{(1-\gamma)}{\gamma}
	G^{(R)}_{\rho} (\kb, \ww)
	\right].
\end{aligned}
\end{align}

Ignoring the interaction terms in $E_{\rho}$ and approximating the dressed Green's function $G^{(R)}_{\rho} (\kb, \ww)$ here with the bare one $\bd_{\rho}^{(R)} (\kb,\ww)$, we arrive at the semiclassical density response function:
\begin{align}\label{eq:AII_ClassiPi}
\begin{aligned}
	\Pi_0 \left( \vex{k} , \ww \right) 
	= \,&	
	-\frac{2}{\pi}
	\frac{ h (1-\gamma) k^2 }{ k^2 - i h (1-\gamma) \lb \ww}.	
\end{aligned}
\end{align}
This expression can be reduced to a more familiar form using $D = 1/(\lambda h)$ 
[Eq.~(\ref{eq:hlbgamDef})],
\begin{align}\label{eq:AII_ClassiPiF}
\begin{aligned}
	\Pi_0 \left( \vex{k} , \ww \right) =
	-
	\kappa
	\frac{ D_c k^2 }{ D_c k^2 - i \ww},	
\end{aligned}
\end{align}
where $D_c$ the charge diffusion constant and $\kappa$ the charge compressibility are given by
\begin{align}\label{eq:DcKappaDef}
\begin{aligned}
	D_c=\frac{D}{1-\gamma},
	\qquad
	\kappa=\frac{2}{\pi} h (1-\gamma).
\end{aligned}
\end{align}

Once the density response function is known, the conductivity can be calculated through
\begin{align}\label{eq:conduct}
\begin{aligned}
	&\sigma (\omega) = \lim _{k \rightarrow 0 } \frac{i \ww}{k^2} \Pi(\kb, \ww),
\end{aligned}
\end{align}
where the current continuity has been used. 
The semiclassical result in Eq.~(\ref{eq:AII_ClassiPiF}) gives the Drude conductivity
\begin{align}
\begin{aligned}
	&\sigma_0 = \frac{2}{\pi} \frac{1}{\lb} = D (2\nu_0), 
\end{aligned}
\end{align}
where $\nu_0$ is the density of states per spin. 

Eq.~(\ref{eq:AII_Pi0}) implies that the quantum correction to the density linear response is proportional to the difference of the dressed and bare retarded Green's functions
\begin{align}\label{eq:AII_deltapi0}
\begin{aligned}
	\delta \Pi \left( \vex{k} , \ww \right)
	=\, &
	\frac{8}{\pi^2} h^2 \left(  \frac{1-\gamma}{\gamma} \right)^2
	\left[  G^{(R)}_{\rho} (\kb,\ww) - \bd_{\rho}^{(R)}(\kb,\ww) \right].  
\end{aligned}
\end{align}
Employing the Dyson equation [Eq.~(\ref{eq:Dyson})], along with the causality structure of the Green's function [Eq.~(\ref{eq:AII_G})] and self energy [Eq.~(\ref{eq:AII_Sigma})], the expression in the square brackets of the equation above can be approximated as
\begin{align}\label{eq:AII_deltaG}
\begin{aligned}
	 G^{(R)}_{\rho} (\kb,\ww) - \bd_{\rho}^{(R)}(\kb,\ww)
	 \approx \, &
	\bd_{\rho}^{(R)}(\kb,\ww) 
	\,
	\Sigma^{(R)}_{\rho} (\kb,\ww) 
	\,
	\bd_{\rho}^{(R)}(\kb,\ww).
\end{aligned}
\end{align} 
In what follows, we take into account the interaction terms in $E_{\rho}$ (see Fig.~\ref{fig:AII_2}), and calculate their contribution to the H.-S.\ field's retarded self energy $\Sigma^{(R)}_{\rho}$. Once the self energy $\Sigma^{(R)}_{\rho}$ is known, the correction to the density response function and conductivity obtains from Eqs.~(\ref{eq:AII_deltapi0}) and (\ref{eq:AII_deltaG}).

\subsection{Self energy}

Figs.~\ref{fig:AII_3} and \ref{fig:AII_4} depict the retarded self energy diagrams of the H.-S.\ field $\rho$ for class AII. Additional diagrams contribute in principle, but their total contribution vanishes (or is negligible compared with the logarithmic correction we are interested in). These additional diagrams appear in Appendix \ref{Sec:App2}.

\subsubsection{Category 1}

\begin{figure}
	\centering
	\includegraphics[width=0.9\linewidth]{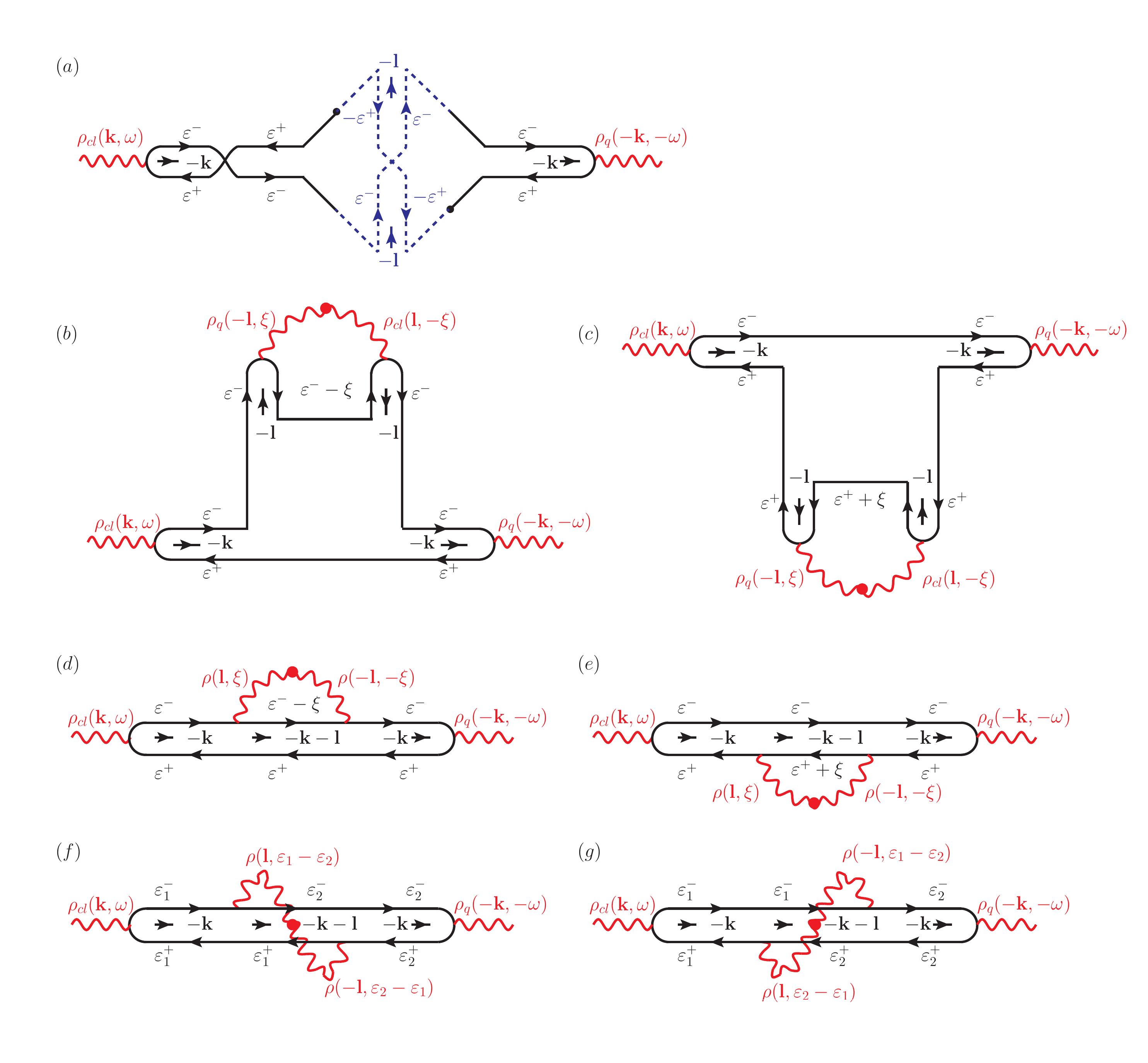}
	\caption{(Color online) Self energy diagrams for class AII: Category 1.
	Diagram (a) is the weak antilocalization correction due to the virtual Cooperon loop. 
	Diagrams (b)--(e) are Altshuler-Aronov (AA) corrections,
	while (f) and (g) renormalize the interaction.
	}
	\label{fig:AII_3}
\end{figure}

All contributions from the diagrams in Fig. \ref{fig:AII_3} can be expressed as 
\begin{align}\label{eq:AII_C1}
\begin{aligned}
	-i \Sigma^{(R)}_{\rho}(\kb,\ww)
	=\,&
	 -4 h^2 \lb   \bd_0 ^2 (k,-\ww)
	\intl{\varepsilon_1,\varepsilon_2} 
	\left( F_{\varepsilon^+_1} -F_{\varepsilon^-_1} \right)
	\Sigma_X(\varepsilon_1^-,\varepsilon_1^+,\varepsilon_2^+,\varepsilon_2^-;-\kb,-\kb),
\end{aligned}
\end{align}
where we have defined $\varepsilon^\pm_{1,2} \equiv \varepsilon_{1,2} \pm \ww/2$. 
$\Sigma_X$ denotes the corresponding self energy of 
the diffuson 
matrix $X$ 
when the density field $\rho$ is integrated out first \cite{FNLSM1983,BelitzKirkpatrick94}.
Its arguments specify the frequency and momentum indices.

The associated self energies $\Sigma_X$ in Figs.~\ref{fig:AII_3}(a)--\ref{fig:AII_3}(e) are diagonal in frequency space, and are given by (in respective order),
\bsub
\label{eq:AII_3ae}
\begin{align}
	\Sigma_X^{(a)}(-\kb,-\ww)
	 =\,& 
	 -\frac{\lb}{2} k^2
	 \intl{\vex{l}}
	 \bd_0(l,-\ww)
	 ,		
	 \label{eq:WAL--AII}\\
	\Sigma_X^{(b)} (-\kb,-\ww)
	=\, & 
	 \frac{i}{4} \pi h \gamma \lb^2  
	\intl{\vex{l},\xi}
	\left[
		\bd_0^{-1}(k,-\ww)\bd_0(l,\xi)\bd_u(l,\xi)+\bd_u(l,\xi)
	\right] 
	\left[ \tanh \left(\frac{\varepsilon^{-}-\xi}{2T}\right)-\tanh \left(\frac{\varepsilon^{-}}{2 T}\right)\right] 
	,
	\\
	\Sigma_X^{(c)} (-\kb,-\ww)
	=\, & 
	\frac{i}{4} \pi h \gamma \lb^2 
	\intl{\vex{l},\xi}
	\left[
	\bd_0^{-1}(k,-\ww) \bd_0(l,\xi)\bd_u(l,\xi)+\bd_u(l,\xi)
	\right] 
	\left[ -\tanh \left(\frac{\varepsilon^{+}+\xi}{2T}\right) +\tanh \left(\frac{\varepsilon^{+}}{2T} \right)\right]
	,
	\\
	\Sigma_X^{(d)} (-\kb,-\ww)
	= \, & 
	\frac{i}{4} \pi  h \gamma \lb ^2 
	\intl{\vex{l},\xi}
	\left\lbrace 
	\begin{aligned}
	&
	\bd_0( \lvert -\kb-\vex{l} \rvert, -\ww-\xi) \frac{\bd_u(l,\xi)}{\bd_0(l,\xi)}
	\left[\tanh \left(\frac{\varepsilon^- -\xi}{2T}\right)+ \coth \left(\frac{\xi}{2 T}\right) \right]
	\\ 
	+ &
	\bd_0(\lvert -\kb-\vex{l} \rvert,-\ww-\xi) \frac{\bd_u(l,-\xi)}{\bd_0(l,-\xi)}
	\left[\tanh \left(\frac{\varepsilon^-}{2T}\right)-\coth \left(\frac{\xi}{2 T}\right)\right] 
	\end{aligned}
	\right\rbrace 
	,
	\\
	\Sigma_X^{(e)} (-\kb,-\ww)
	=\,&
	\frac{i}{4} \pi  h \gamma \lb ^2  \intl{\vex{l},\xi}
	\left\lbrace 
	\begin{aligned}
	&
	\bd_0(\lvert -\kb-\vex{l} \rvert,-\ww-\xi) \frac{\bd_u(l,\xi)}{\bd_0(l,\xi)}
	\left[-\tanh \left(\frac{\varepsilon^+ + \xi}{2T}\right)+\coth \left(\frac{\xi}{2 T}\right) \right]
	\\ 
	+ &
	\bd_0(\lvert -\kb-\vex{l} \rvert,-\ww-\xi) \frac{\bd_u(l,-\xi)}{\bd_0(l,-\xi)}
	\left[-\tanh \left(\frac{\varepsilon^+}{2T}\right)-\coth \left(\frac{\xi}{2 T}\right)\right] 
	\end{aligned}
	\right\rbrace  	
	,	
\end{align}
\esub
where we have defined $\varepsilon_{1,2}=\varepsilon$ and we have omitted the factor $\delta_{\varepsilon_1,\varepsilon_2}$.	
The modulus of frequency in these expressions is cut off at the large limit by the elastic scattering rate $ \Lambda=\tau_{\mathsf{el}}^{-1}$. The momentum integrations, on the other hand, are performed over the whole space, except for Eq.~(\ref{eq:AII_3ae}a). (Alternatively, we could choose the integration scheme where we integrate over $0 <D l^2 < \Lambda$ and $-\infty < \ww < \infty$.)

We evaluate these integrals up to logarithmic accuracy in the ultraviolet cutoff $\Lambda$ by first carrying out an expansion in terms of external frequency $\omega$ and momentum $\kb$. For the higher-order terms in this expansion, the powers of $\bd_{0,u}(l,\xi)$ are larger, leading to a negligible value after integration. For this reason, these terms are omitted.

The diagram in Fig.~\ref{fig:AII_3}(a) corresponds to the weak anti-localization (WAL) correction
due to the virtual Cooperon loop.	
Performing the momentum integration over $0 <D l^2 < \Lambda$, we find $\Sigma_X^{(a)}=- (\lb / 8 \pi )  k^2 \ln (\Lambda/\omega)$. 
In the limit of vanishing external frequency $\omega \rightarrow 0$, the WAL correction must be cut off by dephasing due to inelastic scattering. In Sec.~\ref{sec:dephasing}, we review the calculation of the dephasing rate $\tau_{\phi}^{-1}$ from the AAK equations \cite{AAK}, derived here from the Keldysh sigma model formalism. 	
As a result, $\Sigma_X^{(a)}$ instead takes the form
\begin{align}\label{eq:AII_SigmaWL}
\begin{aligned}
	\Sigma_X^{(WAL)} =\Sigma_X^{(a)}
	=\, &
	- \frac{\lb}{8 \pi}  k^2 \ln \left(\frac{\Lambda}{\tau_{\phi}^{-1}}\right).
\end{aligned}
\end{align}
To obtain Eq.~(\ref{eq:AII_SigmaWL}), it is necessary to replace the bare Cooperon 
in Fig. \ref{fig:AII_3}(a) with the full one [Eq.~(\ref{eq:AII_fullpropagator})], see Fig \ref{fig:AII_5}. 
This gives the formal expression	
\begin{align}\label{eq:AII_WL}
\begin{aligned}
	\Sigma_X^{(a)} 
	=
	- \frac{\lb}{2} 
	\intl{\vex{l}}
	\left[
	\bd_0^{-1}(l,-\ww)
	+
	k^2
	\right] 
	N^{-1}_{\varepsilon_1^-,-\varepsilon_2^+;-\varepsilon_1^+,\varepsilon_2^-}(-\bl,-\bl),			
\end{aligned}
\end{align}
which must be averaged over the thermal fluctuations of the diffusive charge density field $\rho$
(Sec.~\ref{sec:dephasing}).

Diagrams in Figs.~\ref{fig:AII_3}(b)--\ref{fig:AII_3}(e) represent Altshuler-Aronov (AA) corrections. Integrating and summing Eqs.~(\ref{eq:AII_3ae}b)--(\ref{eq:AII_3ae}e) yields
\begin{align}\label{eq:AII_SigmaAA}
\begin{aligned}
	\begin{aligned}
	\Sigma_X^{(AA)} 
	= \, &
	k^2 
	\left\lbrace 
	\frac{\lb}{4 \pi}\left[ 1+\frac{1}{\gamma}\ln (1-\gamma) \right] \ln \left(\frac{\Lambda}{T}\right)
	\right\rbrace 
	 -i h \lb \ww 
	\left[
	\frac{\lb}{4 \pi} \ln (1-\gamma) \ln \left(\frac{\Lambda}{T}\right) 
	+
	\frac{\lb}{8 \pi} \gamma \ln \left(\frac{\Lambda}{T}\right)
	\right] 
	 +
	\Sigma_{\varepsilon}.
	\end{aligned}
\end{aligned}
\end{align}
The AA corrections are automatically cut off by temperature in the infrared, since 
the Bragg condition for carrier scattering off of static Friedel oscillations is met
only at the Fermi surface \cite{AAG,Zala}. 	
In Eq.~(\ref{eq:AII_SigmaAA}), 
$\Sigma_{\varepsilon}$ is a constant term (independent of the external frequency $\omega$ and momentum $k$), and takes the form		
\begin{align}\label{eq:SigEps}
\begin{aligned}
	\Sigma_{\epsilon}
	= 
	\frac{i}{8} \pi  h \gamma \lb ^2 \intl{\vex{l},\xi} 
	\,
	\left[
	\bd_0(l,\xi)+\bd_0(l,-\xi) 
	\right] 
	&
	\left[
	\frac{\bd_u(l,\xi)}{\bd_0(l,\xi)}
	-
	\frac{\bd_u(l,-\xi)}{\bd_0(l,-\xi)}
	\right]
\\
	\times&
	\left[
	2 \coth \left(\frac{\xi}{2T}\right)
	-
	\tanh \left(\frac{\xi+\varepsilon}{2 T}\right)
	- 
	\tanh\left(\frac{\xi-\varepsilon}{2 T}\right)
	\right].
\end{aligned}
\end{align}
This is the ``outscattering rate,'' which is half of the collision integral that enters the semiclassical kinetic equation \cite{AleinerBlanter}. The latter determines the rate of energy relaxation \cite{AAG}. Although the integral expression for $\Sigma_{\epsilon}$ is infrared divergent, it is forbidden from affecting the linear response due to the charge U(1) Ward identity (current conservation). 	

The associated self energies $\Sigma_X$ of diagrams in Figs.~\ref{fig:AII_3}(f) and \ref{fig:AII_3}(g) are off-diagonal in the frequency indices, and take the forms
\bsub
\label{eq:AII_3fg}
\begin{align}
	\Sigma_X^{(f)} 
	=\, &
	- \frac{i}{4} \pi h \gamma \lb ^2   
	 \intl{\vex{l}}
	\left\lbrace 
	\begin{aligned}
	&
	\bd_0(\lvert -\kb-\vex{l} \rvert,-\ww-\varepsilon_1+\varepsilon_2) \frac{\bd_u(l,\varepsilon_1-\varepsilon_2)}{\bd_0(l,\varepsilon_1-\varepsilon_2)}
	\left[\tanh \left(\frac{\varepsilon_2^{-}}{2T}\right)+\coth \left(\frac{\varepsilon_1-\varepsilon_2}{2 T}\right) \right]
	\\ 
	+ &
	\bd_0(\lvert -\kb-\vex{l} \rvert,-\ww-\varepsilon_1+\varepsilon_2) 
	\frac{\bd_u(l,-\varepsilon_1+\varepsilon_2)}{\bd_0(l,-\varepsilon_1+\varepsilon_2)}
	\left[-\tanh \left(\frac{\varepsilon_2^+}{2T}\right)-\coth \left(\frac{\varepsilon_1-\varepsilon_2}{2 T}\right)\right] 
	\end{aligned}
	\right\rbrace 
	, 	
	\\
	\Sigma_X^{(g)} 
	=\, & 
	-\frac{i}{4} \pi h \gamma \lb ^2 
	 \intl{\vex{l}}
	\left\lbrace 
	\begin{aligned}
	&
	\bd_0(\lvert -\kb-\vex{l} \rvert, -\ww+\varepsilon_1-\varepsilon_2) \frac{\bd_u(l,-\varepsilon_1+\varepsilon_2)}{\bd_0(l,-\varepsilon_1+\varepsilon_2)}
	\left[-\tanh \left(\frac{\varepsilon_2^{+}}{2T}\right)
	-\coth \left(\frac{\varepsilon_1-\varepsilon_2}{2 T}\right) \right]
	\\ 
	+ &
	\bd_0(\lvert -\kb-\vex{l} \rvert, -\ww+\varepsilon_1-\varepsilon_2) 
	\frac{\bd_u(l,\varepsilon_1-\varepsilon_2)}{\bd_0(l,\varepsilon_1-\varepsilon_2)}
	\left[\tanh \left(\frac{\varepsilon_2^-}{2T}\right)
	+\coth \left(\frac{\varepsilon_1-\varepsilon_2}{2 T}\right)\right] 
	\end{aligned}
	\right\rbrace . 	
\end{align}
\esub
Changing the integration variable $\varepsilon_2 \rightarrow \xi \equiv \varepsilon_1-\varepsilon_2$
in Eq.~(\ref{eq:AII_C1})
and integrating these self energies over $\xi$,	
we arrive at expressions quite similar to Eqs.~(\ref{eq:AII_3ae}d) and (\ref{eq:AII_3ae}e), 
resulting in 
\begin{align}\label{eq:AII_Sigmaoff}
\begin{aligned}
	\intl{\varepsilon_2} \left( \Sigma_X^{(f)}+\Sigma_X^{(g)} \right) 
	=
	i h \gamma \lb \ww
		\left[ \frac{\lb}{8 \pi}
			\ln \left(\frac{\Lambda}{T}\right)
		\right] 
	-\Sigma_{\varepsilon}.
\end{aligned}
\end{align}
The constant $-\Sigma_{\varepsilon}$ cancels with $\Sigma_{\varepsilon}$ in Eq.~(\ref{eq:AII_SigmaAA}),
as required by the Ward identity.	

Adding Eqs.~(\ref{eq:AII_SigmaWL}), (\ref{eq:AII_SigmaAA}) along with Eq.~(\ref{eq:AII_Sigmaoff}) and inserting the result into Eq.~(\ref{eq:AII_C1}), the diagrams in Fig. \ref{fig:AII_3} altogether give the contribution 
\begin{align}\label{eq:AII_sigma1}
\begin{aligned}
	-i \Sigma_{\rho}^{(R)}(\kb,\omega)
	= \, &
	-\frac{4}{\pi} h^2 \lb \ww
	\bd_0 ^2 (k, -\ww) 
	\left\lbrace 
	\left[ k^2 \delta \lb - i h \lb \ww (-\delta h) \right] 
	+ i h \gamma  \lb \ww
	(-\delta\Gamma)
	\right\rbrace, 
\end{aligned}
\end{align}
where $\delta \lb$, $\delta h$ and $\delta \Gamma$ are defined by
\bsub
\label{eq:AII_d1}
\begin{align}
	\delta \lb 
	\equiv \, &
	- \frac{\lb}{8 \pi} \ln \left(\frac{\Lambda}{\tau_{\phi}^{-1}}\right) 
	+\frac{\lb}{4 \pi} \left[ 1+\frac{1}{\gamma}\ln (1-\gamma) \right] \ln \left(\frac{\Lambda}{T}\right),
	\\
	\delta h 
	\equiv \, &
	- 
	\frac{\lb}{4 \pi} \ln (1-\gamma) \ln \left(\frac{\Lambda}{T}\right) 
	-
	\frac{\lb}{8 \pi} \gamma \ln \left(\frac{\Lambda}{T}\right),		
	\\
	\delta \Gamma 
	\equiv \, &
	-\frac{\lb}{8 \pi} 
	\ln \left( \frac{\Lambda}{T} \right).		 
\end{align}
\esub

\subsubsection{Category 2}

\begin{figure}
	\centering
	\includegraphics[width=0.9\linewidth]{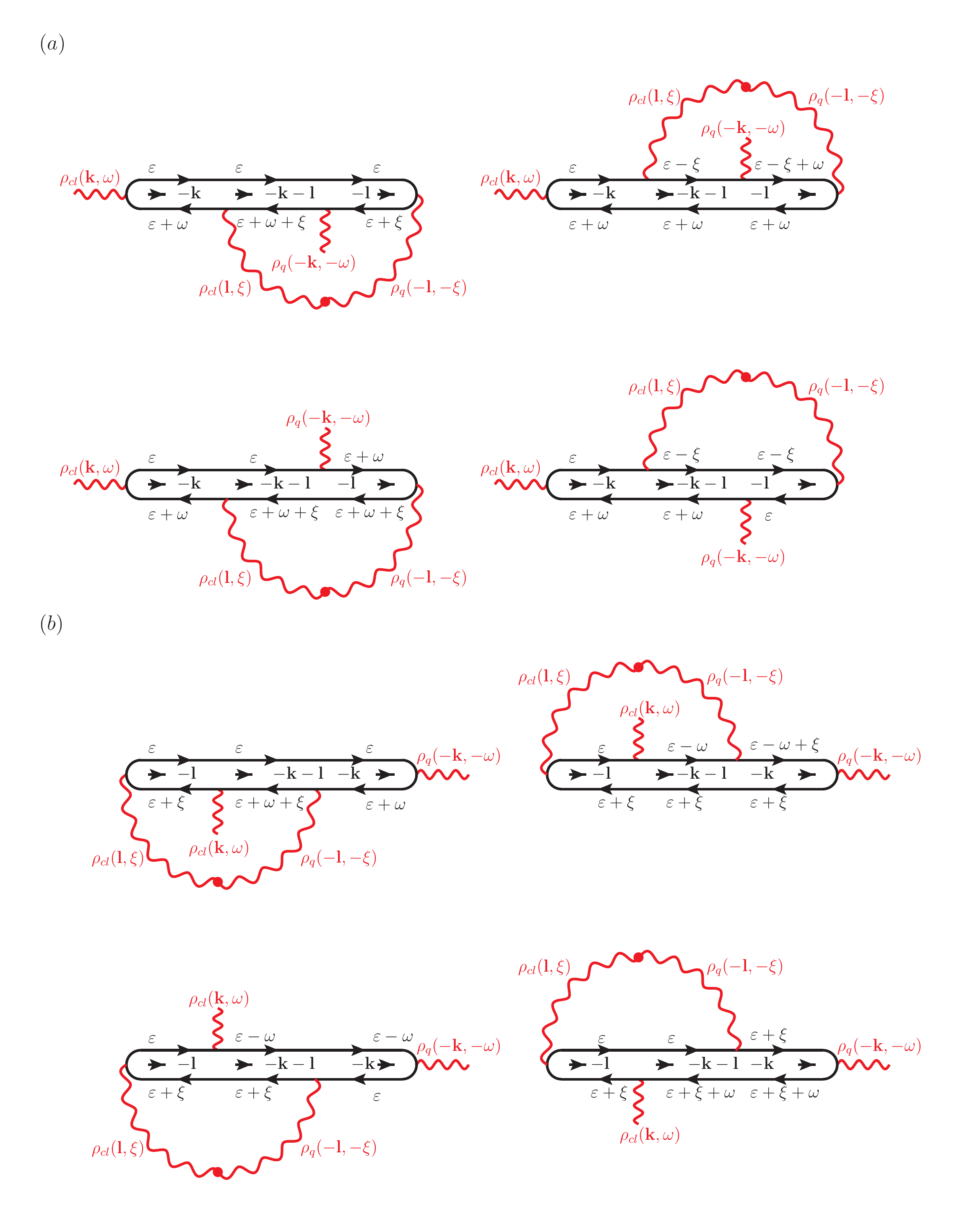}
	\caption{(Color online) Self energy diagrams for class AII: Category 2.
	These diagrams represent the AA wave function renormalization, which determines
	the energy scaling of the density of states.}
	\label{fig:AII_4}
\end{figure}

In Fig.~\ref{fig:AII_4}, we show another group of self energy diagrams with non-negligible amplitudes.
These amplitudes are given by expressions that are identical apart from the distribution function piece. 
Diagrams in Figs.~\ref{fig:AII_4}(a) and \ref{fig:AII_4}(b) respectively give
\bsub
\label{eq:AII_4}
\begin{align}
	\begin{aligned}
	(a)
	=\, 
	\left( -4 h^2 \lb\right) 
	\left( -\frac{i}{4} \pi h \gamma \lb ^2  \right) 
	\bd_0 (k, -\ww)
	\intl{\varepsilon,\vex{l},\xi}
	& 
	\bd_0(\lvert -\kb-\vex{l} \rvert,-\ww-\xi)
	\bd_u (l,-\xi)
	\left( F_{\varepsilon +\ww} -F_{\varepsilon}\right) 
	\\
	 \times &
	\left( 
	F_{\varepsilon+\xi}
	-
	F_{\varepsilon-\xi+\ww}
	+
	F_{\varepsilon+\ww}
	-
	F_{\varepsilon}
	\right), 	
	\end{aligned}
	\\
	\begin{aligned}
	(b)
	=\,
	\left( -4 h^2 \lb\right) 
	\left( -\frac{i}{4} \pi h \gamma \lb ^2  \right) 
	\bd_0 (k, -\ww)
	\intl{\varepsilon,\vex{l},\xi}
	&
	\bd_0(\lvert -\kb-\vex{l} \rvert,-\ww-\xi)
	\bd_u(l,-\xi)
	\left( F_{\varepsilon +\xi} -F_{\varepsilon}\right) 
	\\
	 \times &
	\left( 
	F_{\varepsilon+\ww}
	-
	F_{\varepsilon-\ww+\xi}
	+
	F_{\varepsilon+\xi}
	-
	F_{\varepsilon}
	\right).
	\end{aligned}
\end{align}
\esub 
Combining these two equations and carrying out the integration gives the net contribution from diagrams in Fig.~\ref{fig:AII_4}:
\begin{align}\label{eq:AII_sigma2}
\begin{aligned}
	-i \Sigma_{\rho}^{(R)}(\kb,\omega)
	=\,&
	 -\frac{4}{\pi} h^2 \lb 
	\ww \bd_0 (k, -\ww)
	\left( -\delta z \right),
\end{aligned}
\end{align}
where we have defined $\delta z$ as
\begin{align}\label{eq:AII_d2}
\begin{aligned}
	\delta z 
	\equiv \, &
	\frac{\lb}{4 \pi}  \ln (1-\gamma)
	\ln \left(\frac{\Lambda}{T}\right).		 
\end{aligned}
\end{align}
Eq.~(\ref{eq:AII_d2}) from the diagrams in Fig.~\ref{fig:AII_4}
is the Altshuler-Aronov wave function renormalization (see below), which determines
the energy scaling of the disorder-averaged density of states.

\subsubsection{Results}

Summing up contributions from Figs.~\ref{fig:AII_3} [Eq.~(\ref{eq:AII_sigma1})] and \ref{fig:AII_4} [Eq.~(\ref{eq:AII_sigma2})], we find the total retarded self energy
\begin{align}\label{eq:AII_Sigma3}
\begin{aligned}
	 -i \Sigma_{\rho}^{(R)} (\kb,\omega)
	= \, &
	  -\frac{4}{\pi} h^2 \lb
	\ww \bd_0 ^2 (k, -\ww) 
	\left\lbrace 
		\left[ k^2 \delta \lb - i h \lb \ww (-\delta h) \right] 
		+ i h \gamma  \lb \ww
		 (-\delta\Gamma)
	+
	\bd_0 ^{-1} (k, -\ww) 
	(-\delta z)
	\right\rbrace. 
\end{aligned}
\end{align}
The three terms in the braces come from Figs.~\ref{fig:AII_3}(a)--\ref{fig:AII_3}(e), \ref{fig:AII_3}(f)--\ref{fig:AII_3}(g), and \ref{fig:AII_4}, respectively. The first two correspond to the renormalization of the parameters ($h$, $\lb$) and the interaction coefficient ($\Gamma=h\gamma$), whereas the third term is related to the wave function renormalization of matrix $\Xh$.

To one-loop order, the wave function renormalization $Z$ acquires the form
\begin{align}
	Z= 1+\delta z,
\end{align}	
and the renormalized $h_R$, $\lb_R$ are related to the bare ones by
\begin{align}\label{eq:AII_RG0}
\begin{aligned}
	& h_R= h(1 + \delta h + \delta z),
	\qquad
	\frac{1}{\lb_R}= \frac{1}{\lb} (1 - \delta \lb + \delta z).	
\end{aligned}
\end{align}
In Appendix~\ref{Sec:App1} we prove that, to one loop order,
\begin{align}\label{eq:kappa2}
\begin{aligned}
	h_R (1-\gamma_R) = h (1-\gamma),	
\end{aligned}
\end{align}
from which one can infer the renormalized interaction strength $\gamma_R$.
In a disordered normal metal, this identity holds to all loop orders, meaning the 
charge
compressibility $\kappa$ defined in Eq.~(\ref{eq:DcKappaDef}) does not renormalize \cite{BelitzKirkpatrick94}. This constraint does \emph{not} apply to the non-standard classes \cite{FosterLudwig2}, e.g., the class C superconductor (see Appendix~\ref{Sec:App1}),
since the density of states is typically critical in such systems even in the absence of interactions.	
Using Eqs.~(\ref{eq:AII_d1}),~(\ref{eq:AII_d2}), one may note
\begin{align}\label{eq:kappa1}
\begin{aligned}
	\delta h + \delta z = \gamma \delta \Gamma,		
\end{aligned}
\end{align}
and as a result
\begin{align}
\begin{aligned}
	h_R \gamma_R=h \gamma(1+\delta \Gamma).
\end{aligned}
\end{align}

Utilizing the identity in Eq.~(\ref{eq:kappa1}), Eq.~(\ref{eq:AII_Sigma3}) reduces to
\begin{align}\label{eq:AII_Sigma4}
\begin{aligned}
	-i \Sigma_{\rho}^{(R)} (\kb,\omega)
	= \, &
	-\frac{4}{\pi} h^2 
	\ww \bd_0 ^2 (k, -\ww) k^2 \lb (\delta \lb - \delta z ).
\end{aligned}
\end{align}
The quantum correction to the density response function is
\begin{align}
\begin{aligned}
	\delta \Pi (\kb,\omega)
	=\,
	-i \frac{2}{\pi} h^2 (1-\gamma)^2 \bd_u^2(k,-\ww) \ww k^2
	\lb \left( \delta \lb -\delta z \right)
	=\,
	-i \frac{2}{\pi} h^2 (1-\gamma)^2 \bd_u^2(k,-\ww) \ww k^2
	\left(  \lb_R -\lb \right),
\end{aligned}
\end{align}
and the conductivity correction is
\begin{align}
\begin{aligned}
	\delta \sigma  
	 = \,
	 -\frac{2}{\pi} \frac{\delta \lb - \delta z}{\lb} 
	 = \,
	 \frac{2}{\pi} \left(  \frac{1}{\lb_R} - \frac{1}{\lb} \right).
\end{aligned}
\end{align}
Substituting the explicit forms of $\delta \lb$ and $\delta z$ given by Eqs.~(\ref{eq:AII_d1}) and (\ref{eq:AII_d2}), respectively, we obtain the result in Eq.~(\ref{eq:correction}a).

\section{Class C}\label{sec:C}

\subsection{Spin density linear response}
For the class C superconductor, the spin density linear response function can be obtained in a similar fashion as in Sec.~\ref{sec:AII}.
It is defined as
\begin{align}\label{eq:SpinResp}
\begin{aligned}
	\Pi^{i,j}
	\left( \vex{k} , \ww \right) 
	=\,
	\frac{\delta  n_{s}^{i} (\kb, \ww)}{\delta B_{\mathsf{cl}}^{j}(\kb,\ww)}
	\bigg|_{\Bvcl=0}	
	=\,
	 \frac{i}{2} \frac{\delta ^2 Z[\vex{B}]}{\delta B_{\mathsf{cl}}^{j}(\kb,\ww) \delta B_{\mathsf{q}}^{i} (-\kb, -\ww)}
	\bigg|_{\Bvql=\Bvcl=0},
\end{aligned}
\end{align}
where $Z[\vex{B}]$ is given by Eq.~(\ref{eq:FNLSM_C}). $n_{s}$ here refers to the spin density, and  similar to the charge density $n$, is averaged over the two branches of the Keldysh contour. Superscripts $i$ and $j$ index the spin component.

As in the case of class AII, the first step is to shift the H.-S.\ field, now a vector field, by $\bv_{cl,q} \rightarrow \bv_{cl,q} - \blv_{cl,q}$. Then one can calculate the spin density response function from

\begin{align}
\begin{aligned}
	\Pi ^{i,j} \left( \vex{k} , \ww \right) 
	=\,
	-
	\frac{2}{\pi} h \frac{(1-\gamma)}{\gamma}
	\int \dd \bv \dd \qh  \, e^{-S_q-S_c-S_{\bv}} 
	\left[
	\delta_{i,j}+
	i\frac{4}{\pi} h \frac{(1-\gamma)}{\gamma}
	\bcl^{i} (\kb, \ww)  \bql^{j} (-\kb, -\ww)
	\right],
\end{aligned}
\end{align}
where the actions acquire the forms,
\bsub\label{eq:C_S}
\begin{align}
	S_q
	 =\,& 
	\frac{1}{4 \lambda}
	\intl{\vex{x}}
	\tr\left[\Nabla \qh \cdot \Nabla \qh \right]
	+
	i h
	\intl{\vex{x}}
	\tr\left[\qh (\hat{\omega} + i \eta \tauh^3)\right],
	\\ 
	S_c
	=\,&
	-i h \intl{\vex{x}} \tr 
	\left[
	\left( \bvcl  +\bvql \hat{\tau}^1\right) 
	\cdot \h{\sv} \mf(\hat{\omega}) \hat{q} (\vex{x}) \mf(\hat{\omega})
	\right],
	\\
	S_{\bv}
	=\,&
	-i \frac{4}{\pi} h \frac{1}{\gamma} \intl{\vex{x},t} \bv_{\mathsf{cl}} \cdot \bv_{\mathsf{q}}. 
\end{align}
\esub

\subsection{Parameterization}

For class C, we employ the following $\qh$ matrix parameterization in Keldysh space around the saddle point $\qsp=\tauh^3$, i.e.,
\begin{align}\label{eq:C_q}
	 \qh
	= \, &
	\begin{bmatrix}
		\sqrt{1-\Wh^\dagger \Wh} & \Wh^\dagger \hat{s}^2
		\\
		\hat{s}^2 \Wh                & -\sqrt{1-\hat{s}^2 \Wh \Wh^\dagger \hat{s}^2 }
	\end{bmatrix}
	_{\tau},
\end{align}
where $\Wh$ is now a matrix in the spin as well as frequency spaces and satisfies the condition
\begin{align}\label{eq:C_Wsym1}
\begin{aligned}
	\Wh
	=\,&
	\h{\Sigma}^1 \Wh^{\T}  \h{\Sigma}^1, 
\end{aligned}
\end{align}

Inserting this parameterization into $S_q+S_{c}$ [Eq.~(\ref{eq:C_S})], and expanding in powers of $\Wh$, we obtain the action up to quadratic order in $\Wh$:
\begin{align}\label{eq:C_SQ2}
\begin{aligned}
	S_{W}^{(2)} 
	= \, & 
	\int 
	\Wh^{\dagger}\,^{\alpha,\beta}_{1,2}(\kb_1)
	M^{\beta,\alpha;\sigma,\gamma}_{2,1;4,3}(\kb_1,\kb_2)
	\Wh^{\gamma,\sigma}_{3,4}(\kb_2)
	+
	J^{\dagger}\,^{\beta,\alpha}_{2,1}(\kb) \Wh^{\alpha,\beta}_{1,2}(\kb)
	+
	J^{\beta,\alpha}_{2,1}(\kb) \Wh^{\dagger}\,^{\alpha,\beta}_{1,2}(\kb)	,	
\end{aligned}
\end{align}
where $M$, $J$, and $J^{\dagger}$ (independent of $J$) are now defined as
\bsub\label{eq:c_MNJ}
\begin{align}
	&\begin{aligned}
		M^{\beta,\alpha;\sigma,\gamma}_{2,1;4,3}(\kb_1,\kb_2) 
		\equiv \, &
		\frac{1}{2} \left[ k_1^2 - i h \lb (\ww_1-\ww_2) \right] 
		\delta_{\alpha,\sigma} \delta_{\beta,\gamma}
		\delta_{1,4}\delta_{2,3}
		\delta_{\kb_1,\kb_2}
		\\
		& + i h \lb
		\left[
		\bvcl(\kb_1-\kb_2,\ww_4-\ww_1)
		+
		F_{4} \bvql(\kb_1-\kb_2,\ww_4-\ww_1)
		\right]
		\cdot \sv^{\sigma,\alpha}
		\delta_{\beta,\gamma}\delta_{2,3} \,,
	\end{aligned}	
	\\	 
	& J^{\dagger}\,^{\beta,\alpha}_{2,1}(\kb) 
	\equiv \, 
	i h \sqrt{\lb}
	\left[
	(F_{2}-F_{1})\bvcl(-\kb,\ww_2-\ww_1)
	+(1-F_{1}F_{2})\bvql(-\kb,\ww_2-\ww_1)
	\right] \cdot (\sv s^2)^{\beta,\alpha},
	\\
	& J^{\beta,\alpha}_{2,1}(\kb) 
	\equiv \, 
	i h \sqrt{\lb}
	\bvql(\kb,\ww_2-\ww_1) \cdot (s^2 \sv)^{\beta,\alpha}.		
\end{align}
\esub
Here the superscripts ($\alpha,\beta$, etc.) index the spin space (instead of particle-hole space as in Sec. \ref{sec:AII}). We have used the $\Wh$ matrix's symmetry in Eq.~(\ref{eq:C_Wsym1}) to simplify the action and rescaled $\Wh$ by
\begin{align}\label{eq:Wrescale}
\begin{aligned}
	& \Wh \rightarrow \sqrt{\lb} \Wh,
	\qquad
	& \Wh^{\dagger} \rightarrow \sqrt{\lb} \Wh^{\dagger}.	
\end{aligned}
\end{align}

As before, we retain the quartic term in $S_q$, which takes the form
\begin{align}\label{eq:C_SQ4}
\begin{aligned}
	S_q^{(4)}
	= \, &
	\int \delta_{\kb_1+\kb_3,\kb_2+\kb_4}
	\Wh^{\dagger}\,^{\alpha,\beta}_{1,2} (\kb_1) \Wh ^{\beta,\gamma}_{2,3}(\kb_2)
	\Wh ^{\dagger}\,^{\gamma,\sigma}_{3,4} (\kb_3) \Wh^{\sigma,\alpha}_{4,1}(\kb_4)	
	\\
	&\times \frac{\lb}{8} 
	\left[ 
	-(\kb_1 \cdot \kb_3 +\kb_2 \cdot \kb_4) 
	+\frac{1}{2}(\kb_1 + \kb_3)\cdot (\kb_2 + \kb_4) 
	-i \frac{h}{2} \lb (\ww_1-\ww_2+\ww_3-\ww_4) 
	\right] .	
\end{aligned}
\end{align}

\subsection{Feynman rules}

\begin{figure}
\centering
\includegraphics[width=0.7\linewidth]{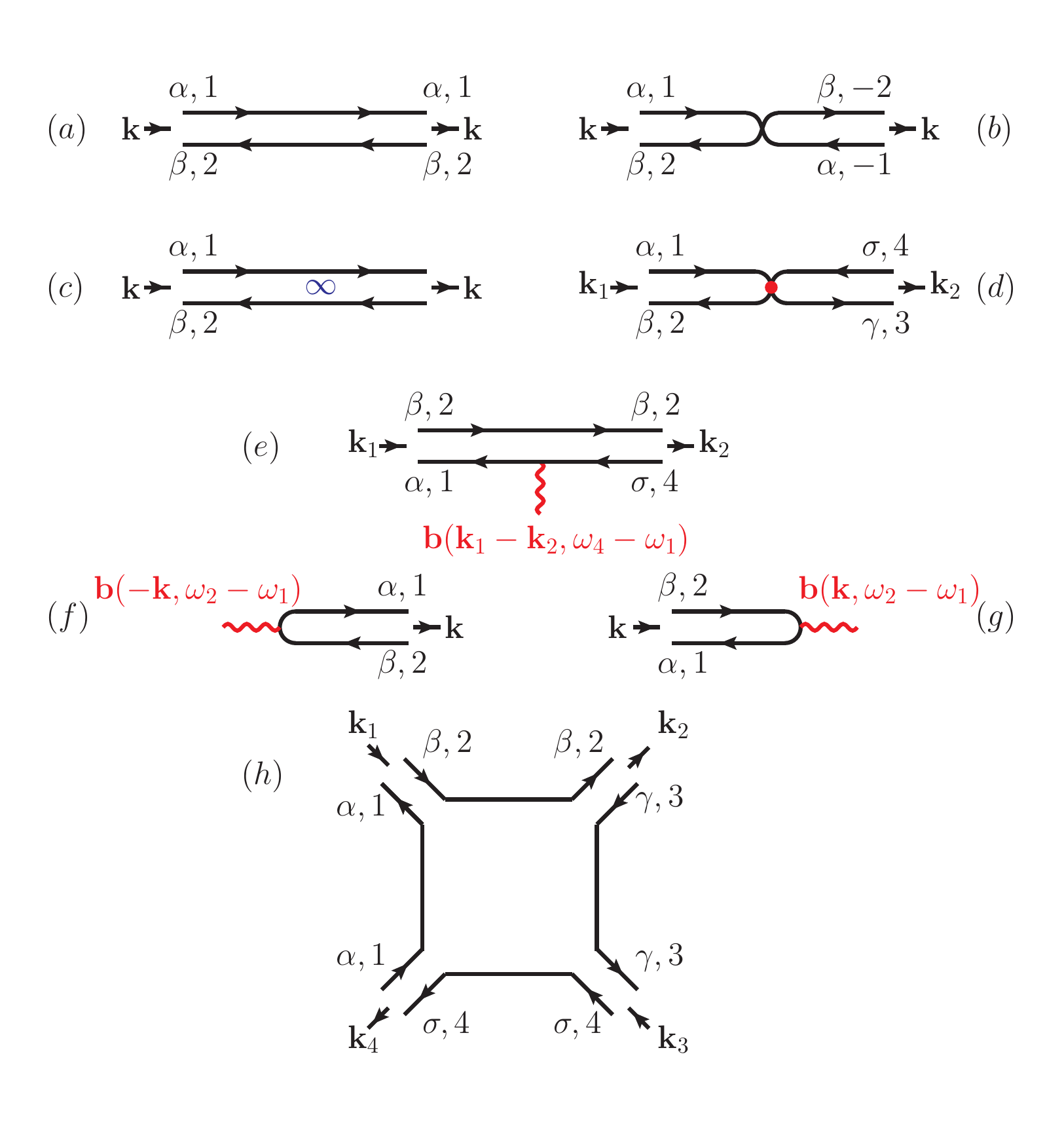}
\caption
	{(Color online) Feynman rules for class C: 
	The bare and full propagators of the matrix field $\Wh$ are illustrated in (c) and (d), respectively. 
	As shown in Eq.~(\ref{eq:C_barePropagator}), the bare propagator consists of two terms with different frequency structure. They are depicted in (a) and (b). 
	(e)--(g) show the interaction vertices coupling between H.-S.\ field $\bv$ and matrix field $\Wh$, while (h) depicts the 4-point diffusion vertex.	
	}
\label{fig:C_1}
\end{figure}

In Fig.~\ref{fig:C_1}, we show the Feynman rules for class C. Throughout this section, we employ the notation in which the solid black line represents matrix field $\Wh$ and the red wavy line stands for H.-S.\ 
(hydrodynamic spin density)
vector field $\bv$.

\subsubsection{Bare propagator}

Without interaction, the $\Wh$ propagator is given by
\begin{equation}\label{eq:C_barePropagator}
	\braket{\Wh^{\alpha,\beta}_{1,2}(\kb)\Wh^{\dagger}\,^{\gamma,\sigma}_{3,4}(\kb)}_{0}
	=\Delta_0(k,\ww_1-\ww_2) 
	\left[
	\delta_{\alpha,\sigma}\delta_{\beta,\gamma}\delta_{1,4}\delta_{2,3}
	+
	\delta_{\alpha,\gamma}\delta_{\beta,\sigma}\delta_{1,-3}\delta_{2,-4}
	\right]. 
\end{equation}
It contains two terms represented respectively by diagrams in Figs.~\ref{fig:C_1}(a) and \ref{fig:C_1}(b).
In Fig. \ref{fig:C_1}(c), we depict the same propagator in Fig. \ref{fig:C_1}(a) but with a ``$\infty$'' symbol in the middle. This diagram is used to represent the sum of two terms in Eq.~(\ref{eq:C_barePropagator}).

\subsubsection{Interaction vertices}

Figs.~\ref{fig:C_1}(e)--\ref{fig:C_1}(g) show 
the interaction vertices arising from action $S_{W}^{(2)}$, with amplitudes given by the following equations, in respective order,
\begin{align}
\begin{aligned}
	(e) 
	=\,&
	-i h \lb 
	\left[
	\bvcl(\kb_1-\kb_2,\ww_4-\ww_1)
	+
	F_{4} \bvql(\kb_1-\kb_2,\ww_4-\ww_1)
	\right]
	\cdot \sv^{\sigma,\alpha}, 	
	\\
	(f) 
	=\, &
	-i h \sqrt{\lb}
	\left[
	(F_{2}-F_{1})\bvcl(-\kb,\ww_2-\ww_1)
	+(1-F_{1}F_{2})\bvql(-\kb,\ww_2-\ww_1)
	\right] \cdot (\sv s^2)^ {\beta,\alpha},
	\\
	(g)
	=\,&
	-i h \sqrt{\lb} \,
	\bvql(\kb,\ww_2-\ww_1) \cdot (s^2 \sv)^{\beta,\alpha}.		
\end{aligned}
\end{align}

\subsubsection{Full propagator}

Incorporating the quadratic interaction term with the diffusion action, we arrive at the full propagator, which is represented by diagram in Fig. \ref{fig:C_1}(d). 
It is given by 
\begin{equation}
	\braket{\Wh^{\alpha,\beta}_{1,2}(\kb_1)\Wh^{\dagger}\,^{\gamma,\sigma}_{3,4}(\kb_2)}
	= \tilde{M}^{-1}\,^{\alpha,\beta;\gamma,\sigma}_{1,2;3,4}(\kb_1,\kb_2),
\end{equation}
where  $\tilde{M}$ is the symmetrized $M$ kernel,
\begin{align}
\begin{aligned}
	\tilde{M}^{\beta,\alpha;\sigma,\gamma}_{2,1;4,3}(\kb_1,\kb_2) 
	\equiv \, &
	\frac{1}{4}
	\left[ 
	M^{\beta,\alpha;\sigma,\gamma}_{2,1;4,3}(\kb_1,\kb_2) 
	+
	M^{\alpha,\beta;\sigma,\gamma}_{-1,-2;4,3}(\kb_1,\kb_2) 
	+
	M^{\beta,\alpha;\gamma,\sigma}_{2,1;-3,-4}(\kb_1,\kb_2) 
	+
	M^{\alpha,\beta;\gamma,\sigma}_{-1,-2;-3,-4}(\kb_1,\kb_2) 
	\right] .
\end{aligned}	
\end{align}
and the matrix inversion is defined by
\begin{align}
\begin{aligned}
	\intl{\ww_3,\ww_4,\kb_2}
	\tilde{M}^{-1}\,^{\beta,\alpha;\sigma,\gamma}_{2,1;4,3}(\kb_1,\kb_2)
	\tilde{M}^{\gamma,\sigma;\alpha',\beta'}_{3,4;1',2'}(\kb_2,\kb'_1)
	=\frac{1}{2}
	\left( 
	\delta_{1,1'}\delta_{2,2'}\delta_{\alpha,\alpha'}\delta_{\beta,\beta'}
	+
	\delta_{1,-2'}\delta_{2,-1'}\delta_{\alpha,\beta'}\delta_{\beta,\alpha'}
	\right) 
	\delta_{\kb_1,\kb_1'}.
\end{aligned}
\end{align}
In this section, we employ the notation that repeated spin indices imply summation.

\subsubsection{4-point diffusion vertex}

The quartic action $S_q^{(4)}$ [Eq.~(\ref{eq:C_SQ4})] gives a 4-point vertex with amplitude
\begin{align}\label{eq:C_4PVert}
\begin{aligned}
	(h)=\,
	- \frac{\lb}{4} 
	\left[ 
	-(\kb_1 \cdot \kb_3 +\kb_2 \cdot \kb_4) 
	+\frac{1}{2}(\kb_1 + \kb_3)\cdot (\kb_2 + \kb_4) 
	-i \frac{h}{2} \lb (\ww_1-\ww_2+ \ww_3-\ww_4) 
	\right]
	\delta_{\kb_1+\kb_3,\kb_2+\kb_4}, 
\end{aligned}
\end{align}
which has been multiplied by a factor of 2 to account for the vertex symmetry. It is shown in Fig. \ref{fig:C_1}(h).

\subsection{Effective theory for H.-S.\ field}

\subsubsection{Effective action}

As in Sec.~\ref{sec:AII}, we integrate out the matrix $\qh$ degree of freedom and develop an effective theory involving only the H.-S.\ field $\bv$. It is described by the partition function
\begin{align}\label{eq:C_Eb}
\begin{aligned}
	Z
	=\,&
	\int \dd \bv \,  e^{-E_{\bv}},
	\\
	E_{\bv} 
	\equiv \, &
	S_{\bv} -\ln \left(  \int \dd \qh \,  e^{-S_q-S_{c} } \right). 	
\end{aligned}
\end{align}
$S_q+S_c$ can be expressed as
\begin{align}\label{eq:C_Sqc}
\begin{aligned}
	S_q+S_c 
	=\, &
	\int (\Wh^\dagger \tilde{M} \Wh + J^\dagger \Wh + \Wh^\dagger J) 
	+
	S_{q}^{(4)} [\Wh^\dagger,\Wh],
\end{aligned}
\end{align}	
where we have rewritten the first term by exploiting $\Wh$'s symmetry.
Combining Eqs.~(\ref{eq:C_Sqc}) and (\ref{eq:C_Eb}), we obtain the approximated effective action 
\begin{align}\label{eq:C_eb}
	E_{\bv} 
	\approx \, &
	S_{\bv}
	-
	\int J^\dagger \tilde{M}^{-1} J
	+
	\tr \ln \tilde{M}
	+
	\braket{ S_{D4}},
\end{align}
where $\braket{ S_{D4}}$ stands for
\begin{align}
\begin{aligned}
	\braket{ S_{D4} }
	\equiv \, &
	\frac{
		\int \dd \Wh^\dagger \dd \Wh 
		\,
		\exp \left[ -\int \Wh^\dagger \tilde{M} \Wh \right] 
		\,
		S_{q}^{(4)} [\Wh^\dagger-J^{\dagger}\tilde{M}^{-1},\Wh-\tilde{M}^{-1}J]
	}
	{
		\int \dd \Wh^\dagger \dd \Wh
		\,
		\exp \left[ -\int \Wh^\dagger \tilde{M} \Wh \right] 
	}.
\end{aligned}
\end{align}	

$E_{\bv}$ is then expanded in powers of $\lb$ (or equivalently $\bv$). 
To the lowest order, the 3rd term in Eq.~(\ref{eq:C_eb}) $\tr \ln \tilde{M}$ is an irrelevant constant, and the 2nd term $-\int J^\dagger \tilde{M}^{-1} J$ can be obtained by simply replacing $\tilde{M}^{-1}$ with the bare propagator [see Fig. \ref{fig:C_2}(a)]. Adding these terms along with $S_{\bv}$ yields the lowest-order effective action 

\begin{align}\label{eq:C_S0}
\begin{aligned}
	E_0 
	= \, &
	-i \frac{4}{\pi} h \frac{1}{\gamma} 
	\int \bvql(-\kb,-\ww) \cdot \bvcl(\kb,\ww)
	\frac{\bd_0 (k, -\ww)}{\bd_u (k, -\ww)}
	\\
	&
	-
	(2 i h)^2 \lb
	\int
	\bvql(-\kb, -\ww) \cdot \bvql(\kb,\ww)
	\frac{\ww}{\pi}
	\coth{\left( \frac{\ww}{2 T}\right) }
	\bd_0 (k, -\ww). 	
\end{aligned}
\end{align}

\subsubsection{Bare propagator}

Comparing $E_0$ in Eq.~(\ref{eq:C_S0}) with the analogous result previously obtained for class AII [see Eq.~(\ref{eq:AII_S0})], one can easily deduce the bare Green's function for the H.-S.\ field $\bv$,
\begin{align}\label{eq:bb}
\begin{aligned}
	\braket{b^i_{a}(\kb,\ww) b^j_{b} (-\kb,-\ww)}_0
	=&
	i	\delta_{i,j}
	\begin{bmatrix}
		\bd_{\rho}^{(K)}(\kb,\ww)  & \bd_{\rho}^{(R)} (\kb,\ww)
		\\ 
		\bd_{\rho}^{(A)} (\kb,\ww)  & 0
	\end{bmatrix},
\end{aligned}
\end{align}
where $\bd_{\rho}$ was defined in Eq.~(\ref{eq:Deltarho}) and 
$a,b\in\left\lbrace \msf{cl},\msf{q} \right\rbrace $. As in the case of class AII, it is represented by a red curvy line with a dot in the middle, see Fig.~\ref{fig:HS}.

\subsubsection{Interaction vertices}

\begin{figure}
	\centering
	\includegraphics[width=0.9\linewidth]{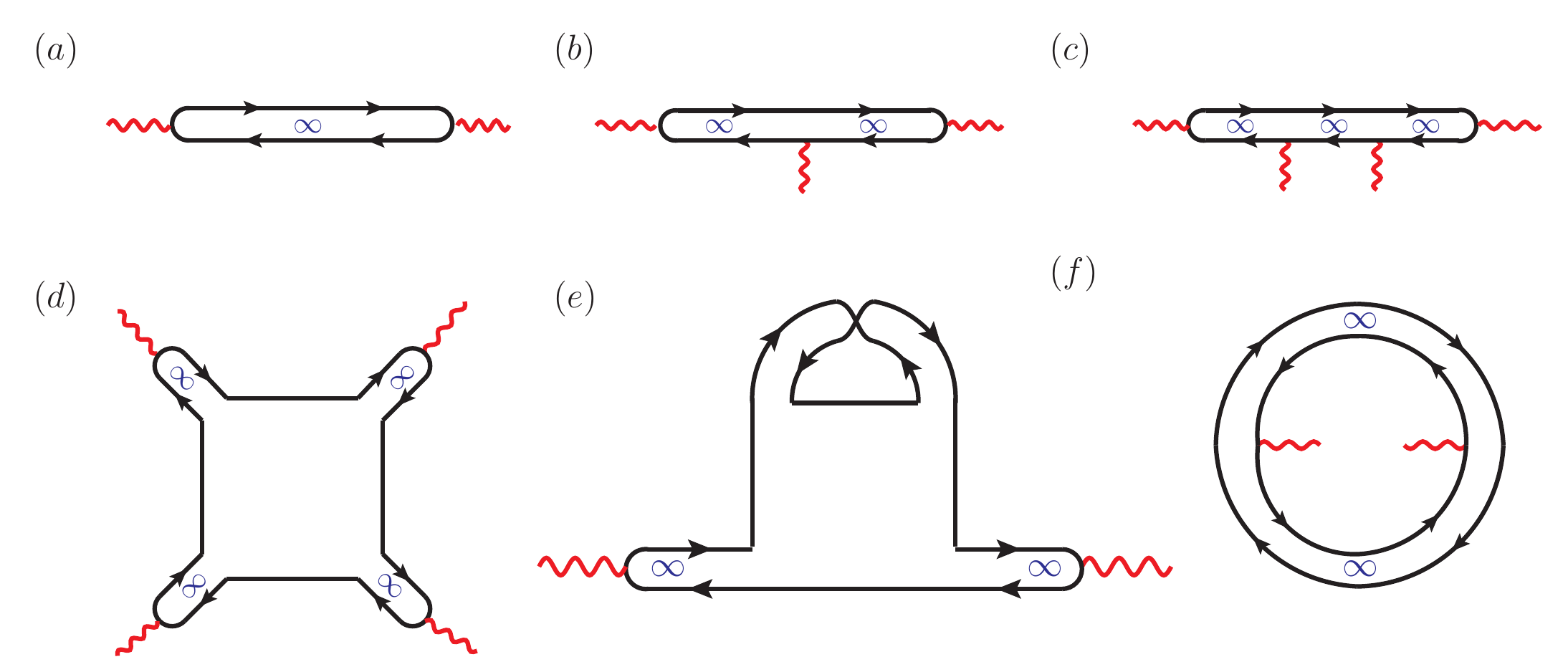}
	\caption{(Color online) Interaction vertices of H.S. field $\bv$ for class C.}
	\label{fig:C_2}
\end{figure}

The interaction vertices arising from the higher-order terms in effective action $E_{\bv}$ are shown in Figs.~\ref{fig:C_2}(b)--\ref{fig:C_2}(f). As we have already explained, their contributions give rise to the quantum correction to linear response function.

\subsubsection{Causality structure of the dressed Green's function and self energy}

The bosonic field's causality structure discussed in Sec.~\ref{sec:AII} also applies here. 
The dressed Green's function $G_{\bv}$ and self energy $\Sigma_{\bv}$ of the H.-S.\ field $\bv$ share the same structure with $G_{\rho}$ [Eq.~(\ref{eq:AII_G})] and $\Sigma_{\rho}$ [Eq.~(\ref{eq:AII_Sigma})], respectively.
Once again, their retarded, advanced, and Keldysh components follow the condition in Eq.~(\ref{eq:FDT}).

\subsubsection{Spin density response and Green's function}

The spin density response function $\Pi^{i,j}$ is determined by the retarded Green's function:
\begin{align}\label{eq:C_Pi1}
\begin{aligned}
	& \Pi^{i,j} \left( \vex{k} , \ww \right) 
	= \,  
	-\frac{2}{\pi} h \frac{(1-\gamma)}{\gamma}
	\left[
	\delta_{i,j}-\frac{4}{\pi} h \frac{(1-\gamma)}{\gamma}
	\left( G^{(R)}_{\bv}\right) \,^{i,j} (\kb, \ww)
	\right].
\end{aligned}
\end{align}
To the zeroth order in $\lb$, $\left( G^{(R)}_{\bv}\right) \,^{i,j} (\kb, \ww)$ becomes the bare propagator $\bd_{\rho}^{(R)}(\kb,\ww) \delta_{i,j}$, and we have
\bsub\label{eq:C_Classi}
\begin{align}
	&\Pi_0^{i,j} \left( \vex{k} , \ww \right) 
	=\, 
	-\delta_{i,j}
	\frac{2}{\pi}
	\frac{ h (1-\gamma) k^2 }{ k^2 - i h (1-\gamma) \lb \ww}
	=\,
	-\delta_{i,j}
	\kappa
	\frac{ D_c k^2 }{ D_c k^2 - i \ww},	
	\\
	&\sigma_0^{i,j} 
	= \, 
	\delta_{i,j} \frac{2}{\pi} \frac{1}{\lb} 
	= \, 
	\delta_{i,j} D (2\nu_0). 
\end{align}
\esub
Here we have used Eq.~(\ref{eq:conduct}) which relates the spin density response function and conductivity.

The quantum correction to $\Pi^{i,j}$ can be calculated using
\begin{align}\label{eq:C_deltapi0}
\begin{aligned}
	\delta \Pi^{i,j}\left( \vex{k} , \ww \right)
	=\, &
	\frac{8}{\pi^2} h^2 \left(  \frac{1-\gamma}{\gamma} \right)^2
	\left[ \left(  G^{(R)}_{\bv} \right) ^{i,j}(\kb,\ww) - \bd_{\rho}^{(R)}(\kb,\ww) \delta_{i,j} \right],  
\end{aligned}
\end{align}
where the correction to the Green's function is approximately
\begin{align}\label{eq:C_deltaG}
\begin{aligned}
	\left( G^{(R)}_{\bv} \right) ^{i,j} (\kb,\ww) - \bd_{\rho}^{(R)}(\kb,\ww) \delta_{i,j}
	\approx \, &
	\bd_{\rho}^{(R)}(\kb,\ww) \left(  \Sigma^{(R)}_{\bv}\right)^{i,j}(\kb,\ww) \, \bd_{\rho}^{(R)}(\kb,\ww). 
\end{aligned}
\end{align}


\subsection{Self energy}

In this section, we evaluate the retarded self energy of the H.-S.\ field $\bv$ at one-loop level, and use the result to compute the quantum correction to spin density response function $\Pi^{i,j}$ and spin conductivity $\sigma^{i,j}$. The relevant self energy diagrams are depicted in Figs.~\ref{fig:C_3}, \ref{fig:C_4} and \ref{fig:C_5}.

\subsubsection{Category 1}

\begin{figure}
	\centering
	\includegraphics[width=0.9\linewidth]{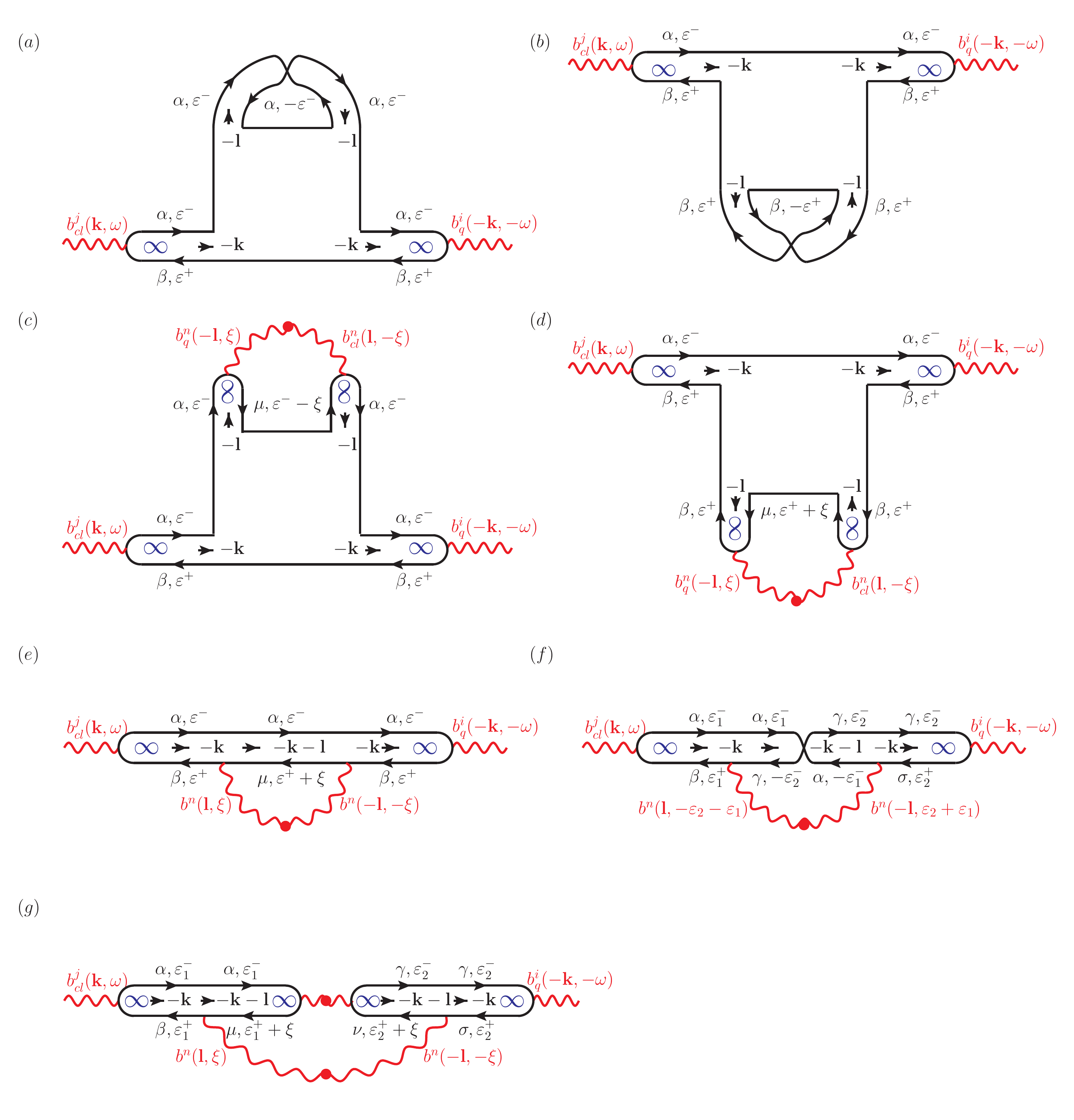}
	\caption{(Color online) Self energy diagrams for class C: Category 1.
	Diagrams (a) and (b) give part of the weak localization correction due to the virtual class C diffuson loop. 
	Diagrams (c)--(e) are Altshuler-Aronov (AA) corrections, while (f) and (g) renormalize the interaction. 	
	}
	\label{fig:C_3}
\end{figure}

The contribution from each diagram in Fig.\ \ref{fig:C_3} can be expressed in terms of $\Sigma_W$ the associated self energy for matrix $\Wh$, i.e.,
\begin{align}\label{eq:C_C1}
\begin{aligned}
	-i \left( \Sigma^{(R)}_{\bv}\right) ^{i,j} (\kb,\ww)
	=\, & 	 
	-4 h^2 \lb 
	(s^j s^2 )^{\beta,\alpha} (s^2 s^i)^{\gamma,\sigma}
	\bd_0^2(k,-\ww)
	\intl{\varepsilon_1,\varepsilon_2}		
	\left( F_{\varepsilon_1^+}  - F_{\varepsilon_1^-} \right)
	\Sigma_W^{\alpha,\beta;\sigma,\gamma}
	(\varepsilon_1^-,\varepsilon_1^+,\varepsilon_2^+,\varepsilon_2^-;-\kb,-\kb).
\end{aligned}
\end{align}
For the diagrams in Figs.~\ref{fig:C_3}(a)--\ref{fig:C_3}(e), the corresponding self energies $\Sigma_W$ are diagonal in energy and spin spaces and acquire the forms,
\bsub
\label{eq:C_3ae}
\begin{align}
	\Sigma_W^{(a)} (-\kb,-\ww)
	=\,&
	- \frac{\lb}{8} \bd_0^{-1}(k,-\ww)
	\intl{\vex{l}}
	\bd_0(l,2\varepsilon^-),	
	\label{eq:WL1--C}	
	\\
	\Sigma_W^{(b)} (-\kb,-\ww)
	=\,&
	- \frac{\lb}{8}  \bd_0^{-1}(k,-\ww)
	\intl{\vex{l}}
	\bd_0(l,-2\varepsilon^+),		
	\label{eq:WL2--C}
	\\
	\Sigma_W^{(c)} (-\kb,-\ww)
	=\,& 
	i\frac{3}{8} \pi h \gamma \lb^2
	\intl{\vex{l},\xi}
	\left[
	\bd_0^{-1}(k,-\ww)\bd_0(l,\xi)\bd_u(l,\xi)+\bd_u(l,\xi)
	\right] 
	\left[ \tanh \left(\frac{\varepsilon^{-}-\xi}{2T}\right)-\tanh \left(\frac{\varepsilon^{-}}{2 T}\right)\right]
	,
	\\
	\Sigma_W^{(d)} (-\kb,-\ww)
	=\,& 
	i\frac{3}{8} \pi h \gamma \lb^2
	\intl{\vex{l},\xi}
	\left[
	\bd_0 ^{-1} (k,-\ww) \bd_0(l,\xi)\bd_u(l,\xi)+\bd_u(l,\xi)
	\right] 
	\left[ -\tanh \left(\frac{\varepsilon^{+}+\xi}{2T}\right) +\tanh \left(\frac{\varepsilon^{+}}{2T} \right)\right]		
	,
	\\
	\Sigma_W^{(e)} (-\kb,-\ww)
	=\, & 
	- i\frac{3}{4} \pi h \gamma \lb ^2 \intl{\vex{l},\xi}
	\left\lbrace 
	\begin{aligned}
		&
		\bd_0(\lvert -\kb-\vex{l} \rvert,-\ww-\xi) \frac{\bd_u(l,\xi)}{\bd_0(l,\xi)}
		\left[\tanh \left(\frac{\varepsilon^+ + \xi}{2T}\right)- \coth \left(\frac{\xi}{2 T}\right) \right]
		\\ &
		+ \bd_0(\lvert -\kb-\vex{l} \rvert,-\ww-\xi) \frac{\bd_u(l,-\xi)}{\bd_0(l,-\xi)}
		\left[\tanh \left(\frac{\varepsilon^+}{2T}\right)+\coth \left(\frac{\xi}{2 T}\right)\right]
	\end{aligned}
	\right\rbrace,	
\end{align}
\esub
where we have neglected the factor $\delta_{\varepsilon_1,\varepsilon_2}\delta_{\alpha,\gamma}\delta_{\beta,\sigma}$ and set  $\varepsilon_1=\varepsilon_2=\varepsilon$.

Inserting Eqs.~(\ref{eq:C_3ae}a)--(\ref{eq:C_3ae}e) into Eq.~(\ref{eq:C_C1}), and carrying out integration by employing the approximation technique introduced in Sec.~\ref{sec:AII}, we obtain the net contributions from diagrams in Figs.~\ref{fig:C_3}(a)--\ref{fig:C_3}(e):
\begin{align}\label{eq:C_sigma1}
\begin{aligned}
	-i \left(  \Sigma^{(R)}_{\bv}\right) ^{i,j} (\kb,\ww)
	=\, &
	-\delta_{i,j} 4 h^2 \lb  
	\bd_0^2(k,-\ww)
	\left\lbrace 
	\frac{\ww}{\pi}
	\left[ 
	k^2 \delta \lb		
	-i h \lb \ww (-\delta h) 	
	\right] 
	+ 3 c_0
	\right\rbrace, 
\end{aligned}
\end{align}
where 
\bsub
\label{eq:C_d1}
\begin{align}
	\delta \lb 
	\equiv \, &		
	- \frac{\lb}{8 \pi} \ln  \left( \frac{\Lambda}{T} \right)
	+ \frac{3 \lb}{4 \pi}\left[ 1+\frac{1}{\gamma}\ln (1-\gamma) \right] \ln \left(\frac{\Lambda}{T}\right),
	\\
	\delta h
	\equiv \, &
	 \frac{\lb}{8 \pi}  \ln  \left( \frac{\Lambda}{T}\right)
	-\frac{3\lb}{4 \pi} \ln (1-\gamma) \ln \left(\frac{\Lambda}{T}\right) 
	-\frac{3\lb}{8 \pi} \gamma \ln \left(\frac{\Lambda}{T}\right),
	\\
	c_0 
	\equiv \, &
	\intl{\varepsilon} (F_{\varepsilon^+}-F_{\varepsilon^-})
	\Sigma_\varepsilon.
\end{align}
\esub
$\Sigma_\varepsilon$ 
is the outscattering rate previously defined 
in Eq.~(\ref{eq:SigEps}).
The first term in $\delta \lb$ and $\delta h$ comes from diagrams in Figs.~\ref{fig:C_3}(a) and \ref{fig:C_3}(b), which represent 
part of the class C weak localization correction. 
Diagrams in Figs.~\ref{fig:C_3}(c)--\ref{fig:C_3}(e) correspond to AA corrections and give rise to the other terms.

Note that unlike class AII, here the pure quantum interference correction at one loop order is directly cut off in the infrared by temperature $T$, independent of dephasing. The derivation is as follows. After inserting Eqs.~(\ref{eq:C_3ae}a)--(\ref{eq:C_3ae}b) into Eq.~(\ref{eq:C_C1}), we arrive at the integral 
\begin{align}
\begin{aligned}
	\intl{\varepsilon}		
	\left( F_{\varepsilon^+}  - F_{\varepsilon^-} \right)
	\intl{\vex{l}} 
	\bd_0(l,\pm 2 \varepsilon^{\mp})	
	=\,&
	\intl{\varepsilon}		
	\left( F_{\varepsilon+\ww}  - F_{\varepsilon} \right)
	\intl{\vex{l}} 
	\bd_0(l, 2 \varepsilon)	
	\approx\,
	\intl{\varepsilon,\vex{l}}		
	\left( \ww \partial_{\varepsilon} F_{\varepsilon}  \right) 
	\bd_0(l, 2 \varepsilon)
	=
	-\ww \intl{\varepsilon,\vex{l}}		
	F_{\varepsilon}  
	\partial_{\varepsilon}	\bd_0(l, 2\varepsilon)
	\\
	=	\,&
	\frac{\ww}{\pi} \frac{1}{4\pi} \ln\left(\frac{\Lambda}{T}\right).
\end{aligned}
\end{align}
Here we have approximated $\left( F_{\varepsilon+\ww}  - F_{\varepsilon} \right)$ by $\left( \ww \partial_{\varepsilon} F_{\varepsilon}  \right) $ and applied an integration by parts. 
The key difference relative to the standard class AII WAL correction [Eq.~(\ref{eq:WAL--AII})] is that
the energy argument of the loop propagator $\bd_0(l)$ in Eqs.~(\ref{eq:WL1--C}) and (\ref{eq:WL2--C}) is 
$\pm 2 \varepsilon^{\mp} = \pm 2 \epsilon -\omega$, not merely the external frequency $-\omega$. 
The subsequent $\varepsilon$-integration regularizes the infrared for any finite $T > 0$ [such that $F(\varepsilon)$ is smooth].
The total energy $\varepsilon$ serves as a ``mass'' for the class C diffuson mode, which is only gapless at $\varepsilon = 0$. 
This is a general feature of pure interference corrections due to nonstandard class quantum diffusion modes \cite{BTDLC05,KYG01}. 
In class C (which features broken time-reversal symmetry), additional localizing corrections arise at all one-body energies at two-loop order, due to the unitary class diffuson. These and all higher order corrections due to the Wigner-Dyson class modes must (by contrast) be cut by dephasing. In the next section, we will consider the effect of restoring time-reversal, which promotes class C to class CI. As a result, a WL correction due to the orthogonal class AI Cooperon appears that is also cut by dephasing.

For the remaining diagrams in Fig.\ \ref{fig:C_3}, i.e. Figs.~\ref{fig:C_3}(f) and \ref{fig:C_3}(g), the associated $\Sigma_{W}$ are no longer diagonal in frequency and spin spaces and are given by, respectively,
\bsub
\label{eq:C_3fg}
\begin{align}
	\,&
	\begin{aligned}	
		\Sigma_W^{(f)}
		=\,& -\frac{i}{4} \pi h \gamma \lb ^2 
		\sum_{n=1}^{3} (s^{n})^{\sigma,\alpha}  (s^{n})^{\gamma,\beta}
		\\
		&\times
		\intl{\vex{l}}
		\left\lbrace 
		\begin{aligned}
			&
			\bd_0(\lvert -\kb-\vex{l} \rvert,-\ww+\varepsilon_1+\varepsilon_2) \frac{\bd_u(l,-\varepsilon_1-\varepsilon_2)}{\bd_0(l,-\varepsilon_1-\varepsilon_2)}
			\left[-\tanh \left(\frac{\varepsilon_2^{-}}{2T}\right)+\coth \left(\frac{\varepsilon_1+\varepsilon_2}{2 T}\right) \right]
			\\ 
			+ & \bd_0(\lvert -\kb-\vex{l} \rvert,-\ww+\varepsilon_1+\varepsilon_2) 
			\frac{\bd_u(l,\varepsilon_1+\varepsilon_2)}{\bd_0(l,\varepsilon_1+\varepsilon_2)}
			\left[\tanh \left(\frac{\varepsilon_2^+}{2T}\right)-\coth \left(\frac{\varepsilon_1+\varepsilon_2}{2 T}\right)\right] 
		\end{aligned}
		\right\rbrace ,
	\end{aligned}	
	\\	
	 \, &
	\begin{aligned}
		\Sigma_W^{(g)}
		=\,&
		 -\frac{i}{4} \pi h \gamma \lb ^2 (- i \pi h \gamma \lb) 
		\sum_{n=1}^{3} (s^{n})^{\sigma,\nu}  (s^{n})^{\mu,\beta} \
		\left( \delta_{\alpha,\gamma} \delta_{\mu,\nu} + \delta_{\alpha,\nu} \delta_{\gamma,\mu}\right) 
		\\
		&\times
		\intl{\vex{l},\xi}
		\bd_0(\lvert -\kb-\vex{l} \rvert,-\ww-\xi)\bd_u(\lvert -\kb-\vex{l} \rvert,-\ww-\xi) 
		\left[\tanh \left(\frac{\varepsilon_2^{+}+\xi}{2T}\right)-\tanh \left(\frac{\varepsilon_2^-}{2 T}\right) \right]
		\\
		 & \quad \times 
		\left\lbrace 
		\frac{\bd_u(l,\xi)}{\bd_0(l,\xi)}
		\left[\tanh \left(\frac{\varepsilon_1^{+}+\xi}{2T}\right)-\coth \left(\frac{\xi}{2 T}\right) \right]
		+
		\frac{\bd_u(l,-\xi)}{\bd_0(l,-\xi)}
		\left[\tanh \left(\frac{\varepsilon_2^+}{2T}\right)+\coth \left(\frac{\xi}{2 T}\right)\right] 
		\right\rbrace.
	\end{aligned}		
\end{align}
\esub
Their combined contribution can be written as 
\begin{align}\label{eq:C_sigma2}
\begin{aligned}	
	-i \left(  \Sigma^{(R)}_{\bv}\right) ^{i,j} (\kb,\ww)
	=\,
	&
	-\delta_{i,j} 4  h^2 \lb  \bd_0 ^2 (k,-\ww)
	 \left[ \frac{\ww}{\pi} (i h \lb \gamma \omega ) (-\delta \Gamma) + c_0 \right], 
\end{aligned}
\end{align}	
where
\begin{align}\label{eq:C_d2}
\begin{aligned}	
	\delta \Gamma
	\equiv \, &
	\frac{\lb}{8 \pi} \ln \left(\frac{\Lambda}{T}\right)
	+\frac{\lb}{2 \pi} \frac{\gamma}{(1-\gamma) } \ln \left(\frac{\Lambda}{T}\right).		
\end{aligned}
\end{align}
Note that the terms proportional to $c_0$ in Eq.~(\ref{eq:C_sigma1}) and Eq.~(\ref{eq:C_sigma2}) do not cancel each other (unlike the case in class AII). 
We know that all such terms must cancel in the final result due to the spin SU(2) Ward identity. 
In what follows, we do not keep track of these terms involving the outscattering rate $\Sigma_{\varepsilon}$ 
(which would give a ``mass'' to the spin polarization function).

\subsubsection{Category 2}

\begin{figure}
	\centering
	\includegraphics[width=0.95\linewidth]{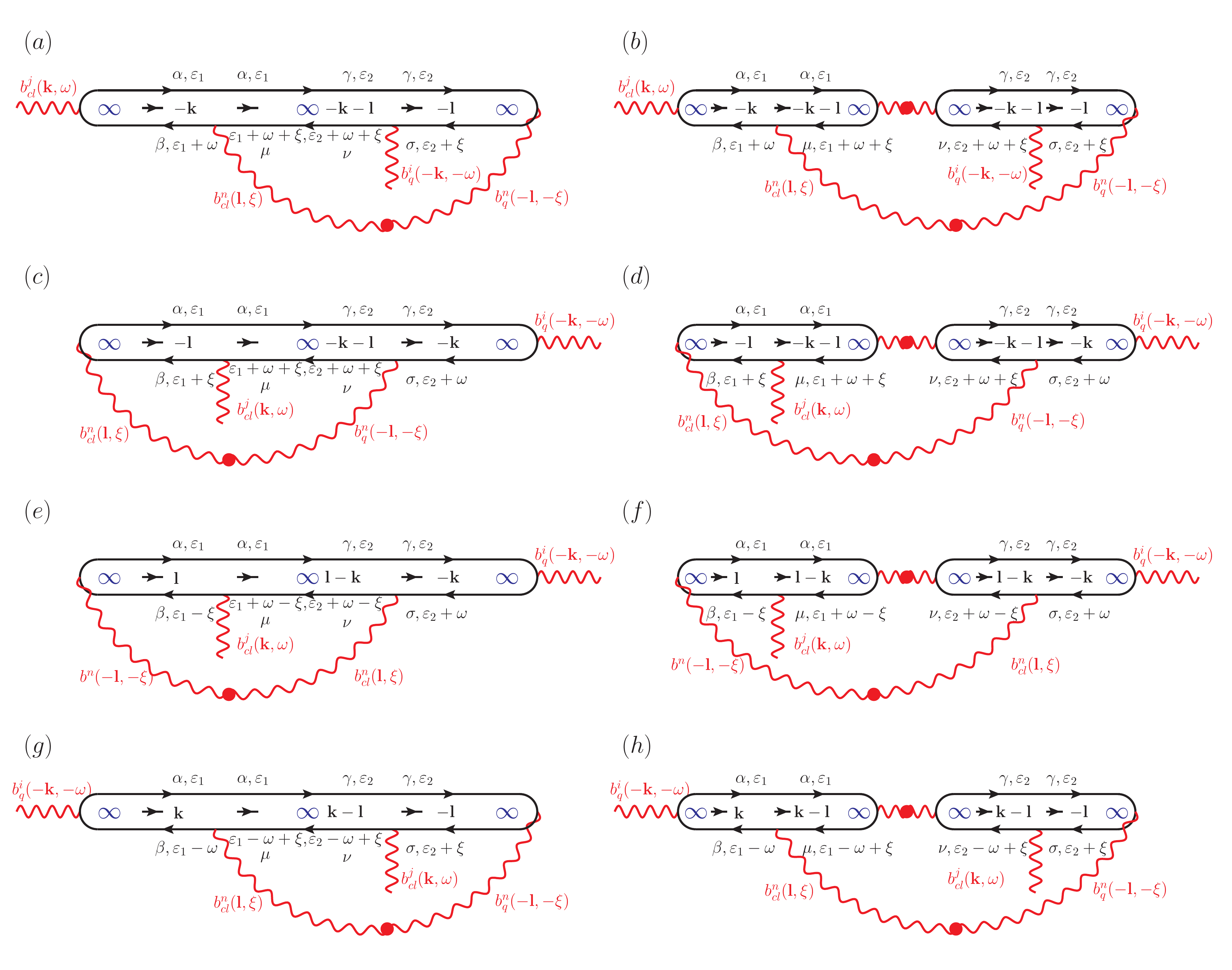}
	\caption{(Color online) Self energy diagrams for class C: Category 2.
	These diagrams give part of the AA wave function renormalization.
	}
	\label{fig:C_4}
\end{figure}

Diagrams in Figs.~\ref{fig:C_4}(a)--\ref{fig:C_4}(h) give, in respective order, the following contributions:
\bsub
\label{eq:C_4}
\begin{align}
	\, &
	\begin{aligned}
		(a)
		=\, &
		-4 \delta_{i,j} h^2 \lb 
		\bd_0 (k, -\ww)
		\\
		& \times 
		(i 2 \pi h \gamma \lb ^2) 	
		\intl{\vex{l},\xi} 
		\bd_0(\lvert -\kb-\vex{l} \rvert,-\ww-\xi)
		\bd_u(l,-\xi) 
		\intl{\varepsilon_1}
		\left( F_{\varepsilon_1 +\ww} -F_{\varepsilon_1}\right)
		\frac{1}{4}(F_{\varepsilon_1+\xi} + 3F_{-\varepsilon_1-\ww}),
	\end{aligned}
	\\	
	\, &
	\begin{aligned}	
		(b)
		=\,&
		-4 \delta_{i,j} h^2 \lb 
		\bd_0 (k, -\ww)	 
		\\
		& \times (i 2 \pi h \gamma \lb ^2)	
		\intl{\vex{l},\xi} 
		\bd_0(\lvert -\kb-\vex{l} \rvert,-\ww-\xi) \bd_u(\lvert -\kb-\vex{l} \rvert,-\ww-\xi)
		\bd_u(l,-\xi) 
		\intl{\varepsilon_1}
		\left( F_{\varepsilon_1 +\ww} -F_{\varepsilon_1}\right)
		\\
		&\times (-i \pi h \gamma \lb) \intl{\varepsilon_2}
		F_{\varepsilon_2+\xi} 
		(F_{\varepsilon_2+\ww+\xi}-F_{\varepsilon_2}),	
	\end{aligned}
	\\
	\, &
	\begin{aligned}	
		(c)
		=\, &
		-4 \delta_{i,j} h^2 \lb 
		\bd_0 (k, -\ww) 
		\\
		& \times (i 2 \pi h \gamma \lb ^2)
		\intl{\vex{l},\xi} 
		\bd_0(\lvert -\kb-\vex{l} \rvert,-\ww-\xi)
		\bd_u(l,-\xi) 
		\intl{\varepsilon_1}
		\left( F_{\varepsilon_1 +\xi} -F_{\varepsilon_1}\right)
		\frac{1}{4}(F_{\varepsilon_1+\ww} + 3F_{-\varepsilon_1-\xi}),
	\end{aligned}	
	\\
	\, &
	\begin{aligned}	
		(d)
		=\,&
		-4 \delta_{i,j}  h^2 \lb 
		\bd_0 (k, -\ww) 
		\\
		& \times (i 2 \pi h \gamma \lb ^2)
		\intl{\vex{l},\xi} 
		\bd_0(\lvert -\kb-\vex{l} \rvert,-\ww-\xi) \bd_u(\lvert -\kb-\vex{l} \rvert,-\ww-\xi)
		\bd_u(l,-\xi) 
		\intl{\varepsilon_1}
		\left( F_{\varepsilon_1 +\xi} -F_{\varepsilon_1}\right)
		\\
		&\times
		(-i \pi h \gamma \lb)
		\intl{\varepsilon_2}
		F_{\varepsilon_2+\ww} 
		(F_{\varepsilon_2+\ww+\xi}-F_{\varepsilon_2}),
	\end{aligned}	
	\\
	\, &
	\begin{aligned}	
		(e)
		=\,&
		-4 \delta_{i,j}  h^2 \lb 
		\bd_0 (k, -\ww) 
		\\
		& \times (i 2 \pi h \gamma \lb ^2)
		\intl{\vex{l},\xi,\varepsilon_1} 
		\bd_0(\lvert -\kb+\vex{l} \rvert,-\ww+\xi)
		\bd_0(l,\xi) 
		\\
		& \times	
		\left[
		\frac{\bd_u (l,-\xi)}{\bd_0 (l,-\xi)} 
		\left(1- F_{\varepsilon_1 -\xi} F_{\varepsilon_1}\right)
		+
		\left( 
		\frac{\bd_u (l,-\xi)}{\bd_0 (l,-\xi)}
		-\frac{\bd_u (l,\xi)}{\bd_0 (l,\xi)}
		\right) 
		\coth \left(\frac{\xi}{2T}\right)
		\left( F_{\varepsilon_1 -\xi} -F_{\varepsilon_1}\right)
		\right] ,
	\end{aligned}
	\\
	\, &
	\begin{aligned}	
		(f)
		=\,&
		-4 \delta_{i,j}  h^2 \lb 
		\bd_0 (k, -\ww) 
		\\
		& \times (i 2 \pi h \gamma \lb ^2)
		\intl{\vex{l},\xi,\varepsilon_1} 
		\bd_0(\lvert -\kb+\vex{l} \rvert,-\ww+\xi) \bd_u(\lvert -\kb+\vex{l} \rvert,-\ww+\xi)
		\bd_0(l,\xi) 
		\\
		& \times	
		\left[
		\frac{\bd_u (l,-\xi)}{\bd_0 (l,-\xi)} 
		\left(1- F_{\varepsilon_1 -\xi} F_{\varepsilon_1}\right)
		+
		\left( 
		\frac{\bd_u (l,-\xi)}{\bd_0 (l,-\xi)}
		-\frac{\bd_u (l,\xi)}{\bd_0 (l,\xi)}
		\right) 
		\coth \left(\frac{\xi}{2T}\right)
		\left( F_{\varepsilon_1 -\xi} -F_{\varepsilon_1}\right)
		\right] 
		\\	
		&\times (-i \pi h \gamma \lb) \intl{\varepsilon_2} (F_{\varepsilon_2 +\ww-\xi}-F_{\varepsilon_2}),
	\end{aligned}	
	\\
	\, &
	\begin{aligned}	
		(g)
			=\,&
			-4 \delta_{i,j} h^2 \lb 
			\bd_0 (k, \ww) 
			\\
			 & \times (i 2 \pi h \gamma \lb ^2)
			\intl{\vex{l},\xi} 
			\bd_0(\lvert \kb-\vex{l} \rvert,\ww-\xi)
			\bd_u(l,-\xi) 
			\intl{\varepsilon_1}
			\left( 1-F_{\varepsilon_1 -\ww} F_{\varepsilon_1} \right), 
	\end{aligned}	
		\\
	\, &
	\begin{aligned}		
		(h)
		=\,&
		-4 \delta_{i,j}  h^2 \lb 
		\bd_0 (k, \ww) 
		\\
		& \times (i 2 \pi h \gamma \lb ^2)
		\intl{\vex{l},\xi} 
		\bd_0(\lvert \kb-\vex{l} \rvert,\ww-\xi) \bd_u(\lvert \kb-\vex{l} \rvert,\ww-\xi)
		\bd_u(l,-\xi) 
		\intl{\varepsilon_1}\left( 1-F_{\varepsilon_1 -\ww} F_{\varepsilon_1} \right) 
		\\
		&\times
		(-i \pi h \gamma \lb)
		\intl{\varepsilon_2}
		(F_{\varepsilon_2-\ww+\xi}-F_{\varepsilon_2}).			
	\end{aligned}
\end{align}
\esub
We expand the integrals here in terms of external frequency $\ww$ and momentum $\kb$, and find that it is sufficient to retain only the leading-order terms. Up to logarithmic accuracy, Eqs.~(\ref{eq:C_4}g) and (\ref{eq:C_4}h) vanish, and the summation of the remaining equations in Eq.~(\ref{eq:C_4}) assumes the form
\begin{align}\label{eq:C_sigma3}
\begin{aligned}	
	-i \left(  \Sigma^{(R)}_{\bv}\right) ^{i,j} (\kb,\ww)
	=\,&
	- \delta_{i,j} 4 h^2 \lb  \bd_0 (k, -\ww)  \frac{\ww}{\pi} (-\delta z_1),	
\end{aligned}
\end{align}
where
\begin{align}\label{eq:C_d3}
\begin{aligned}
		\delta z_1
		\equiv \, &
		\frac{ 3\lb}{4 \pi} \ln (1-\gamma) \ln \left(\frac{\Lambda}{T}\right)
		+ 
		\frac{\lb}{ \pi} \frac{\gamma}{(1 - \gamma)} \ln \left(\frac{\Lambda}{T}\right).
\end{aligned}
\end{align}

\subsubsection{Category 3}

\begin{figure}
	\centering
	\includegraphics[width=0.9\linewidth]{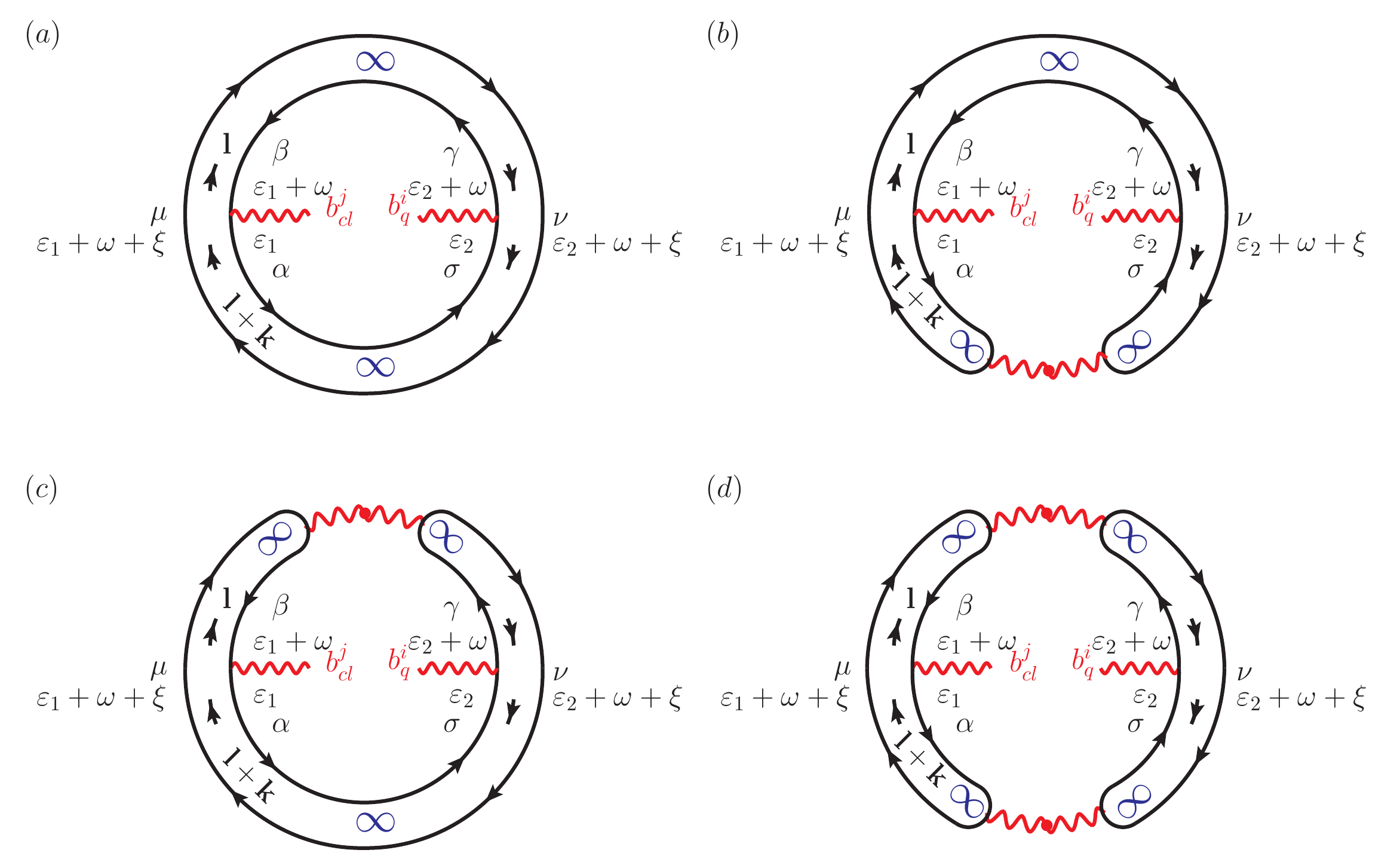}
	\caption{(Color online) Self energy diagrams for class C: Category 3.
	These diagrams give the remainder of the wave function renormalization.
	(a) is a pure class C quantum interference correction to the spin conductivity
	(weak localization)  
	and to the density of states. 
	(b)--(d) are AA corrections. 		
	}
	\label{fig:C_5}
\end{figure}
Fig.~\ref{fig:C_5} depicts another group of diagrams that give significant contribution to the retarded self energy. 
We find analogous diagrams for class AII (see Appendix \ref{Sec:App2}) whose net contributions vanish.  
In the present case, the amplitudes acquire the following forms:
\bsub
\label{eq:C_5}
\begin{align}
	\,&
	\begin{aligned}
		(a)
		=\,&
		-2 \delta_{i,j} h^2 \lb^2
		\intl{\vex{l},\varepsilon}
		\left[ 
		\begin{aligned}
			&\bd_0(\lvert \kb+\vex{l} \rvert,-2\varepsilon-\ww)
			\bd_0(l,-2\varepsilon-2\ww)
			F_{\varepsilon}
			\\
			+
			&\bd_0(\lvert \kb+\vex{l} \rvert,-2\varepsilon)
			\bd_0(l,-2\varepsilon -\ww)
			F_{\varepsilon}	
		\end{aligned}
		\right],
	\end{aligned}
	\\
	\,&
	\begin{aligned}
	(b)
	=\, &
	-8 \delta_{i,j} h^2 \lb ^2
	\intl{\vex{l},\xi} 
	\bd_0(\lvert \kb+\vex{l} \rvert,\ww+\xi) \bd_u(\lvert \kb+\vex{l} \rvert,\ww+\xi)
	\bd_0(l,\xi) 
	\\
	& \times
	(- i \pi h \gamma \lb)
	\intl{\varepsilon_1}(F_{\varepsilon_1}-F_{\varepsilon_1+\ww+\xi})
	\frac{1}{4}(3 F_{\varepsilon_1}-F_{\varepsilon_1+2\ww+\xi} ),
	\end{aligned}	
	\\
	\,&
	\begin{aligned}
	(c)
	=\,&
	-8 \delta_{i,j} h^2 \lb^2
	\intl{\vex{l},\xi}
	\bd_0(\lvert \kb+\vex{l} \rvert,\ww+\xi)
	\bd_0(l,\xi) \bd_u(l,\xi)
	\\
	& \times
	(-i \pi h \gamma \lb)\intl{\varepsilon_2}	F_{\varepsilon_2} 	(F_{\varepsilon_2+\ww}-F_{\varepsilon_2+\ww+\xi}),
	\end{aligned}
	\\
	\,&
	\begin{aligned}
	(d)
	=\,& 
	-8 \delta_{i,j} h^2 \lb ^2
	 \intl{\vex{l},\xi}
	\bd_0(\lvert \kb+\vex{l} \rvert,\ww+\xi) \bd_u(\lvert \kb+\vex{l} \rvert,\ww+\xi)
	\bd_0(l,\xi) \bd_u(l,\xi)
	\\
	& \times
	(- i \pi h \gamma \lb)^2
	\intl{\varepsilon_2}F_{\varepsilon_2} 	(F_{\varepsilon_2+\ww}-F_{\varepsilon_2+\ww+\xi})
	\,
	\intl{\varepsilon_1}(F_{\varepsilon_1}-F_{\varepsilon_1+\ww+\xi}).		
	\end{aligned}
\end{align}
\esub
Evaluating the integrals and adding the results, one finds the net contribution from the diagrams in Fig. \ref{fig:C_5}:
\begin{align}\label{eq:C_sigma4}
\begin{aligned}	
	-i \left(  \Sigma^{(R)}_{\bv}\right) ^{i,j} (\kb,\ww)
	= \, &
	\delta_{i,j} i \frac{4}{\pi}  h (-\delta z_2),
	\\
\end{aligned}
\end{align}
where	
\begin{align}\label{eq:C_d4}
\begin{aligned}		
	\delta z_2 
	\equiv \, &
	 \frac{\lb}{8 \pi}
	\ln \left(\frac{\Lambda}{T}\right)
	+
	\frac{\lb}{2  \pi} \frac{\gamma}{(1-\gamma)}
	\ln \left(\frac{\Lambda}{T}\right)	.			
\end{aligned}
\end{align}
Here Eq.~(\ref{eq:C_5}a) contributes to the WL correction. It is easy to see that, as with Eq.~(\ref{eq:C_3ae}a), it gives a logarithmic correction whose infrared cutoff is temperature $T$.

To evaluate the diagrams in Figs.~\ref{fig:C_3}--\ref{fig:C_5}, we have used the following identities:
	\begin{align}
	\begin{aligned}
	& \sum_{n=1}^{3} \left(s^{n}\right)^{\sigma,\mu} \left( s^{n}\right)^{\mu,\beta}=3\delta_{\beta,\sigma},
	\qquad	
	\tr[s^i s^j] =2 \delta_{i,j},
	\qquad
	\tr[s^i] =0,
		\\
	& \sum_{n=1}^{3} \left( s^{2}s^{n} \right)^{\alpha,\mu} \left( s^{n}s^{2} \right)^{\mu,\gamma}=3\delta_{\alpha,\gamma},
	\qquad  \quad \,
	\sum_{n=1}^{3} \left( s^{2}s^{n}\right)^{\mu,\beta} \left( s^{n}s^{2}\right)^{\sigma,\mu}=3\delta_{\beta,\sigma},
	\\
	&\sum_{n=1}^3 \tr \left[ \left(  s^{n} s^j s^2 \right) \left( s^2 s^i s^{n} \right)^{\T} \right] 
	=2 \delta_{i,j}	,
	\qquad
	\sum_{n=1}^3 \tr [s^j s^{n} s^i s^{n}] = -2\delta_{i,j},
	\\
	&\sum_{n=1}^3 \tr [(s^{n} s^j s^2) (s^2 s^{n} s^i )^{\T}] =-6\delta_{i,j},
	\qquad 
	\sum_{n=1}^3 \tr [(s^{j} s^{n} s^2) (s^2 s^{i} s^{n} )^{\T}] =-6\delta_{i,j}.
	\end{aligned}	
	\end{align}


\subsubsection{Results}

Adding Eqs.~(\ref{eq:C_sigma1}), (\ref{eq:C_sigma2}), (\ref{eq:C_sigma3}) as well as (\ref{eq:C_sigma4}), we arrive at the overall retarded self energy  
\begin{align}\label{eq:C_SigT}
\begin{aligned}
	& -i \left(  \Sigma^{(R)}_{\bv}\right)^{i,j} (\kb,\ww) = 
	-\delta_{i,j} \frac{4}{\pi}  h^2 \lb
	\bd_0^2 (k,-\ww)
	\left[ 
	\begin{aligned}
		& \ww k^2 
		\left( \delta \lb -\delta z_1 +2 \delta z_2 \right) 
		\\ &
		 -i h \lb \ww^2
		\left( -\delta h+\gamma \delta \Gamma -\delta z_1 + \delta z_2 \right) 
		 +
		\frac{k^4}{i h \lb} (-\delta z_2)
	\end{aligned}
	\right].	 					
\end{aligned}
\end{align}
To one-loop order, the wave function renormalization $Z$ for the field $\qh$ and the renormalized parameters $h_R$, $\lambda_R$ and $\gamma_R$ are given by
\begin{align}\label{eq:C_RG0}
\begin{aligned}
	& 
	Z= 1 +  \delta z_1 -2 \delta z_2,
	\qquad
	h_R= h(1 + \delta h + \delta z_1 -2 \delta z_2),
	\qquad
	\frac{1}{\lb_R}= \frac{1}{\lb} (1 - \delta \lb + \delta z_1 - 2 \delta z_2),
	\qquad
	\gamma_R = \gamma.	
\end{aligned}
\end{align}
The derivation of the last equality 
(the nonrenormalization of the interaction $\gamma$)	
is shown in Appendix~\ref{Sec:App1}.
Using Eqs.~(\ref{eq:C_d1}), (\ref{eq:C_d2}), (\ref{eq:C_d3}) and (\ref{eq:C_d4}), 
one may verify that the second term 
inside the brackets on the right-hand side of 
Eq.~(\ref{eq:C_SigT}), proportional to $-i h \lambda \omega^2$, vanishes as required by current conservation.	

We then find the quantum correction to the spin density response function 
\begin{align}\label{eq:C_dPi}
\begin{aligned}
	\delta \Pi^{i,j}
	(\kb,\omega)
	=& - \delta_{i,j} i\frac{2}{\pi} h^2 (1-\gamma)^2 \lb	
	\bd_u^2(k,-\ww)
	\left[ 
	\ww k^2 
	\left( \delta \lb -\delta z_1 +2 \delta z_2 \right) 
	+
	\frac{k^4}{i h \lb} (-\delta z_2)
	\right], 
\end{aligned}
\end{align}
and to the spin conductivity 	
\begin{align}\label{eq:C_dsigma}
\begin{aligned}
	\delta \sigma^{i,j}  
	=\,
	-\delta_{i,j}\frac{2}{\pi} \frac{1}{\lb} 
	\left( \delta \lb -\delta z_1 +2 \delta z_2\right) 	
	=\,
	\delta_{i,j}\frac{2}{\pi} \left( \frac{1}{\lb_R}-\frac{1}{\lb}\right).	
\end{aligned}
\end{align}
Eq.~(\ref{eq:C_dsigma}) can be written as Eq.~(\ref{eq:correction}b)
using the explicit forms of $\delta \lb$ , $\delta z_1$ and  $\delta z_2$.

\section{Class CI \label{sec:CI}}

As a last example, we consider the noninteracting class CI superconductor where both the spin-rotational and time-reversal symmetries are preserved \cite{CRealization1}. Its nonlinear sigma model can be easily obtained from class AI. The partition function $Z[\vex{B}]$ is given by
\bsub
\begin{align}\label{eq:CI_Z}
	Z[\vex{B}]
	=\, &
	\int \dd \qh \exp \left( -S_q-S_c-S_{\vex{B}} \right), 
	\\
	S_q
	=\,& 
	\frac{1}{4 \lambda}
	\intl{\vex{x}}
	\tr\left[\Nabla \qh^{\dagger} \cdot \Nabla \qh \right]
	+
	\frac{i h}{2}
	\intl{\vex{x}}
	\tr\left[ (\hat{\omega} + i \eta \tauh^3) (\qh +\qh^{\dagger})\right],
	\\ 
	S_c
	=\,&
	-\frac{i h}{2}
	\intl{\vex{x}} \tr 
	\left[
	\left( \Bvcl  +\Bvql \hat{\tau}^1\right) \cdot \h{\sv} 
	\mf(\hat{\omega})\left( \hat{q} + \hat{q}^{\dagger} \right)  \mf(\hat{\omega})
	\right],
	\\
	S_{\vex{B}}
	=\,&
	-i \frac{4}{\pi} h \intl{\vex{x},t} \Bvcl \cdot \Bvql, 
\end{align}
\esub
where $\qh$ is a matrix in Keldysh, spin, as well as frequency spaces, and subject to the conditions
\begin{align}\label{eq:CI_qsym}
	\qh^{\dagger} \qh=\,1,
	\qquad
	-\h{s}^2 \h{\tau}^1 \h{\Sigma}^1 \qh^{\T} \h{s}^2 \h{\tau}^1 \h{\Sigma}^1 =\, \qh^{\dagger}.
\end{align}
$\qsp=\hat{\tau}^3$ is the saddle point.
Note that there is no H.-S.\ field in the theory of the noninteracting system. However, the external field $\vex{B}$ couples to the matrix field $\qh$ in the same way as the H.-S.\ field does in the interacting case.

\subsection{Parameterization}

To calculate the spin response function, we first apply the transformation:
\begin{align}
	\qh \rightarrow \hat{\tau}^3 \qh,
	\qquad
	\qh^{\dagger} \rightarrow \qh^{\dagger} \hat{\tau}^3, 
\end{align}
which brings the saddle point to the identity and transforms the actions $S_q$ and $S_c$ to
\bsub\label{eq:CI_Sqc}
\begin{align}
	S_q
	=\,& 
	\frac{1}{4 \lambda}
	\intl{\vex{x}}
	\tr\left[\Nabla \qh^{\dagger} \cdot \Nabla \qh \right]
	+
	\frac{i h}{2}
	\intl{\vex{x}}
	\tr\left[ (\hat{\omega}\tauh^3 + i \eta ) (\qh +\qh^{\dagger})\right],
	\\ 
	S_c
	=\,&
	-\frac{i h}{2}
	\intl{\vex{x}} \tr 
	\left[
	\left( \Bvcl  +\Bvql \hat{\tau}^1\right) \cdot \h{\sv} 
	\mf(\hat{\omega})\left( \tauh^3\hat{q} + \hat{q}^{\dagger}\tauh^3 \right)  \mf(\hat{\omega})
	\right].
\end{align}
\esub
Moreover, the constraints of $\qh$ now become
\begin{align}\label{eq:CI_qsym1}
	\qh^{\dagger} \qh=\,1,
	\qquad
	\h{s}^2 \h{\tau}^1 \h{\Sigma}^1 \qh^{\T} \h{s}^2 \h{\tau}^1 \h{\Sigma}^1 =\, \qh^{\dagger}.
\end{align}

Given that $\qh$ is not Hermitian, a parameterization different from that in class AII and C is used:
\begin{align}\label{eq:CI_q}
	\qh
	= \, &
	\exp \left( i \Wh \right) 
	= \,
	1 + i \Wh -\frac{1}{2} \Wh^2 -\frac{i}{6} \Wh^3  +\frac{1}{24} \Wh^4 +...,
\end{align}
where $\Wh$ follows the conditions
\begin{align}\label{eq:CI_Wsym1}
\begin{aligned}
	\Wh=\,& \Wh^{\dagger},
	\qquad
	\h{s}^2 \h{\tau}^1 \h{\Sigma}^1 \Wh^{\T} \h{s}^2 \h{\tau}^1 \h{\Sigma}^1 =\, -\Wh	. 
\end{aligned}
\end{align}
To be more specific, the second condition given above means

\begin{align}\label{eq:CI_Wsym2}
	& \Wh^{a,b;\alpha,\beta}_{1,2}=\, s_{\alpha,\beta} \Wh^{-b,-a;-\beta,-\alpha}_{-2,-1},	 
\end{align}
where the sign factor $s_{\alpha,\beta}$ is defined by
\begin{align}\label{eq:CI_sgn}
	&s_{\alpha,\beta}
	=
	\begin{cases}
	-1,& \alpha=\beta,
	\\
	1, & \alpha \neq \beta.
	\end{cases}	
\end{align}
Here $\left\lbrace a,b\right\rbrace $, $\left\lbrace \alpha,\beta \right\rbrace $ and $\left\lbrace 1,2 \right\rbrace$ index the Keldysh, spin and frequency spaces, respectively. $-a$ and  $-\alpha$ are defined such that 
$(\hat{\tau}^3)^{-a,-a}=-(\hat{\tau}^3)^{a,a}$ and
$(\hat{s}^3)^{-\alpha,-\alpha}=-(\hat{s}^3)^{\alpha,\alpha}$. 
For example, if $a=1$ and $\alpha=\uparrow$, then we have $-a=2$ and $-\alpha=\downarrow$.

Substituting the parameterization given by Eq.~(\ref{eq:CI_q}) into Eq.~(\ref{eq:CI_Sqc}), and rescaling $\Wh$ as in Eq.~(\ref{eq:Wrescale}), we arrive at the action in terms of $\Wh$.
Up to quadratic order in $\Wh$, it can be expressed as 
\begin{align}\label{eq:CI_SQ2}
\begin{aligned}
	S_{W}^{(2)} 
	= \, & 
	\int 
	\Wh^{1,2;\alpha,\beta}_{1,2}(-\kb_1)
	M^{\beta,\alpha;\sigma,\gamma}_{2,1;4,3}(\kb_1,\kb_2)
	\Wh^{2,1;\gamma,\sigma}_{3,4}(\kb_2)
	+
	J^{\dagger}\,^{\beta,\alpha}_{2,1}(\kb) \Wh^{2,1;\alpha,\beta}_{1,2}(\kb)
	+
	J^{\beta,\alpha}_{2,1}(\kb) \Wh^{1,2;\alpha,\beta}_{1,2}(-\kb)
	\\
	+\,&
	\suml{a,b}
	\int 
	\frac{1}{2}
	\Wh^{a,a;\alpha,\beta}_{1,2}(-\kb_1)
	(N_{a,b})\,^{\beta,\alpha;\sigma,\gamma}_{2,1;4,3}(\kb_1,\kb_2)
	\Wh^{b,b;\gamma,\sigma}_{3,4}(\kb_2),		
\end{aligned}
\end{align}
where $M$, $N$, $J$, and $J^{\dagger}$ are defined as follows
\bsub\label{eq:CI_MNJ}
\begin{align}
&\begin{aligned}
	M^{\beta,\alpha;\sigma,\gamma}_{2,1;4,3}(\kb_1,\kb_2) 
	\equiv \, & 
	\frac{1}{2} \left[ k_1^2 - i h \lb (\ww_1-\ww_2) \right] 
	\delta_{\alpha,\sigma} \delta_{\beta,\gamma}
	\delta_{1,4}\delta_{2,3}
	\delta_{\kb_1,\kb_2}
	\\
	& +i h \lb
	\left[ 
	\Bvcl(\kb_1-\kb_2,\ww_4-\ww_1) 
	+F_4 \Bvql(\kb_1-\kb_2,\ww_4-\ww_1)
	\right] 
	\cdot \sv^{\sigma,\alpha}	
	\delta_{2,3}\delta_{\beta,\gamma},	
	\end{aligned}	
	\\	 
	& J^{\dagger}\,^{\beta,\alpha}_{2,1}(\kb) 
	\equiv \, 
	h \sqrt{\lb}
	\left[
	(F_{2}-F_{1})\Bvcl(-\kb,\ww_2-\ww_1)
	+(1-F_{1}F_{2})\Bvql(-\kb,\ww_2-\ww_1)
	\right] \cdot \sv^{\beta,\alpha},
	\\
	& J^{\beta,\alpha}_{2,1}(\kb) 
	\equiv\,
	- h \sqrt{\lb}
	\Bvql(\kb,\ww_2-\ww_1) \cdot \sv^{\beta,\alpha},
	\\
	&\begin{aligned}
	(N_{a,b})^{\beta,\alpha;\sigma,\gamma}_{2,1;4,3}(\kb_1,\kb_2) 
	\equiv \,&
	\delta_{a,b}\delta_{a,1}
	\left\lbrace 
	\begin{aligned}
	& \frac{1}{2} \left[ k_1^2 - i h \lb (\ww_1+\ww_2) \right]
	\delta_{\alpha,\sigma} \delta_{\beta,\gamma}
	\delta_{1,4}\delta_{2,3}
	\delta_{\kb_1,\kb_2}			 
	\\
	+&i h \lb
	\left[ 
	\Bvcl(\kb_1-\kb_2,\ww_4-\ww_1) 
	+
	F_4 \Bvql(\kb_1-\kb_2,\ww_4-\ww_1)
	\right] 
	\cdot \sv^{\sigma,\alpha}
	\delta_{2,3}\delta_{\beta,\gamma}
	\end{aligned}
	\right\rbrace 
	\\
	\,& +
	\delta_{a,b}\delta_{a,2}
	\left\lbrace 
	\begin{aligned}
	&\frac{1}{2} \left[ k_1^2 + i h \lb (\ww_1+\ww_2) \right] 
	\delta_{\alpha,\sigma} \delta_{\beta,\gamma}
	\delta_{1,4}\delta_{2,3}
	\delta_{\kb_1,\kb_2}			
	\\
	+&i h \lb
	\left[ 
	-\Bvcl(\kb_1-\kb_2,\ww_4-\ww_1) 
	+F_1 \Bvql(\kb_1-\kb_2,\ww_4-\ww_1)
	\right] 
	\cdot \sv^{\sigma,\alpha}	
	\delta_{2,3}\delta_{\beta,\gamma}
	\end{aligned}
	\right\rbrace. 	
\end{aligned}		
\end{align}
\esub
We also retain the cubic term $S_W^{(3)}$ and quartic term  $S_W^{(4)}$ which arise from $S_c$ [Eq.~(\ref{eq:CI_Sqc}b)] and $S_q$ [Eq.~(\ref{eq:CI_Sqc}a)], respectively. They are given by
\bsub
\begin{align}\label{eq:CI_SQ3}
\begin{aligned}
	S_W^{(3)}
	=\,&
	-\frac{\lb}{6} 
	\int \delta_{\kb_1+\kb_2+\kb_3,\kb}
	J^{\dagger}\,^{\beta,\alpha}_{2,1}(\kb)
	\\& \times
	\left[ 
	\begin{aligned}
	\Wh^{2,1;\alpha,\alpha'}_{1,1'}(\kb_1)
	\Wh^{1,2;\alpha',\beta'}_{1',2'}(\kb_2)
	\Wh^{2,1;\beta',\beta}_{2',2}(\kb_3)
	+
	\Wh^{2,2;\alpha,\alpha'}_{1,1'}(\kb_1)
	\Wh^{2,1;\alpha',\beta'}_{1',2'}(\kb_2)
	\Wh^{1,1;\beta',\beta}_{2',2}(\kb_3)
	\\
	+
	\Wh^{2,2;\alpha,\alpha'}_{1,1'}(\kb_1)
	\Wh^{2,2;\alpha',\beta'}_{1',2'}(\kb_2)
	\Wh^{2,1;\beta',\beta}_{2',2}(\kb_3)
	+
	\Wh^{2,1;\alpha,\alpha'}_{1,1'}(\kb_1)
	\Wh^{1,1;\alpha',\beta'}_{1',2'}(\kb_2)
	\Wh^{1,1;\beta',\beta}_{2',2}(\kb_3)		
	\end{aligned}
	\right] 
	\\
	\,& 
	-\frac{\lb}{6} 
	\int \delta_{\kb_1+\kb_2+\kb_3,-\kb}
	J^{\beta,\alpha}_{2,1}(\kb)
	\\& \times
	\left[ 
	\begin{aligned}
	\Wh^{1,2;\alpha,\alpha'}_{1,1'}(\kb_1)
	\Wh^{2,1;\alpha',\beta'}_{1',2'}(\kb_2)
	\Wh^{1,2;\beta',\beta}_{2',2}(\kb_3)
	+
	\Wh^{1,1;\alpha,\alpha'}_{1,1'}(\kb_1)
	\Wh^{1,2;\alpha',\beta'}_{1',2'}(\kb_2)
	\Wh^{2,2;\beta',\beta}_{2',2}(\kb_3)
	\\
	+
	\Wh^{1,2;\alpha,\alpha'}_{1,1'}(\kb_1)
	\Wh^{2,2;\alpha',\beta'}_{1',2'}(\kb_2)
	\Wh^{2,2;\beta',\beta}_{2',2}(\kb_3)	
	+
	\Wh^{1,1;\alpha,\alpha'}_{1,1'}(\kb_1)
	\Wh^{1,1;\alpha',\beta'}_{1',2'}(\kb_2)
	\Wh^{1,2;\beta',\beta}_{2',2}(\kb_3)	
	\end{aligned}
	\right], 	
\end{aligned}	
\end{align}
\begin{align}\label{eq:CI_SQ4}
\begin{aligned}
	S_W^{(4)}
	=\,
	\frac{ \lb}{16 }
	\int
	\delta_{\kb_1+\kb_2+\kb_3+\kb_4,\vex{0}}&	
	\left[ 
	\begin{aligned}
	& -\left( \kb_1 \cdot \kb_3 + \kb_2 \cdot \kb_4 \right) 
	-\frac{1}{2}\left( \kb_1 + \kb_3 \right)\cdot \left(  \kb_2 + \kb_4 \right)
	-\frac{1}{3} \left( k_1^2+k_2^2+k_3^2+k_4^2 \right) 
	\\
	& +\frac{1}{6} i h \lb ( \ww_1 +  \ww_2 +\ww_3 +\ww_4)
	\end{aligned}
	\right] 	
	\\
	\times &
	\left[ 
	\begin{aligned}
	&
	2
	\Wh^{1,2;\alpha,\beta}_{1,-2}(\kb_1)
	\Wh^{2,1;\beta,\gamma}_{-2,3}(\kb_2)
	\Wh^{1,2;\gamma,\sigma}_{3,-4}(\kb_3)
	\Wh^{2,1;\sigma,\alpha}_{-4,1}(\kb_4)
	\\
	+&	
	\Wh^{1,1;\alpha,\beta}_{1,2}(\kb_1)
	\Wh^{1,1;\beta,\gamma}_{2,3}(\kb_2)
	\Wh^{1,1;\gamma,\sigma}_{3,4}(\kb_3)
	\Wh^{1,1;\sigma,\alpha}_{4,1}(\kb_4)
	\\
	+&	
	\Wh^{2,2;\alpha,\beta}_{-1,-2}(\kb_1)
	\Wh^{2,2;\beta,\gamma}_{-2,-3}(\kb_2)
	\Wh^{2,2;\gamma,\sigma}_{-3,-4}(\kb_3)
	\Wh^{2,2;\sigma,\alpha}_{-4,-1}(\kb_4)
	\\
	+&
	4
	\Wh^{1,1;\alpha,\beta}_{1,2}(\kb_1)
	\Wh^{1,1;\beta,\gamma}_{2,3}(\kb_2)
	\Wh^{1,2;\gamma,\sigma}_{3,-4}(\kb_3)
	\Wh^{2,1;\sigma,\alpha}_{-4,1}(\kb_4)
	\\	
	+&	
	4	
	\Wh^{2,2;\alpha,\beta}_{-1,-2}(\kb_1)
	\Wh^{2,2;\beta,\gamma}_{-2,-3}(\kb_2)
	\Wh^{2,1;\gamma,\sigma}_{-3,4}(\kb_3)
	\Wh^{1,2;\sigma,\alpha}_{4,-1}(\kb_4)
	\\	
	+&
	4	
	\Wh^{1,1;\alpha,\beta}_{1,2}(\kb_1)
	\Wh^{1,2;\beta,\gamma}_{2,-3}(\kb_2)
	\Wh^{2,2;\gamma,\sigma}_{-3,-4}(\kb_3)
	\Wh^{2,1;\sigma,\alpha}_{-4,1}(\kb_4)
	\end{aligned}			
	\right]. 
\end{aligned}
\end{align}
\esub
In what follows, we denote their summation as $S_W^{(3,4)}$.

\subsection{Feynman rules}

\subsubsection{Bare propagator}
Ignoring the coupling of the matrix field $\Wh$ to the external field $\vex{B}$, the quadratic action acquires the form
\begin{align}
\begin{aligned}
	S_W^{(2)}
	=\,&
	\suml{a,b}
	\int
	\Wh^{a,b;\alpha,\beta}_{1,2}(-\kb)
	\Wh^{b,a;\beta,\alpha}_{2,1}(\kb)
	\frac{1}{4}
	 \left[
	  k^2 
	  - i h \lb   
	  \left( \z_a \ww_1 +\z_b \ww_2\right) 
	  \right]   ,
	\end{aligned}
\end{align}
where $\z_a$ denotes $(\hat{\tau}^3)^{a,a}$.
We obtain the bare propagator attributed to this action 
\begin{align}
\begin{aligned}
	\braket{\Wh^{a,b;\alpha,\beta}_{1,2}(\kb)
		\Wh\,^{c,d;\gamma,\sigma}_{3,4}(-\kb)}_{0}
	=\,
	&
	\delta_{\alpha,\sigma}\delta_{\beta,\gamma}
	\delta_{a,d}\delta_{b,c}\delta_{1,4}\delta_{2,3}
	\Delta_0(k,-\z_a\ww_1-\z_b\ww_2) 
	\\
	&+
	s_{\alpha,\beta}
	\delta_{\alpha,-\gamma}\delta_{\beta,-\sigma}
	\delta_{a,-c}\delta_{b,-d}\delta_{1,-3}\delta_{2,-4}  
	\Delta_0(k,-\z_a\ww_1-\z_b\ww_2) .
\end{aligned}			   			
\end{align}
Here $\Delta_0$ is defined in Eq.~(\ref{eq:Delta0}).
After substituting the explicit value of the Keldysh indices into this equation, we arrive at:
\bsub\label{eq:CI_Wpropagator}
\begin{align}
	\braket{\Wh^{2,1;\alpha,\beta}_{1,2}(\kb)
		\Wh\,^{1,2;\gamma,\sigma}_{3,4}(-\kb)}_{0}
	=\,
	&
	\left[ 
	\delta_{\alpha,\sigma}\delta_{\beta,\gamma}
	\delta_{1,4}\delta_{2,3} 
	+
	s_{\alpha,\beta}
	\delta_{\alpha,-\gamma}\delta_{\beta,-\sigma}
	\delta_{1,-3}\delta_{2,-4}  
	\right] 
	\Delta_0(k,\ww_1-\ww_2) ,	
	\\
	\braket{\Wh^{1,1;\alpha,\beta}_{1,2}(\kb)
		\Wh\,^{1,1;\gamma,\sigma}_{3,4}(-\kb)}_{0}
	=\,
	&
	\delta_{\alpha,\sigma}\delta_{\beta,\gamma}
	\delta_{1,4}\delta_{2,3} 
	\Delta_0(k,-\ww_1-\ww_2) ,	
	\\
	\braket{\Wh^{2,2;\alpha,\beta}_{1,2}(\kb)
		\Wh\,^{2,2;\gamma,\sigma}_{3,4}(-\kb)}_{0}
	=\,
	&
	\delta_{\alpha,\sigma}\delta_{\beta,\gamma}
	\delta_{1,4}\delta_{2,3} 
	\Delta_0(k,\ww_1+\ww_2) ,
	\\
	\braket{\Wh^{1,1;\alpha,\beta}_{1,2}(\kb)
		\Wh\,^{2,2;\gamma,\sigma}_{3,4}(-\kb)}_{0}
	=\,
	& 
	s_{\alpha,\beta}
	\delta_{\alpha,-\gamma}\delta_{\beta,-\sigma}
	\delta_{1,-3}\delta_{2,-4}   
	\Delta_0(k,-\ww_1-\ww_2) .				   			
\end{align}
\esub
\begin{figure}
	\centering
	\includegraphics[width=0.8\linewidth]{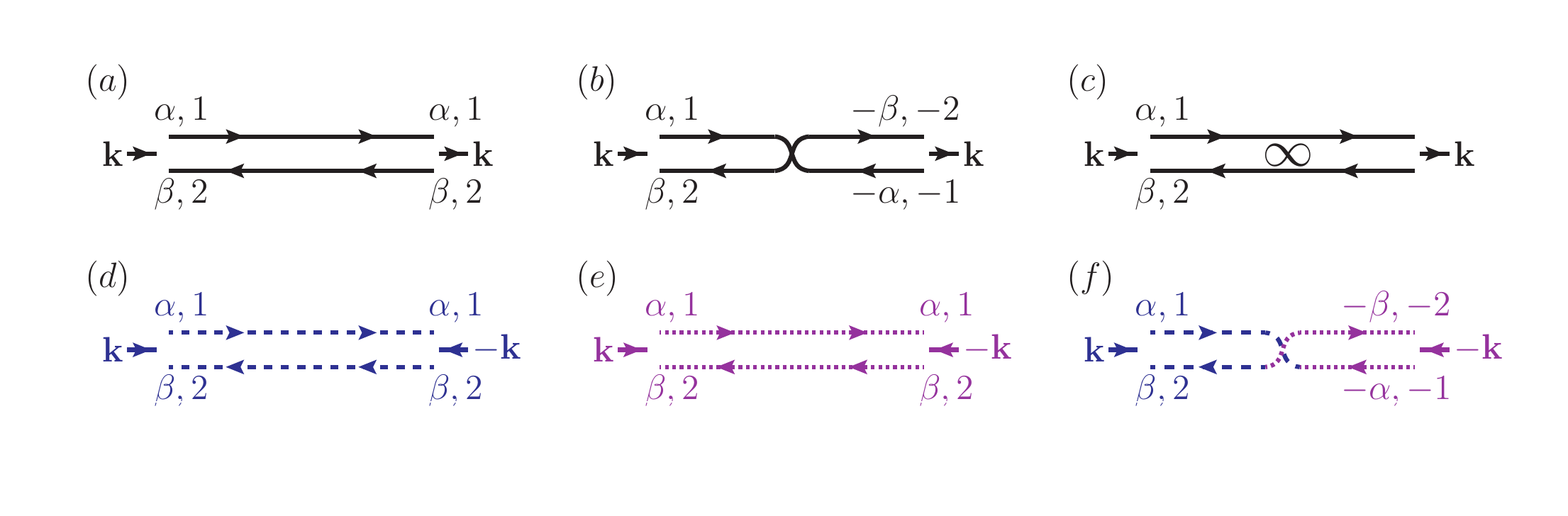}
	\caption{
	(Color online) Class CI bare propagators. (a) and (b) correspond to the two terms in the propagator $\braket{\Wh^{2,1}\Wh^{1,2}}_{0}$ [Eq.~(\ref{eq:CI_Wpropagator}a)]. Their sum is represented by (c). (d), (e) and (f) are associated with 
	$\braket{\Wh^{1,1}\Wh^{1,1}}_{0}$  [ Eq.~(\ref{eq:CI_Wpropagator}b)] , 
	$\braket{\Wh^{2,2}\Wh^{2,2}}_{0}$  [Eq.~(\ref{eq:CI_Wpropagator}c)],
	and $\braket{\Wh^{1,1}\Wh^{2,2}}_{0}$ [Eq.~(\ref{eq:CI_Wpropagator}d)], respectively.
}
	\label{fig:CI_1}
\end{figure}

These terms are represented diagrammatically in Fig.~\ref{fig:CI_1},
where solid black lines indicate $W^{2,1}$ and  $W^{1,2}$,
while dashed blue (doted purple) ones correspond to $W^{1,1}$($W^{2,2}$).
Here the superscripts of $W$ are indices in the Keldysh space.
$W^{2,1}$ and  $W^{1,2}$ are distinguished by the direction of the short arrows. 
Among the diagrams appearing in Fig.~\ref{fig:CI_1},
(a) and (b) represent the two terms in Eq.~(\ref{eq:CI_Wpropagator}a), 
while (d), (e) and (f) correspond to, in respective order, Eqs.~(\ref{eq:CI_Wpropagator}b), (\ref{eq:CI_Wpropagator}c), and (\ref{eq:CI_Wpropagator}d). 
As in class C, we use two parallel lines with symbol $\infty$ in between to denote the sum of two terms in Eq.~(\ref{eq:CI_Wpropagator}a) [see Fig.~\ref{fig:CI_1}(c)].
Note that diagrams in Figs.~\ref{fig:CI_1}(b) and (f) contain a sign factor $s_{\alpha,\beta}$ assuming a value of $-1$ when the spin indices are the same, and $+1$ otherwise.

\subsubsection{Interaction vertices}

\begin{figure}
	\centering
	\includegraphics[width=0.8\linewidth]{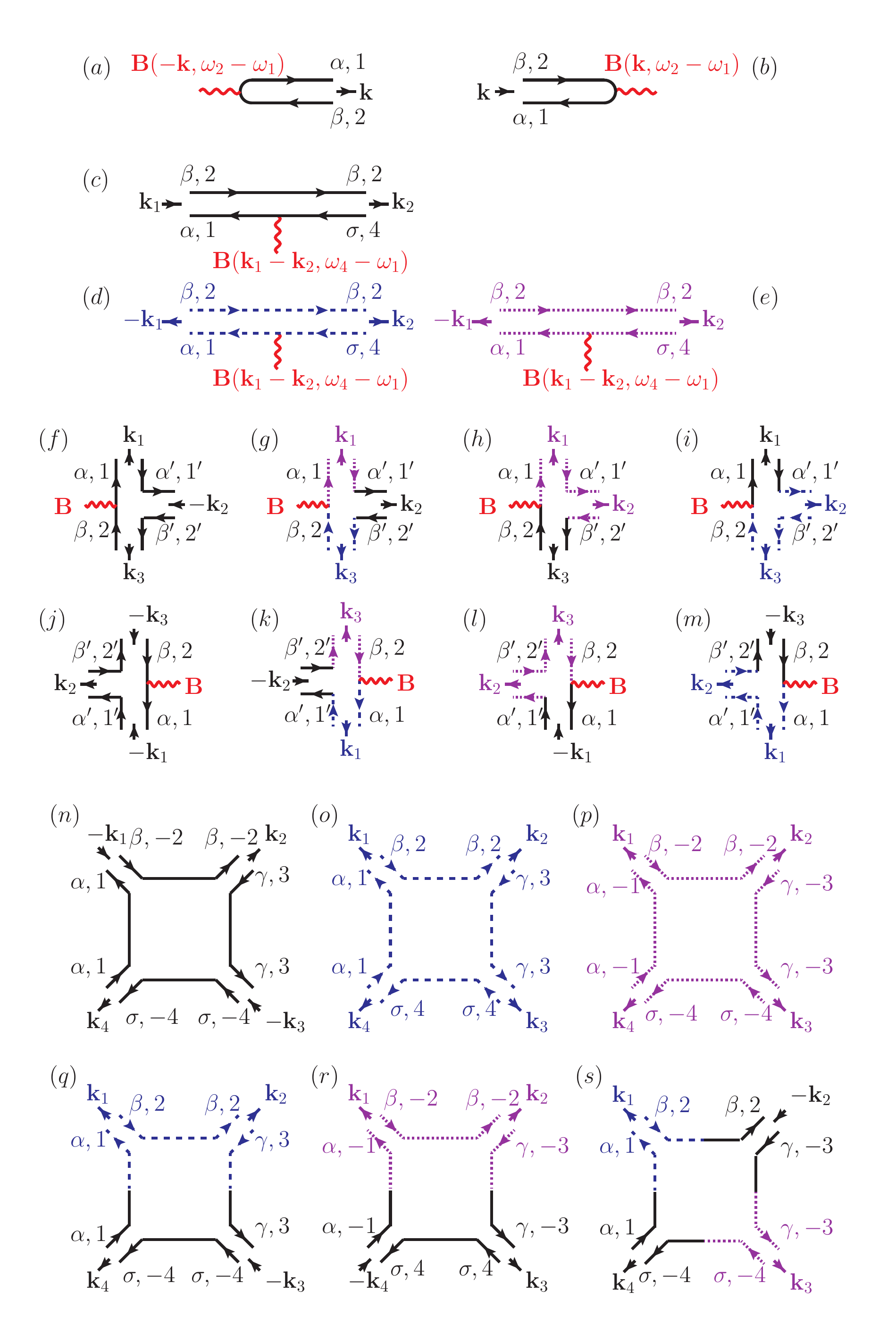}
	\caption{
	(Color online) Class CI interaction and diffusion vertices.
}
	\label{fig:CI_2}
\end{figure}

In Figs.~\ref{fig:CI_2} (a)--(n), we depict the interaction vertices coupling the matrix field $\Wh$ and external fields $\vex{B}$. As with the H.-S.\ field in previous section, here we use red wavy line to denote $\vex{B}$.

The amplitudes of these interaction vertices are given as follows:
\begin{align}
\begin{aligned}
	(a) 
	=\, &
	-h \sqrt{\lb}
	\left[
	(F_{2}-F_{1})\Bvcl(-\kb,\ww_2-\ww_1)
	+(1-F_{1}F_{2})\Bvql(-\kb,\ww_2-\ww_1)
	\right] \cdot \sv^{\beta,\alpha},
	\\
	(b)
	=\,&
	 h \sqrt{\lb}
	\Bvql(\kb,\ww_2-\ww_1) \cdot \sv^{\beta,\alpha}	,		
	\\	
	(c) 
	=\,&
	-i h \lb
	\left[ 
	\Bvcl(\kb_1-\kb_2,\ww_4-\ww_1) 
	+F_4 \Bvql(\kb_1-\kb_2,\ww_4-\ww_1)
	\right] \cdot \sv^{\sigma,\alpha},	
	\\
	(d)
	=\,&
	-\frac{i h \lb}{2}
	\left[ 
	\Bvcl(\kb_1-\kb_2,\ww_4-\ww_1) 
	+
	F_4 \Bvql(\kb_1-\kb_2,\ww_4-\ww_1)
	\right] 
	\cdot \sv^{\sigma,\alpha},
	\\
	(e)
	=\,&
	-\frac{i h \lb}{2}
	\left[ 
	-\Bvcl(\kb_1-\kb_2,\ww_4-\ww_1) 
	+F_1 \Bvql(\kb_1-\kb_2,\ww_4-\ww_1)
	\right] 
	\cdot \sv^{\sigma,\alpha},				
	\\
	(f)=\, &(g)= (h) = (i) 
	\\ 
	=\, &
	\frac{h \lb^{3/2}}{6}
	\left[
	(F_{2}-F_{1})\Bvcl(-\kb,\ww_2-\ww_1)
	+(1-F_{1}F_{2})\Bvql(-\kb,\ww_2-\ww_1)
	\right] \cdot \sv^{\beta,\alpha} \delta_{\kb_1+\kb_2+\kb_3,\kb} \, ,
	\\
	(j)=\, &(k)= (l) = (m) 
	\\
	=\,&
	-\frac{h \lb^{3/2}}{6}
	\Bvql(\kb,\ww_2-\ww_1) \cdot \sv^{\beta,\alpha} \delta_{\kb_1+\kb_2+\kb_3,-\kb}\,	.		
\end{aligned}
\end{align}

\subsubsection{4-point diffusion vertices}

The remaining diagrams in Figs.~\ref{fig:CI_2}, i.e., (n)--(s), illustrate the 4-point diffusion vertices from $S_W^{(4)}$. They share the same amplitude
\begin{align}
\begin{aligned}
	-\frac{ \lb}{4}
	\left[ 
	\begin{aligned}
	& -\left( \kb_1 \cdot \kb_3 + \kb_2 \cdot \kb_4 \right) 
	-\frac{1}{2}\left( \kb_1 + \kb_3 \right)\cdot \left(  \kb_2 + \kb_4 \right)
	-\frac{1}{3} \left( k_1^2+k_2^2+k_3^2+k_4^2 \right) 
	\\
	& +\frac{1}{6} i h \lb ( \ww_1 +  \ww_2 +\ww_3 +\ww_4)
	\end{aligned}
	\right] \delta_{\kb_1+\kb_2+\kb_3+\kb_4,\vex{0}} \, .
\end{aligned}
\end{align}
Here, to account for the vertex symmetry, the amplitude of diagram~(n) has been multiplied by a factor of 2, while that of (o) and (p) have been multiplied by 4.

\subsection{Spin response}

The spin density response function $\Pi^{i,j}(\kb,\ww)$ can be obtained by taking derivatives of the partition function $Z[\vex{B}]$ with respect to the external fields,
see Eq.~(\ref{eq:SpinResp}).
We rewrite $Z[\vex{B}]$ in Eq.~(\ref{eq:CI_Z}) as
\begin{align}\label{eq:CI_ZZ}
\begin{aligned}
	Z[\vex{B}]
	=\,
	\int \dd \Wh
	\exp
	\left\lbrace 
	\begin{aligned}
	&
	 -\int (\Wh^{1,2} \tilde{M} \Wh^{2,1} + J^\dagger \Wh^{2,1} + \Wh^{1,2} J) 
	 - \suml{a,b} \int \frac{1}{2} \Wh^{a,a} \tilde{N}_{a,b} \Wh^{b,b} 
	\\& 
	- S_{W}^{(3,4)} [\Wh^{1,2},\Wh^{2,1},\Wh^{1,1},\Wh^{2,2}]
	-S_{\vex{B}}
	\end{aligned}
	\right\rbrace,
\end{aligned}
\end{align}	
where $\tilde{M}$ and $\tilde{N}$ are the symmetrized $M$ and $N$, respectively. 

Integrating out $\Wh$, we find
\begin{align}\label{eq:CI_LOGZ}
	\ln Z[\vex{B}] 
	\approx \, &
	-S_{\vex{B}}
	+
	\int J^\dagger \tilde{M}^{-1} J
	-
	\tr \ln \tilde{M}
	-
	\frac{1}{2}	\tr \ln \tilde{N}
	-
	\braket{ S_{W}^{(3,4)}},
\end{align}
where $\braket{ S_{W}^{(3,4)}}$ denotes
\begin{align}
\begin{aligned}
	\braket{ S_{W}^{(3,4)}}
	\equiv \, &
	\frac{
		\int \dd \Wh 
		\,
		e^ 
		{-\int \Wh^{1,2} \tilde{M} \Wh^{2,1}  
		-\suml{a,b} \int \frac{1}{2} \Wh^{a,a} \tilde{N}_{a,b} \Wh^{b,b} 
		 }
		\,
		S_{W}^{(3,4)} 
		\left[\Wh^{1,2}-J^{\dagger}\tilde{M}^{-1},\Wh^{2,1}-\tilde{M}^{-1}J,
		\Wh^{1,1},\Wh^{2,2}
		\right] 
	}
	{
		\int \dd \Wh
		\,
		e^{  -\int \Wh^{1,2} \tilde{M} \Wh^{2,1}  
		-\suml{a,b} \int \frac{1}{2} \Wh^{a,a} \tilde{N}_{a,b} \Wh^{b,b} 
		} 
	}.
\end{aligned}
\end{align}

\subsubsection{Semiclassical result}

\begin{figure}
	\centering
	\includegraphics[width=0.8\linewidth]{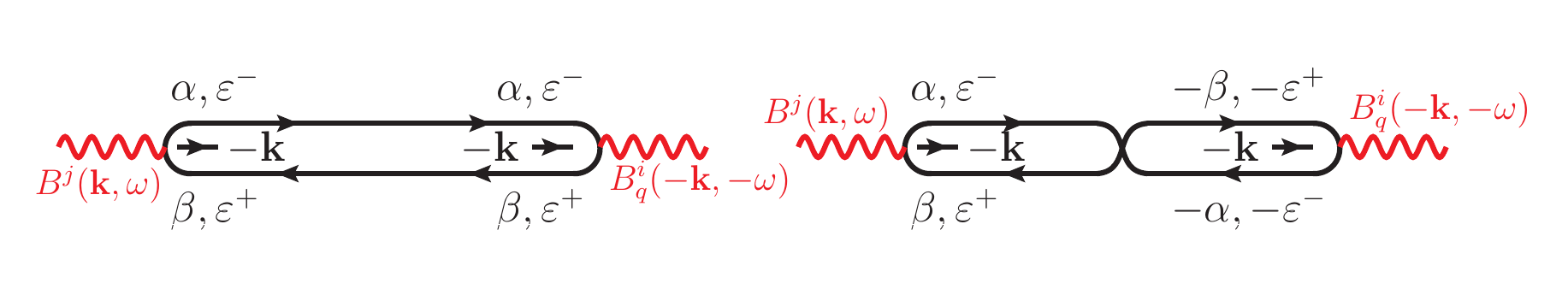}
	\caption{
		(Color online) Lowest order diagrams in $\ln Z$.
	}
	\label{fig:CI_0}
\end{figure}

To the lowest order in the perturbation parameter $\lb$, the 3rd and 4th terms in Eq.~(\ref{eq:CI_LOGZ}) are two inessential constants independent of $\vex{B}$, while the 2nd term illustrated in Fig.~\ref{fig:CI_0} gives nonzero contribution. Using the identity 
$(\hat{s}^i)^{-\beta,-\alpha}s_{\alpha,\beta}=(\hat{s}^i)^{\alpha,\beta}$, it is straightforward to show that contributions from the two diagrams in Fig.~\ref{fig:CI_0} are identical. 

Combining all these terms, we arrive at
\begin{align}\label{eq:CI_LOGZ0}
\begin{aligned}
	\ln Z[\vex{B}]
	= \, &
	i \frac{4}{\pi} h 
	\int \Bvql(-\kb,-\ww) \cdot \Bvcl(\kb,\ww)
	k^2 \bd_0 (k, -\ww)
	\\
	&
	-
	4 h^2 \lb
	\int
	\Bvql(-\kb, -\ww) \cdot \Bvql(\kb,\ww)
	\frac{\ww}{\pi}
	\coth{\left( \frac{\ww}{2 T}\right) }
	\bd_0 (k, -\ww)
	+ O(\lb),
\end{aligned}
\end{align}
which leads to the semiclassical spin response results:
\bsub\label{eq:CI_Classi}
\begin{align}
	&\Pi_0^{i,j} \left( \vex{k} , \ww \right) 
	=\, 
	-\delta_{i,j}
	\frac{2}{\pi} h
	\frac{  k^2 }{ k^2 - i h \lb \ww}
	=\,
	-\delta_{i,j}
	(2\nu_0)
	\frac{ D k^2 }{ D k^2 - i \ww},	
	\\
	&\sigma_0^{i,j} 
	= \, 
	\delta_{i,j} \frac{2}{\pi} \frac{1}{\lb} 
	= \, 
	\delta_{i,j} D (2\nu_0). 
\end{align}
\esub
\subsubsection{Quantum correction}

\begin{figure}
	\centering
	\includegraphics[width = 0.95\linewidth]{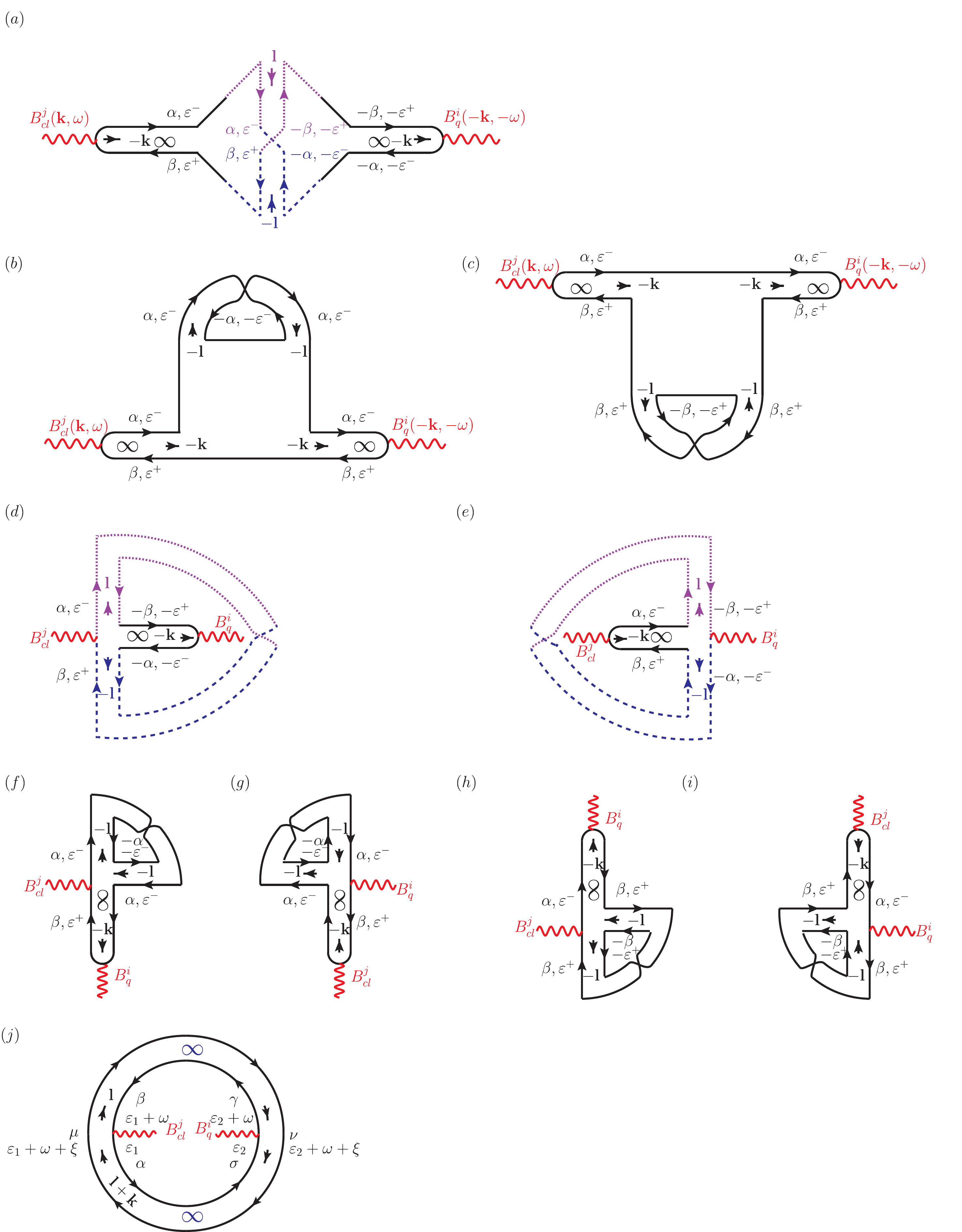}
	\caption{(Color online) Class CI linear response diagrams.}
	\label{fig:CI_3}
\end{figure}

The evaluation of the quantum correction to spin response requires higher order terms of the form of $\Bcl^j(\kb,\ww)\Bq^i(-\kb,-\ww)$ in $\ln Z$.
The corresponding nonvanishing diagrams are shown in Fig.~\ref{fig:CI_3}.

Diagrams in Figs.~\ref{fig:CI_3}(a)--(i) are from the term $-\braket{S_{W}^{(3,4)}}$ in Eq.~(\ref{eq:CI_LOGZ}). Their contributions to $\ln Z$ are as follows:
\bsub\label{eq:CI_1}
\begin{align}
	&\begin{aligned}
	(a)
	=\, & 	 
	-\delta_{i,j} 4 h^2 \lb 
	\bd_0^2(k,-\ww)
	\left[ 
	\left( k^2 + 2 i h \lb \ww \right) 
	\frac{\lb}{6} 
	\intl{\varepsilon}		
	\left( F_{\varepsilon^+}  - F_{\varepsilon^-} \right)
	\intl{\vex{l}} 
	\bd_0(l,-\ww)	
	\right] ,	
	\end{aligned}
	\\
	&\begin{aligned}
	(b)
	=\, & 	 
	-\delta_{i,j} 4 h^2 \lb 
	\bd_0^2(k,-\ww)
	\left[ 
	\bd_0^{-1}(k,-\ww) 
	\frac{\lb}{12} 
	\intl{\varepsilon}		
	\left( F_{\varepsilon^+}  - F_{\varepsilon^-} \right)		
	\intl{\vex{l}} 
	\bd_0(l,2\varepsilon^-)	
	\right] ,	
	\end{aligned}	
	\\
	&\begin{aligned}
	(c)
	=\, & 	 
	-\delta_{i,j} 4 h^2 \lb 
	\bd_0^2(k,-\ww)
	\left[ 	
	\bd_0^{-1}(k,-\ww) 
	\frac{\lb}{12} 
	\intl{\varepsilon}		
	\left( F_{\varepsilon^+}  - F_{\varepsilon^-} \right)		
	\intl{\vex{l}} 
	\bd_0(l,-2\varepsilon^+)	
	\right] ,
	\end{aligned}	
	\\
	&\begin{aligned}
	(d)=(e)
	=\, & 	 
	-\delta_{i,j} 4 h^2 \lb 
	\bd_0(k,-\ww)
	\left[ 
	\frac{\lb}{6} 
	\intl{\varepsilon}		
	\left( F_{\varepsilon^+}  - F_{\varepsilon^-} \right)
	\intl{\vex{l}} 
	\bd_0(l,-\ww)	
	\right] ,	
	\end{aligned}	
	\\
	&\begin{aligned}
	(f)=(g)
	=\, & 	 
	-\delta_{i,j} 4 h^2 \lb 
	\bd_0(k,-\ww)
	\left[ 
	-\frac{\lb}{6} 
	\intl{\varepsilon}		
	\left( F_{\varepsilon^+}  - F_{\varepsilon^-} \right)
	\intl{\vex{l}} 
	\bd_0(l,2\varepsilon^-)	
	\right] 	,
	\end{aligned}	
	\\
	&\begin{aligned}
	(h)=(i)
	=\, & 	 
	-\delta_{i,j} 4 h^2 \lb 
	\bd_0(k,-\ww)
	\left[ 
	-\frac{\lb}{6} 
	\intl{\varepsilon}		
	\left( F_{\varepsilon^+}  - F_{\varepsilon^-} \right)
	\intl{\vex{l}} 
	\bd_0(l,-2\varepsilon^+)	
	\right]  ,
	\end{aligned}	
\end{align}
\esub
where, as before, $\varepsilon^{\pm}$ is defined as $\varepsilon^{\pm}=\varepsilon \pm \ww/2$ .
For notational simplicity, here we have omitted the factor $\Bcl^j(\kb,\ww)\Bq^i(-\kb,-\ww)$.

We notice that two different types of integral appear in these equations. 
For Eqs.~(\ref{eq:CI_1}a) and~(\ref{eq:CI_1}d), the integrand contains external-frequency-dependent propagator $\bd_0(l,-\ww)$. As a result, these integrals bring a factor of $\ln (\frac{\Lambda}{\ww})$. In the presence of interactions, the corresponding 
dc spin conductance	
correction is cut off in the infrared limit by the dephasing rate, as in class AII.
On the other hand, for the remaining equations, propagator $\bd_0(l,\mp\varepsilon^{\pm})$ depends on the integration variable $\varepsilon$. From a similar argument in class C, the corresponding contribution is cut off in the infrared directly by temperature, instead of the dephasing rate.

Combining all these equations in Eq.~(\ref{eq:CI_1}), the net contribution from the diagrams in Fig.~\ref{fig:CI_3}(a)--(i) is
\begin{align}\label{eq:CI_sigma1}
	\frac{\delta^2 \ln Z}{\delta\Bcl \delta\Bq}
	=\, & 	 
	-\delta_{i,j} 4 h^2 \lb 
	\bd_0^2(k,-\ww) \frac{\ww}{\pi}
	\left[  k^2 \delta \lb -i h \lb \ww (-\delta h) 
	+\bd_0^{-1}(k,-\ww) (-\delta z_1)
	\right], 	 
\end{align}
where	
\bsub\label{eq:CI_d1}
\begin{align}
	\delta \lb \equiv\,&
	\frac{\lb}{24 \pi}\ln\left(\frac{\Lambda}{\ww}\right)
	+
	\frac{\lb}{24\pi}\ln\left(\frac{\Lambda}{T}\right),
	\\
	\delta h \equiv\,&
	\frac{\lb}{12\pi}\ln\left(\frac{\Lambda}{\ww}\right)
	-\frac{\lb}{24\pi}\ln\left(\frac{\Lambda}{T}\right),
	\\
	\delta z_1 \equiv\,&
	-\frac{\lb}{12 \pi}\ln\left(\frac{\Lambda}{\ww}\right)
	+\frac{ \lb}{6 \pi}\ln\left(\frac{\Lambda}{T}\right).
\end{align}
\esub

Fig.~\ref{fig:CI_3}(j) depicts diagram arising from $-\tr \ln \tilde{M}$ in Eq.~(\ref{eq:CI_LOGZ}), with amplitude
\begin{align}
	&\begin{aligned}
	(j)
	=\,&
	-\delta_{i,j} 2 h^2 \lb^2
	\intl{\vex{l},\varepsilon}
	\left[ 
	\begin{aligned}
	&\bd_0(l,-2\varepsilon-2\ww)\bd_0(\lvert \kb+\vex{l} \rvert,-2\varepsilon-\ww)F_{\varepsilon}
	\\
	+
	&\bd_0(l,-2\varepsilon-\ww)\bd_0(\lvert \kb+\vex{l} \rvert,-2\varepsilon)F_{\varepsilon}
	\end{aligned}	
	\right]. 				
	\end{aligned}			
\end{align}
A straightforward calculation shows that diagram in Fig.~\ref{fig:CI_3}(j) gives contribution
\begin{align}\label{eq:CI_sigma2}
\begin{aligned}	
	\frac{\delta^2 \ln Z}{\delta\Bcl \delta\Bq}=\delta_{i,j} i \frac{4}{\pi}  h (-\delta z_2),
\end{aligned}
\end{align}
where	
\begin{align}\label{eq:CI_d2}
\begin{aligned}		
	\delta z_2 
	\equiv \, &
	\frac{\lb}{8 \pi}
	\ln\left(\frac{\Lambda}{T}\right).	
\end{aligned}
\end{align} 

Using Eqs.~(\ref{eq:SpinResp}), (\ref{eq:CI_sigma1}) and (\ref{eq:CI_sigma2}), we obtain the one-loop quantum correction to the spin response function
\begin{align}\label{eq:CI_dPi}
\begin{aligned} 
	\delta \Pi^{i,j} (\kb,\ww)
	=\,&
	-i \delta_{i,j} \frac{2}{\pi}  h^2 \lb
	\bd_0^2 (k,-\ww)
	\left[ 
	\begin{aligned}
	& \ww k^2 
	\left( \delta \lb -\delta z_1 +2 \delta z_2 \right) 
	-i h \lb \ww^2
	\left( -\delta h -\delta z_1 + \delta z_2 \right) 
	+
	\frac{k^4}{i h \lb} (-\delta z_2)
	\end{aligned}
	\right]
	\\
	=\,&
	 - i \delta_{i,j} \frac{2}{\pi} h^2 \lb	
	\bd_0^2(k,-\ww)
	\left[ 
	\ww k^2 
	\left( \delta \lb -\delta z_1 +2 \delta z_2 \right) 
	+
	\frac{k^4}{i h \lb} (-\delta z_2)
	\right], 					
\end{aligned}
\end{align}
where in the last equality, we have used $-\delta h -\delta z_1+\delta z_2=0$, proved by substituting the explicit form of these variables in Eqs.~(\ref{eq:CI_d1}) and (\ref{eq:CI_d2}).

From this result we acquire the relative quantum correction to the spin conductivity for the noninteracting class CI superconductor:
\begin{align}\label{eq:CI_dsigma}
\begin{aligned}
	\frac{\delta \sigma^{i,j}}{ \sigma_{0}^{i,j}} 
	=\, &
	-
	\left( \delta \lb -\delta z_1 +2 \delta z_2\right) 		
	\\
	= \, &
	-\frac{\lb}{8 \pi}  \ln \left(\frac{\Lambda}{\ww}\right)
	-\frac{\lb}{8 \pi}  \ln \left(\frac{\Lambda}{T}\right).
\end{aligned}
\end{align} 
It consists of two terms: the first logarithmic correction is cut off in the infrared limit by the external frequency $\ww$, while the second one is cut off by temperature $T$.


\section{Weak (anti)localization and phase relaxation}\label{sec:dephasing}

In this section, we investigate the dephasing time by evaluating the WAL correction using two different approaches, both carried out in the symplectic metal. 
The first approach is similar to the one employed by Altshuler, Aronov, and Khmelnitsky (AAK)~\cite{AAK};
in the second approach, we employ a standard perturbation technique widely used for the evaluation of the dephasing time~\cite{Blanter,FukuyamaAbrahams,Santos}.

\subsection{AAK approach}\label{sec:AAK}

\subsubsection{Equation of the Cooperon in the presence of the density field in the space-time representation}\label{sec:CEQ}

In the space-time representation, $S_Y^{(2)}$ the quadratic action for the Cooperon matrix field $\Yh$ takes the form [see Eq.~(\ref{eq:AII_SQ2}b) for its momentum-frequency version]
\begin{align}\label{eq:SY}
\begin{aligned}
	S_Y^{(2)}[\Yh^{\dagger}, \Yh]
	=& 
	\intl{\rb,t_1,t_2}
	\Yh^{\dagger}_{t_1,t_2} (\rb)
	\left\lbrace  
	-
	\nabla ^2
	+
	 h \lb
	\left(   \partial_{t_1} -   \partial_{t_2} \right) 
	+
	i h \lb
	\left[  \rcl(\rb,t_1) -\rcl(\rb,t_2) \right] 
	\right\rbrace  	 
	\Yh_{t_2,t_1}(\rb).	
\end{aligned}
\end{align}
Here we have disregarded the H.-S.\ field's quantum component $\rhoq$.
Using Eq.~(\ref{eq:SY}), we find, in the presence of density field $\rcl$, the full Cooperon propagator 
\begin{align}
\begin{aligned}
	C_{t_1,t_2;t_2',t_1'}(\rb,\rb')
	\equiv
	\braket{ \Yh_{t_1,t_2}(\rb) \Yh^{\dagger}_{t_2',t_1'}(\rb') },
\end{aligned}
\end{align}	
obeys the equation
\begin{align}\label{eq:CooperonEQ1}
\begin{aligned}
	&
	\left\lbrace 
	-
	\nabla ^2
	+
	 h \lb
	\left(   \partial_{t_2} -   \partial_{t_1} \right) 
	+
	i h \lb
	\left[  \rcl(\rb,t_2) -\rcl(\rb,t_1) \right] 
	\right\rbrace 
	 C_{t_1,t_2;t_2',t_1'}(\rb,\rb')	
	=
	\delta(t_1-t_1')\delta(t_2-t_2')\delta(\rb-\rb').
\end{aligned}
\end{align}	
Following Ref.~\cite{AAK}, we employ a change of variables
\begin{align}\label{eq:eta}
\begin{aligned}
	&t=\frac{t_1+t_2}{2},
	\qquad
	t'=\frac{t_1'+t_2'}{2},	
	\qquad
	\eta=t_2-t_1,
	\qquad
	\eta'=t_2'-t_1',	
	\\
	&
	C_{\eta,\eta'}^{t,t'}(\rb,\rb')
	=\,
	C_{t_1,t_2;t_2',t_1'}(\rb,\rb'),
\end{aligned}
\end{align}	
after which the equation for the Cooperon [Eq.~(\ref{eq:CooperonEQ1})] reduces to
\begin{align}\label{eq:CooperonEQ2}
\begin{aligned}
	&
	\left\lbrace 
	-
	\nabla ^2
	+
	2 h \lb
	 \partial_{\eta} 
	+
	i h \lb
	\left[  \rcl\left(\rb,t+\frac{\eta}{2}\right) - \rcl\left(\rb,t-\frac{\eta}{2}\right) \right] 
	\right\rbrace 
	 C_{\eta,\eta'}^{t,t'}(\rb,\rb')
	=
	\delta(t-t')\delta(\eta-\eta')\delta(\rb-\rb').
\end{aligned}
\end{align}	
Note that  $t$ appears as a parameter here and thus the solution can be represented as 
\begin{align}\label{eq:Ct}
\begin{aligned}
	C_{\eta,\eta'}^{t,t'}(\rb,\rb')	
	=
	C_{\eta,\eta'}^{t}(\rb,\rb') \delta(t-t'),	
\end{aligned}
\end{align}	
where $C_{\eta,\eta'}^{t}(\rb,\rb')$ follows
\begin{align}\label{eq:CooperonEQ}
\begin{aligned}
	&
	\left\lbrace  
	-
	\nabla ^2
	+
	2 h \lb
	 \partial_{\eta} 
	+
	i h \lb
	\left[  \rcl\left(\rb,t+\frac{\eta}{2}\right) -\rcl\left(\rb,t-\frac{\eta}{2}\right) \right] 
	\right\rbrace 
	 C_{\eta,\eta'}^{t}(\rb,\rb')
	=
	\delta(\eta-\eta')\delta(\rb-\rb').
\end{aligned}
\end{align}

\subsubsection{WAL Correction}\label{sec:CWL}

\begin{figure}
	\centering
	\includegraphics[width=0.6\linewidth]{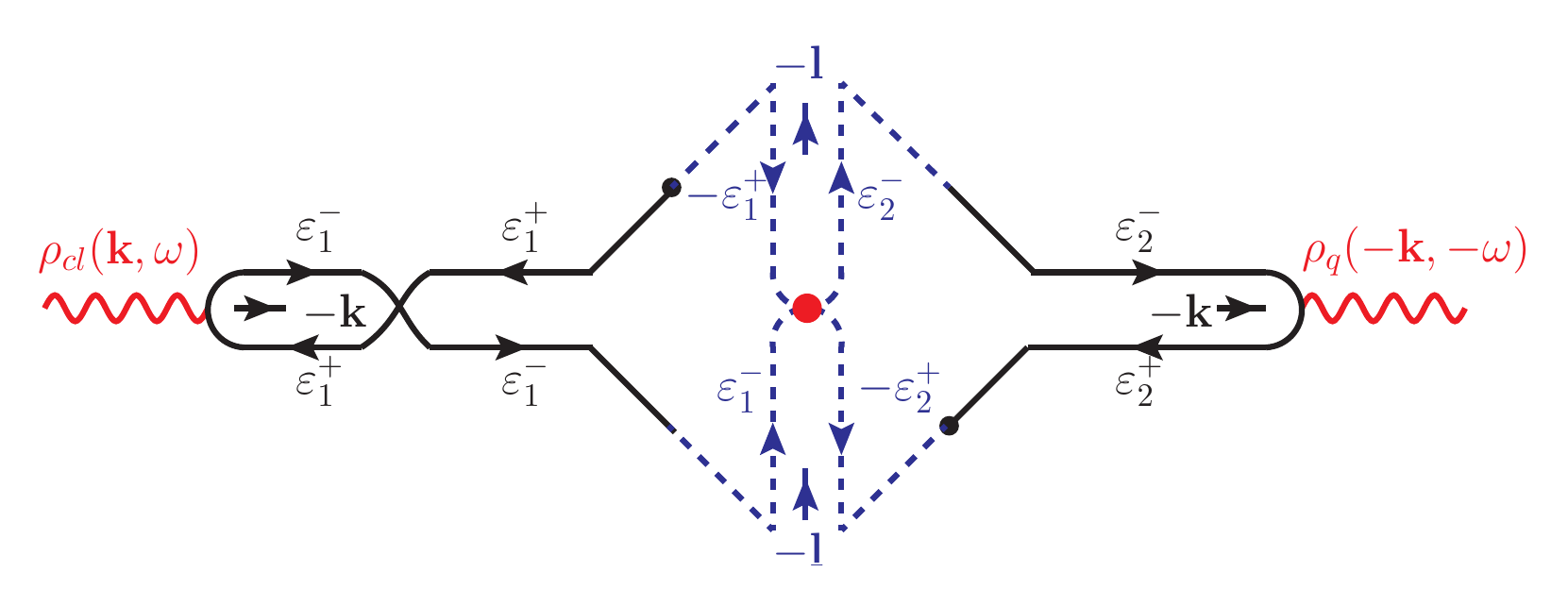}
	\caption{(Color online) Class AII WAL diagram with the full Cooperon propagator.}
	\label{fig:AII_5}
\end{figure}
 
As mentioned in Sec.~\ref{sec:AII}, to recover the correct infrared cutoff of the WAL correction to conductivity, inclusion of higher-order diagrams is needed. Replacing the bare Cooperon in Fig.~\ref{fig:AII_3}(a) with the full one (see Fig.~\ref{fig:AII_5}), we obtain the associated retarded self energy of the density field $\rho$:
\begin{align}\label{eq:SigWL}
\begin{aligned}
	- i \Sigma_{\rho}^{(R)}(\kb,\ww)
	=&
	-4 h^2 \lb
	\bd_0 ^2 (k,-\ww)
	\left( \frac{-\lb}{2} \right)  
	\intl{\varepsilon_1,\varepsilon_2,\vex{l}} 
	\left( F_{\varepsilon_1^+} -F_{\varepsilon_1^-} \right)	 
	\left[
	\bd_0^{-1}(l,-\ww)
	+
	k^2
	\right] 
	\braket{C_{\e_1^-,-\e_2^+;-\e_1^+,\e_2^-}(-\vex{l},-\vex{l})}_{\rho},
\end{aligned}
\end{align}
where  $\e_{1,2}^{\pm}$, as before, stands for $\e_{1,2} \pm \ww/2$. 
The Cooperon propagator entering this equation is averaged over thermal density fluctuations. To simplify the calculation, we directly set the frequency and momentum indices according to the special properties of the averaged Cooperon. Moreover, an extra factor of
the space-time volume should appear on the left hand side of Eq.~(\ref{eq:SigWL}) but is neglected for simplicity.

Particle number conservation demands that the density response function $\Pi(\kb,\ww)$ vanishes as $k \rightarrow 0$. Therefore, we argue the term proportional to $\bd_0^{-1}(l,-\ww)$ in Eq.~(\ref{eq:SigWL}) does not contribute and focus on the remaining terms.

\begin{figure}
	\centering
	\includegraphics[width=0.3\linewidth]{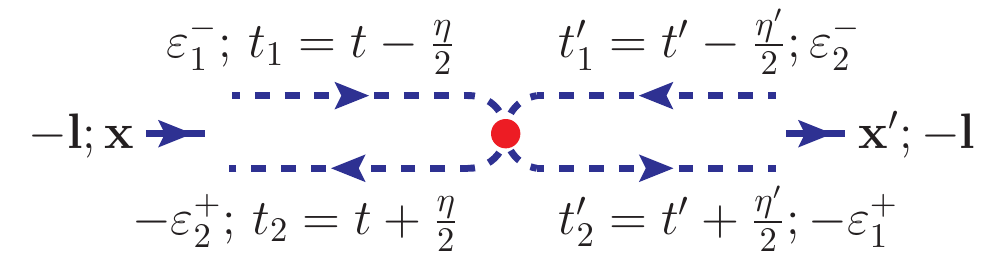}
	\caption{(Color online) Full Cooperon propagator.}
	\label{fig:DE_0}
\end{figure}

We then rewrite Eq.~(\ref{eq:SigWL}) in terms of the Cooperon in the space-time domain.
After the Fourier transform, the integral in Eq.~(\ref{eq:SigWL}) can be expressed as
\begin{align}\label{eq:SigWL1}
\begin{aligned}
	& \intl{\varepsilon_1,\varepsilon_2,\vex{l}} 
	\left( F_{\varepsilon_1^+} -F_{\varepsilon_1^-} \right)	  
	C_{\e_1^-,-\e_2^+;-\e_1^+,\e_2^-}(-\vex{l},-\vex{l})
	\\
	= 
	&\intl{\varepsilon_1}
	\intl{t_1,t_2,t_1',t_2',\rb,\rb'} 
	\left( F_{\varepsilon_1^+} -F_{\varepsilon_1^-} \right)	  
	C_{t_1,t_2;t_2',t_1'}(\rb,\rb')
	\delta(\rb-\rb') \delta(t_2 -t_1')
	e^{i \e_1^- t_1- i \e_1^+ t_2' + i \ww t_2 }.
\end{aligned}
\end{align}
Changing the variables according to Eq.~(\ref{eq:eta}) and using Eq.~(\ref{eq:Ct}) (see Fig.~\ref{fig:DE_0}), Eq.~(\ref{eq:SigWL1}) is further reduced to
\begin{align}\label{eq:SigWL2}
\begin{aligned}
	&2  \intl{\varepsilon_1}
	\intl{t,\eta,\rb} 
	\left( F_{\varepsilon_1^+} -F_{\varepsilon_1^-} \right)	  
	C_{\eta,-\eta}^{t}(\rb,\rb)
	e^{i \ww \eta}	
	=\,
	\frac{2}{\pi} \ww
	\intl{t,\eta,\rb}  
	C_{\eta,-\eta}^{t}(\rb,\rb)
	e^{i \ww \eta}.		
\end{aligned}
\end{align}

We then substitute Eq.~(\ref{eq:SigWL2}) into Eq.~(\ref{eq:SigWL}). Averaging over the density fluctuations removes the dependence of $C_{\eta,-\eta}^{t}(\rb,\rb)$ on $t$ and $\rb$. The corresponding integration over these variables cancels with the extra factor 
of the space-time volume, 
and as a result one obtains:
\begin{align}\label{eq:SigWL3}
\begin{aligned}
	- i \Sigma_{\rho}^{(R)}(\kb,\omega)=\,&
	-\frac{4}{\pi} h ^2 \lb \ww
	\bd_0 ^2 (k,-\ww) k^2 
	\delta \lb_{\msf{WAL}},
\end{aligned}
\end{align}
where
\begin{align}
	\delta \lb_{\msf{WAL}} =\, &
	-\lb \intl{\eta}  
	\braket{C_{\eta,-\eta}^{t}(\rb,\rb)}_{\rho}
	e^{i \ww \eta}.	
\end{align}
From Eqs.~(\ref{eq:conduct})--(\ref{eq:AII_deltaG}) and (\ref{eq:SigWL3}), 
the WAL correction to the dc conductivity can be expressed through the Cooperon as
\begin{align}\label{eq:dccorrection}
	\frac{\delta \sigma_{\msf{WAL}}}{\sigma_0} = &
	\lb \intl{\eta}  
	\braket{C_{\eta,-\eta}^{t}(\rb,\rb)}_{\rho},	
\end{align}
where $C_{\eta,-\eta}^{t}(\rb,\rb)$ is the solution of Eq.~(\ref{eq:CooperonEQ}).

\subsubsection{Cooperon solution in the form of a path integral}\label{sec:CPath}

Multiplying both sides of Eq. (\ref{eq:CooperonEQ}) by $1/(2h\lb) = D/2$ [Eq.~(\ref{eq:hlbgamDef})]
gives
\begin{align}\label{eq:Schrodinger}
\begin{aligned}
	\left\lbrace 
	\partial_{\eta} 
	-
	\frac{D}{2}
	\nabla ^2
	+
	 \frac{i}{2}
	\left[  \rcl\left(\rb,t+\frac{\eta}{2}\right) -\rcl\left(\rb,t-\frac{\eta}{2}\right) \right] 
	\right\rbrace 
	C_{\eta,\eta'}^{t}(\rb,\rb')
	=\,&
	\frac{D}{2}\delta(\eta-\eta')\delta(\rb-\rb'),
\end{aligned}
\end{align}	
which can be considered the imaginary time Schr$\ddot{o}$dinger equation for a particle of mass $1/D$ in the presence of 
the 
stochastically fluctuating field
\begin{align}
	\rho_t(\rb,\eta)
	\equiv\, &
	 \frac{i}{2}
	\left[ \rcl\left(\rb,t+\frac{\eta}{2}\right) -\rcl\left(\rb,t-\frac{\eta}{2}\right) \right].
\end{align}
Its solution in an arbitrary density field $\rhocl$ can be expressed in terms of a path integral \cite{AAK, AleinerBlanter, Feynman}:
\begin{align}\label{eq:pathsol}
\begin{aligned}
	C_{\eta,\eta'}^{t}(\rb,\rb')=
	\frac{D}{2}
	\intl{\rb(\eta')=\rb'}^{\rb(\eta)=\rb} 
	{\cal D} \rb(\tau)
	\exp
	\left( 
	-\intl{\eta'}^{\eta}
	d\tau
	\left\lbrace 
	\frac{1}{2 D} \dot{\rb}^2(\tau)
	+
	\frac{ i }{2}
	\left[ \rcl\left(\rb(\tau),t+\frac{\tau}{2}\right) -\rcl\left(\rb(\tau),t-\frac{\tau}{2}\right) \right]
	\right\rbrace 
	\right) .
\end{aligned}	
\end{align}
To obtain the WAL correction to conductivity, one needs to average the solution in Eq.~(\ref{eq:pathsol}) over the fluctuations of the density field $\rhocl$ whose correlator is given by $i \Delta_{\rho}^{(K)}$ in Eq.~(\ref{eq:Deltarho}), and then substitute the averaged Cooperon propagator into Eq.~(\ref{eq:dccorrection}).
In the limit where the exchange energy $\ww \ll T$, the Keldysh Green's function $i \Delta_{\rho}^{(K)}(\kb,\ww)$ can be approximated as
\begin{align}\label{eq:Gcl}
\begin{aligned}
	i \Delta_{\rho}^{(K)} (\kb,\ww)
	\approx \,& 
	T  \frac{\gamma^2}{\kappa}
	\left(
	\frac{1}{D_c k^2 + i \ww}
	+
	\frac{1}{D_c k^2 - i \ww}		
	\right), 	
\end{aligned}
\end{align}
whose space-time expression has the form
\begin{align}\label{eq:Gclxt}
\begin{aligned}
	i \Delta_{\rho}^{(K)} (\rb,t)
	\approx\, &
	T  \frac{\gamma^2}{\kappa}
	\left( \frac{1}{4 \pi D_c |t|} \right) 
	\exp \left( {-\frac{\rb^2}{4 D_c |t| }} \right). 
\end{aligned}
\end{align}

Eq.~(\ref{eq:Gcl}) is a valid assumption because  
processes with exchange energy $\ww \ll T$ give the major contribution to the dephasing time \cite{altshuler1981}.
It is worth mentioning that Eq.~(\ref{eq:Gcl}) shows that the density fluctuations are themselves diffusive.
These fluctuations of the H.-S.\ field destroy the phase coherence and cut off the weak (anti)localization.
So the system serves as its own heat bath, as expected for the ergodic delocalized phase we have investigated.

\subsection{Self consistent calculation}\label{sec:SelfCons}

In the following, we employ a different approach to examine the higher order processes that are responsible for the dephasing of the WAL correction. 
Instead of expressing the dressed Cooperon in the form of a path integral, we write it as a partial summation of a diagrammatic series. 
Moreover, we take into account the correction from the insertion of the four-point diffusion vertex, besides the interaction vertex coupling matrix field $\Yh$ and the H.-S.\ field $\rho$. 
However, we will show below that the correction from the vertex of the former type can be neglected.
The techniques we use here to treat the WAL and phase relaxation were employed before in several papers~\cite{Blanter,FukuyamaAbrahams,Santos}, but not in the framework of the FNL$\sigma$M.

The WAL correction to the self energy can still be represented by Eq.~(\ref{eq:SigWL}), although the Cooperon propagator $C_{\e_1^-,-\e_2^+;-\e_1^+,\e_2^-}(-\vex{l},-\vex{l})$ entering this formula no longer equals $N^{-1}_{\e_1^-,-\e_2^+;-\e_1^+,\e_2^-}(-\vex{l},-\vex{l})$ [see Eq.~(\ref{eq:AII_MNJ})] due to the inclusion of the four-point diffusion vertex.  It satisfies the equation
\begin{align}\label{eq:CDyson}
	\braket{C}_{\rho}
	=\, &
	C_0 + C_0 \Sigma_Y \braket{C}_{\rho},
\end{align}
with $C_0$ and $\Sigma_Y$ being the bare Cooperon propagator and the irreducible self energy of the matrix field $\Yh$, respectively.

\begin{figure}
	\centering
	\includegraphics[width=0.7\linewidth]{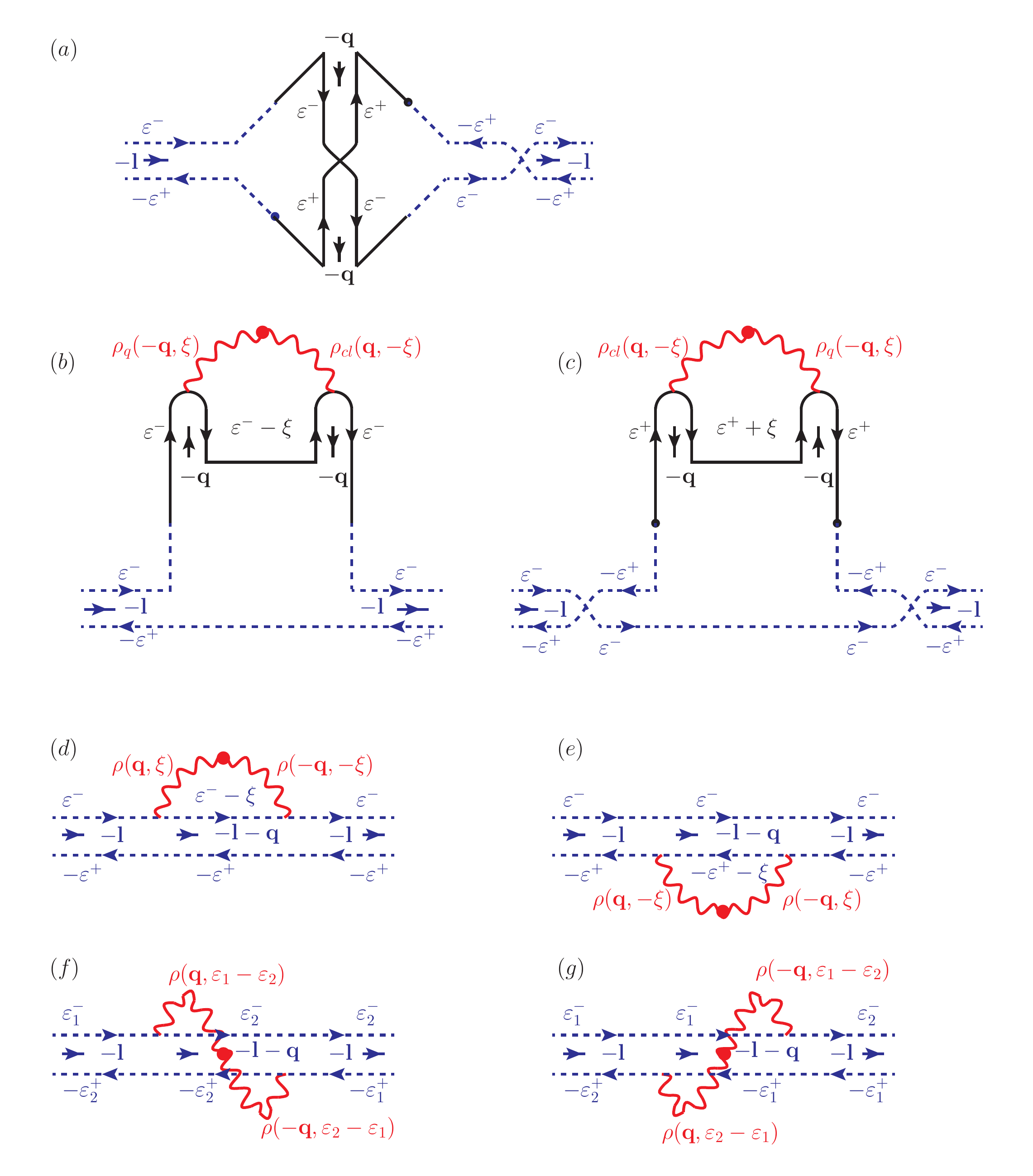}
	\caption{(Color online) The lowest order self energy diagrams for the Cooperon propagator.
	}
	\label{fig:DE_1}
\end{figure}

In Fig.~\ref{fig:DE_1}, we show the lowest order self energy diagrams for the Cooperon.  We neglect $\Sigma_Y$ given by Figs.~\ref{fig:DE_1}(f) and \ref{fig:DE_1}(g) which are off-diagonal in the frequency space, and focus on the diagonal ones, i.e., those depicted in Figs.~\ref{fig:DE_1}(a)--(e).
The associated amplitude of diagram in Fig.~\ref{fig:DE_1}(i), $i\in \left\lbrace a,b,...,e \right\rbrace $, has the form
\begin{align}
\begin{aligned}
	\Sigma_{Y}^{(i)}(-\vex{l},-\ww)
	=\, \Sigma_{X}^{(i)}(-\vex{l},-\ww),
\end{aligned}
\end{align}
where the explicit expression of $\Sigma_{X}^{(i)}$ is given by Eq.~(\ref{eq:AII_3ae}).
For each self energy term $\Sigma_X\,({\varepsilon^-,\varepsilon^+;\varepsilon^+,\varepsilon^-})$
in the diffuson channel, there is a corresponding $\Sigma_Y\,({\varepsilon^-,-\varepsilon^+;-\varepsilon^+,\varepsilon^-})$ which shares the same expression, see Sec.~\ref{sec:AII}.
We have already evaluated these terms in Sec.~\ref{sec:AII}, and found their summation can be expressed as follows
\begin{align}\label{eq:SigY}
\begin{aligned}
	\Sigma_{Y}(-\vex{l},-\ww)=l^2 \delta \lb - i h \lb \ww (-\delta h) + \Sigma_{\varepsilon} ,
\end{aligned}
\end{align}
where $\delta \lb$, $\delta h$ and $\Sigma_{\varepsilon}$ were defined in Eqs.~(\ref{eq:AII_d1}a), (\ref{eq:AII_d1}b) and (\ref{eq:SigEps}), respectively.
Note, however, that the logarithmic denominator of the first term in $\delta \lb$ is not $\tau_{\phi}^{-1}$ as in Eq.~(\ref{eq:AII_d1}a) but $\omega$.

The first two terms in Eq.~(\ref{eq:SigY}) are of linear order in $\lb$ and lead to renormalization of the diffusion parameters: $h$, $\lb$. Since we are only interested in the WAL correction to lowest order in $\lb$, these two terms are neglected. 

\begin{figure}
	\centering
	\includegraphics[width=0.7\linewidth]{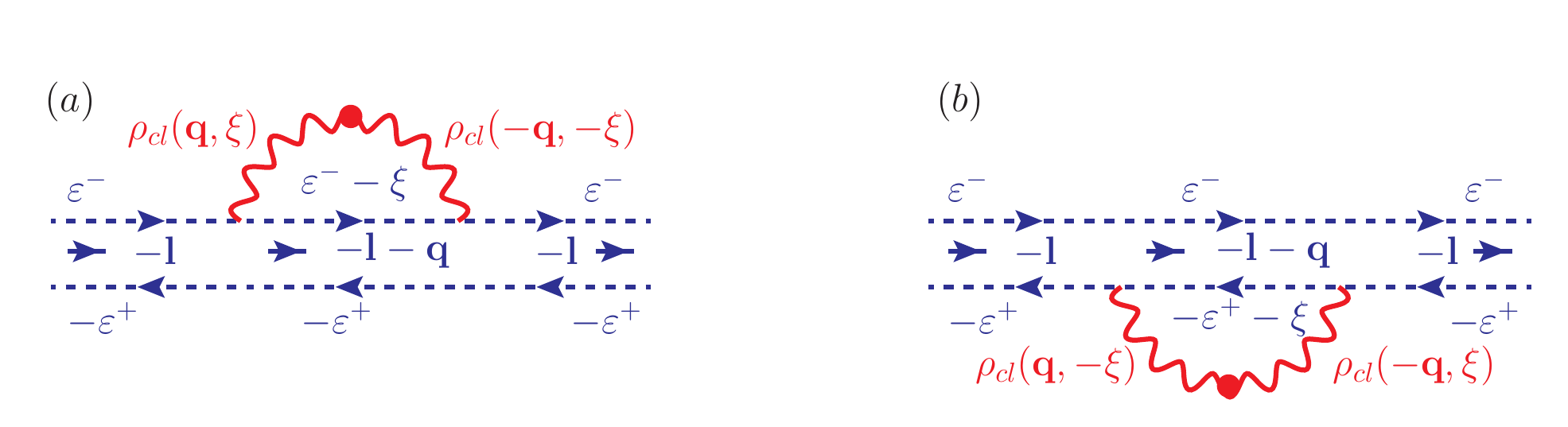}
	\caption{(Color online) Effective self energy diagrams for the Cooperon.
	}
	\label{fig:DE_2}
\end{figure}

The last term $\Sigma_{\varepsilon}$, on the other hand, diverges in the infrared limit and cannot be simply discarded.  
To evaluate Eq.~(\ref{eq:SigEps}), one can make use of the following approximation
\begin{align}
		2 \coth \left(\frac{\xi}{2T}\right)
		-\tanh \left(\frac{\xi+\varepsilon}{2 T}\right)
		- \tanh\left(\frac{\xi-\varepsilon}{2 T}\right)
		\approx&\,
		2 \coth \left(\frac{\xi}{2T}\right)
		-2\tanh \left(\frac{\xi}{2 T}\right)
\nonumber\\
		\approx&\,	
		\left\lbrace 
		\begin{aligned}
			& 2 \coth \left(\frac{\xi}{2T}\right) \approx \frac{4T}{\xi},
			 \quad \lvert \xi \rvert <T,
			\\
			& 0, 
			 \quad \lvert \xi \rvert >T,
		\end{aligned}
		\right. 
\end{align}
in the limit $T \gg \varepsilon$. 
Therefore, the leading contribution to the integral comes from the region $\lvert \xi \rvert<T$. 
Associated processes with exchange energy $ \lvert \xi \rvert>T$ can be ignored, while for processes with $ \lvert \xi \rvert < T$, those carrying a factor of $\coth (\frac{\xi}{2T})$ give the most singular contribution and need to be retained.
Note also that the $\coth (\frac{\xi}{2T})$ term in Eq.~(\ref{eq:SigEps}) comes from the correlator $\braket{\rhocl (\vex{q},\xi) \, \rhocl (-\vex{q},-\xi)}$ [i.e., the Keldysh Green's function $i \Delta_{\rho}^{(K)}(\vex{q},\xi)$].
We then arrive at the conclusion that, the net contribution given by the diagonal self energy diagrams in Figs.~\ref{fig:DE_1}(a)--(e) can be approximated by that from Fig.~\ref{fig:DE_2} where the exchange energy is restricted to the range $\lvert \xi \rvert<T$.

This explains the assumption we employed in Sec.~\ref{sec:AAK}, where we disregard the quantum component $\rho_q$ and processes with exchange energy larger than $T$. In particular, the Cooperon propagator entering the WAL correction can be represented by the one in the classical H.-S.\ field $\rhocl$ with characteristic frequency smaller than $T$. 
Fig.~\ref{fig:DE_2} can be considered as a diagrammatic interpretation of Eq.~(\ref{eq:pathsol}) [or Eq.~(\ref{eq:CooperonEQ})], and gives the first few leading order terms in the perturbation expansion of the Cooperon.

To address the problem of the infrared divergence in $\Sigma_{\varepsilon}$, we include diagonal self energy diagrams with more than one pair of H.-S.\ field propagator (or equivalently, interaction line). We employ the self-consistent Born approximation (SCBA) by substituting the bare Cooperon propagator in Fig.~\ref{fig:DE_2} with the renormalized one, and obtain the self-consistent equation for $\tau_D \equiv -h\lb/\Sigma_{\varepsilon}$:
\begin{align}\label{eq:tD1}
\begin{aligned}
	\tau_D^{-1}
	= 
	-\frac{i}{2} \pi   \gamma \lb  
	\int_{0}^{T} \frac{d \xi}{2 \pi} 
	\intl{\vex{l}} 
	\,
	\left[
	\tilde{\bd}_0(l,\xi)+\tilde{\bd}_0(l,-\xi) 
	\right] 
	\left[
	\frac{\bd_u(l,\xi)}{\bd_0(l,\xi)}
	-
	\frac{\bd_u(l,-\xi)}{\bd_0(l,-\xi)}
	\right] 
	\coth \left(\frac{\xi}{2T}\right),
\end{aligned}
\end{align}
where 
$\tilde{\bd}_0(l,\xi) \equiv 1/(l^2 + i h \lb \xi + h \lb \tau_D^{-1})$
gives the renormalized Cooperon propagator. 
The self energy $\Sigma_{\varepsilon}=-h\lb\tau_D^{-1}$ evaluated within the SCBA is a partial summation of the infinite-order diagrammatic series wherein diagrams with crossed ``interaction'' lines are ignored.
It is easy to check that this integral does not diverge in the infrared limit anymore, and one would get the weak antilocalization correction of the form in Eq.~(\ref{eq:AII_SigmaWL}) with infrared cutoff $\tau_D^{-1}$.

Up to logarithmic accuracy,	Eq.~(\ref{eq:tD1}) is equivalent to
	\begin{align}\label{eq:tD2}
	\begin{aligned}
	\tau_D^{-1}
	= 
	-\frac{i}{2} \pi   \gamma \lb  
	\int_{\tau_D^{-1}}^{T} \frac{d \xi}{2 \pi} 
	\intl{\vex{l}} 
	\,
	\left[
	\bd_0(l,\xi)+\bd_0(l,-\xi) 
	\right] 
	\left[
	\frac{\bd_u(l,\xi)}{\bd_0(l,\xi)}
	-
	\frac{\bd_u(l,-\xi)}{\bd_0(l,-\xi)}
	\right] 
	\coth \left(\frac{\xi}{2T}\right),
\end{aligned}
\end{align}
which leads to the following equation after a straightforward calculation:
\begin{align}\label{eq:tauDResult}
	\tau_D^{-1}
	= 
	\frac{\lb }{4 }
	\frac{\gamma^2}{(2-\gamma)}
	 T 
	\ln \left( \frac{T}{\tau_D^{-1}}\right). 
\end{align}
This result is consistent with the one obtained in Ref.~\cite{Zala-Deph}. 
Eq.~(\ref{eq:tauDResult}) also obtains via the lowest order cumulant expansion in the path
integral Eq.~(\ref{eq:pathsol}), when self-consistency is imposed in the infrared ``by hand'' \cite{ChakravartySchmid}.


\section{Prospects for the ergodic-MBL transition as a ``dephasing catastrophe'' \label{sec:dephasing_C}}

The possibility to approach the ergodic-MBL transition in 2D from the ergodic side (at a many-body 
mobility edge corresponding to temperature $T_{MBL}$) is a primary motivation for
this work. We argue that a key attribute of such a temperature-tuned transition is
the \emph{failure of dephasing} of quantum conductance corrections as $T \rightarrow T_{MBL} > 0$, when 
approached from above. Conversely, we argue that dephasing of quantum interference corrections to dc transport 
is equivalent to the condition that a system serves as its own heat bath, 
making transport classical and ergodic on the longest scales. 

For a system with localizing quantum conductance corrections, the failure of dephasing means
that quantum coherence is achieved across arbitrarily large length scales at finite energy density. 
It also means that localizing quantum interference corrections swamp out AA corrections at
all orders in perturbation theory, since the former diverge (in two dimensions) 
in the infrared as $\tau_\phi^{-1} \rightarrow 0$,
while AA corrections remain finite even at the transition $T = T_{MBL}$. 

To make this idea concrete, consider the one-loop class CI corrections in Eq.~(\ref{eqCI}).  
As discussed
in Sec.~\ref{sec:Keldresults},
there are two WL corrections in this case: the standard orthogonal class
correction cut by dephasing $\tau_\phi^{-1} > 0$, and the nonstandard class correction cut by 
temperature; each term contributes ``half'' of the total WL correction. 
Like the nonstandard correction, the third term (AA correction) is also directly cut by temperature. 
According to the standard self-consistent solution for $\tau_\phi^{-1}(D,\gamma)$ 
in Eqs.~(\ref{DephasingRate--INTRO}) and (\ref{eq:tauDResult})
\cite{ChakravartySchmid,AAG}, 
the dephasing rate vanishes only at zero temperature. If instead there is a many-body mobility edge,
then $\tau_\phi^{-1}$ goes to zero at this energy density, and the first term in Eq.~(\ref{eqCI}) would
diverge, signaling localization. 

\begin{figure}[h!]
\centering
\includegraphics[width=0.9\linewidth]{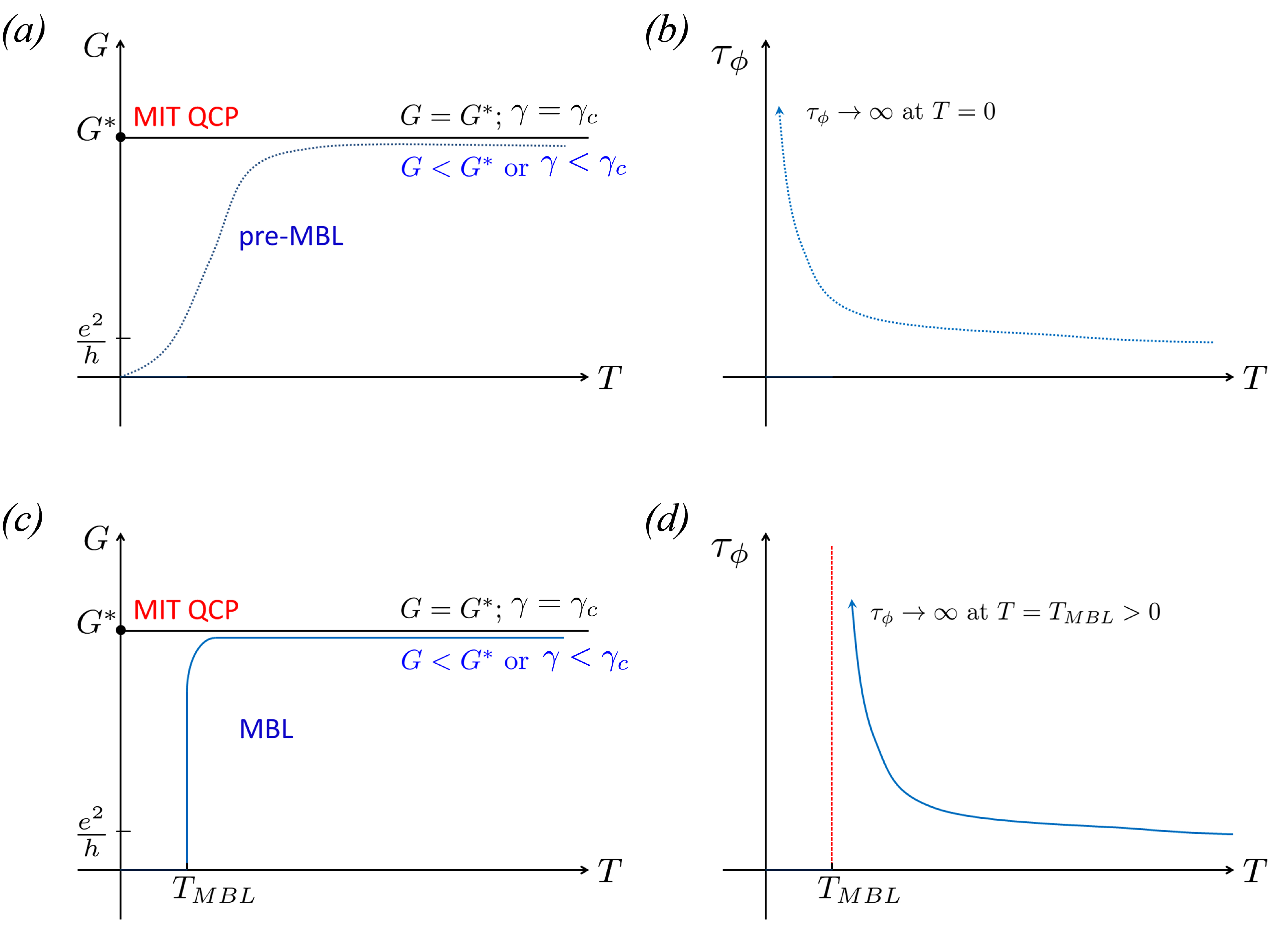}
\caption{Dephasing scenarios for an ergodic-MBL transition in 2D that is continuously connected
to a zero temperature, interacting \cite{BelitzKirkpatrick94} metal-insulator transition quantum critical point (``MIT QCP''). 
A system with weak localizing (WL) quantum conductance corrections in the absence of interactions can be tuned
to the threshold of delocalization at zero temperature via a competing Altshuler-Aronov (AA) correction,
for example in class C \cite{CRealization1,CMomo,CLuca}. 
If we treat the interaction coupling $\gamma$ as a strictly marginal parameter (true to one loop,
possibly to three loops \cite{momothesis,Hikami1992}), then the critical conductance $G^*$
at the zero temperature transition can be made arbitrarily large relative to $e^2/h$. 
In this figure for familiarity we use units appropriate to electrical conductivity, although in the 
superconductor quasiparticle realization of class C this should be replaced with the spin conductivity. 
Figures (a) and (b) present the conductance $G$ and dephasing time $\tau_\phi$ in the 
``pre-MBL'' scenario, meaning the expectation if MBL does not occur. 
For $G$ tuned slightly less than $G^*$ or the interaction 
strength $\gamma$ tuned slightly less than the value $\gamma_c$ needed to delocalize the system at zero temperature,
the pre-MBL scenario has $G$ and $1/\tau_\phi$ vanishing only at zero temperature. 
Figures (c) and (d) instead show the MBL scenario, whereby $G$ and $1/\tau_\phi$ 
vanish at a finite $T_{MBL} > 0$. By tuning $G$ or $\gamma$ sufficiently close
to their critical values, the MBL transition can be continuously deformed to 
the zero temperature MIT, $T_{MBL} \rightarrow 0$.
The one-loop WL and AA corrections for class C are given by Eq.~(\ref{eqC}). 
Since the former arises due to the special nonstandard class diffusion modes near
zero energy, it is automatically cut by temperature \cite{BTDLC05}. 
The dephasing rate $\tau_\phi^{-1}$ would enter at two-loop order, where
the localizing unitary class diffuson correction 
\cite{BelitzKirkpatrick94}
appears at all one-body
energies. 
In this paper we perform explicit calculations only to one loop, so the 
confirmation or refutation of the scenario pictured in (c) and (d) is 
left to future work.  
}
\label{fig:DephScen}
\end{figure}

There are potential practical and conceptual problems with this description. 
The most obvious practical problem 
is that even if the dimensionless bare conductance (defined at the scale of the elastic mean free path) is 
large, the divergence of $\tau_\phi$ means that WL corrections become comparably big close
to the putative ergodic-MBL transition. Then it appears necessary to calculate ever higher order
corrections to capture the physics close to the transition. 
However, class C [Eq.~(\ref{eqC}), \cite{momothesis,CMomo,CLuca}] provides a scenario in 
which control might be possible without calculating to arbitrarily high order.
The point is that the interaction strength $\gamma$ may be marginal to at least three-loop order \cite{momothesis,CMomo,CLuca}. 
Assume that this is the case, so that $\gamma$ can be treated as a constant. 
Then one can tune class C to a zero temperature metal-insulator transition (MIT) 
with \emph{arbitrarily large critical conductance}. 
The key idea is to choose the interaction $\gamma$ so as to exactly balance (e.g.) the
one- and two-loop WL corrections \cite{CMomo,CLuca}; note that the AA correction is 
antilocalizing
for $\gamma < 0$,
which is the physical sign choice for direct exchange-mediated spin-spin interactions. 
Part of the correction that arises at two loops obtains from the unitary Wigner-Dyson class diffusons that appear
at all one-body energies \cite{BelitzKirkpatrick94}, 
and must therefore be cut by dephasing at finite temperature. 
Then, by tuning the interaction slightly below the required threshold, the 
$T = 0$ MIT becomes an arbitrarily low temperature $T_{MBL} > 0$ transition. 
It is reasonable to expect that some aspects of both transitions look the same,
as approached from above, since both are characterized by a diverging
dephasing time $\tau_\phi \rightarrow \infty$. A key question is whether there is a well-defined
average conductance at the MBL transition that deforms smoothly to the large critical conductance
as we tune $T_{MBL} \rightarrow 0$.
Because the one-loop WL correction in class C is directly cut by temperature instead of dephasing, 
a test of this idea requires a two-loop calculation in class C, which we defer to future work. 
The MBL scenario described above is also explicated in Fig.~\ref{fig:DephScen}. 

Conceptual problems with the ergodic-MBL transition and/or the many-body mobility
edge include the following. 
It is possible that many-body mobility edges and/or MBL do not exist in more than 
one spatial dimension due to mobile ``hot bubbles,'' i.e.\ rare ergodic regions \cite{MBLThermalFluct,MBLThermalFluct2}.
We note that a precise formulation of finite temperature response theory as presented in this paper
may allow one to test this scenario, by looking for rare dephasing events that always 
succeed in suppressing quantum interference on the largest scales. 
Another potential difficulty with class C and other realizations of the 10-fold way
is the presence of a nonabelian continuous symmetry [spin SU(2) in the case of classes C and CI].
Arguments have been made \cite{NoNonAbMBL} that such a symmetry is incompatible with MBL. 
However, it is perfectly possible to have a continuous symmetry on the ergodic side that becomes
spontaneously broken in the insulator, whether the latter is a zero-temperature 
Anderson-Mott or finite-temperature MBL phase. Indeed, this is one interpretation
of the ``magnetic instability'' in the spin SU(2) symmetric, interacting orthogonal
class metal at zero temperature \cite{FNLSM1983,BelitzKirkpatrick94,FP2005}.

\section{Acknowledgments}

We thank Vadim Oganesyan for helpful discussions. 
This work was supported by the Welch Foundation Grant No.~C-1809 and by NSF CAREER Grant No.~DMR-1552327
(Y.L.\ and M.S.F.), and by 
NSF CAREER Grant No.~DMR-1653661 and the Wisconsin Alumni Research Foundation (A.L.).

\appendix

\section{Renormalization of the interaction strength}\label{Sec:App1}

The renormalization of the diffusion parameters $h$ and $\lb$ can be obtained directly from the previous calculation, but this does not apply to the interaction strength $\gamma$. In this appendix, we calculate the renormalization of the interaction strength and show that for the symplectic class AII metal $h(1-\gamma)=h_R (1-\gamma_R)$, while for class C superconductor $\gamma_R=\gamma$.

For notational simplicity, we define $d \lb$, $d h$ and $d \gamma$ by
\begin{align}
\begin{aligned}
	\lb_R= \lb (1 + d \lb),	
	\qquad
	h_R= h(1 +dh),
	\qquad
	\gamma_R = \gamma (1 + d \gamma).	
\end{aligned}
\end{align}
From Eq.~(\ref{eq:C_RG0}), we have, to the leading order,
\begin{align}
\begin{aligned}
	d \lb = \delta \lb - \delta z_1 + 2 \delta z_2,
	\qquad
	d h =  \delta h + \delta z_1 -2 \delta z_2.
\end{aligned}
\end{align}
This applies to class C as well as class AII, which can be considered as a special case where $\delta z_2=0$ and $\delta z_1=\delta z$ [see Eq.~(\ref{eq:AII_RG0})].

We notice that 
\begin{align}\label{eq:App1--P1}
\begin{aligned}
	\delta \Pi (\kb,\ww)
	=\, & 
	\Pi (\kb,\ww| \lb_R, h_R, \gamma_R) -\Pi (\kb,\ww| \lb, h, \gamma)
	\\
	=\, & 
	- 	
	\frac{2}{\pi} k^2 
	\bd_u^2(k,-\ww)	
	\left\lbrace 
	\left[  h d h-  h \gamma\left(d \gamma + d h\right)  \right] k^2 
	+
	i h^2 (1- \gamma)^2 \lb \ww d \lb 
	\right\rbrace, 			
\end{aligned} 
\end{align}
where $\Pi (\kb,\ww| \lb, h, \gamma)$ is given by Eq.~(\ref{eq:AII_ClassiPi}).  $\Pi (\kb,\ww| \lb_R, h_R, \gamma_R)$ can be obtained by replacing the variables with renormalized ones.
Comparing this equation with Eq.~(\ref{eq:C_dPi}), we have 
\begin{align}\label{eq:App1--P2}
\begin{aligned}
	h_R(1-\gamma_R) - h(1-\gamma)
	= h  (1-\gamma) dh - h\gamma d\gamma
	=\,&
	h\left(  1-\gamma\right) ^2 (-\delta z_2). 					
\end{aligned}
\end{align}
In Eqs.~(\ref{eq:App1--P1}) and (\ref{eq:App1--P2}), we retain terms to first order in 
$d \lambda$, $d h$, and $d \gamma$, all of which are $\mathcal{O}(\lambda)$. 	
Eq.~(\ref{eq:App1--P2})
implies that $d \gamma$ can be evaluated from
\begin{align}\label{eq:dgamma}
\begin{aligned}
	d\gamma
	=\,&
	\frac{1-\gamma}{\gamma}
	\left[ \delta h + \delta z_1 - \left(  1+\gamma\right) \delta z_2\right].	 					
\end{aligned}
\end{align}
For class AII, $\delta z_2=0$ means that $h(1-\gamma)$ does not renormalize.
This is the statement that the charge compressibility 
[Eq.~(\ref{eq:DcKappaDef})]
is preserved in an interacting Wigner-Dyson class system, although the tunneling density of states receives AA corrections \cite{BelitzKirkpatrick94}. 	
On the other hand, for non-standard class C, this statement does not hold as $\delta z_2 \neq 0$. Substituting the explicit forms of $\delta h$ [Eq.~(\ref{eq:C_d1})], $\delta z_1$ [(\ref{eq:C_d3})] and $\delta z_2$ [Eq.~(\ref{eq:C_d4})] into Eq(\ref{eq:dgamma}) leads to $d\gamma=0$, and as a result $\gamma_R=\gamma$.

\begin{figure}
	\centering
	\includegraphics[width=0.7\linewidth]{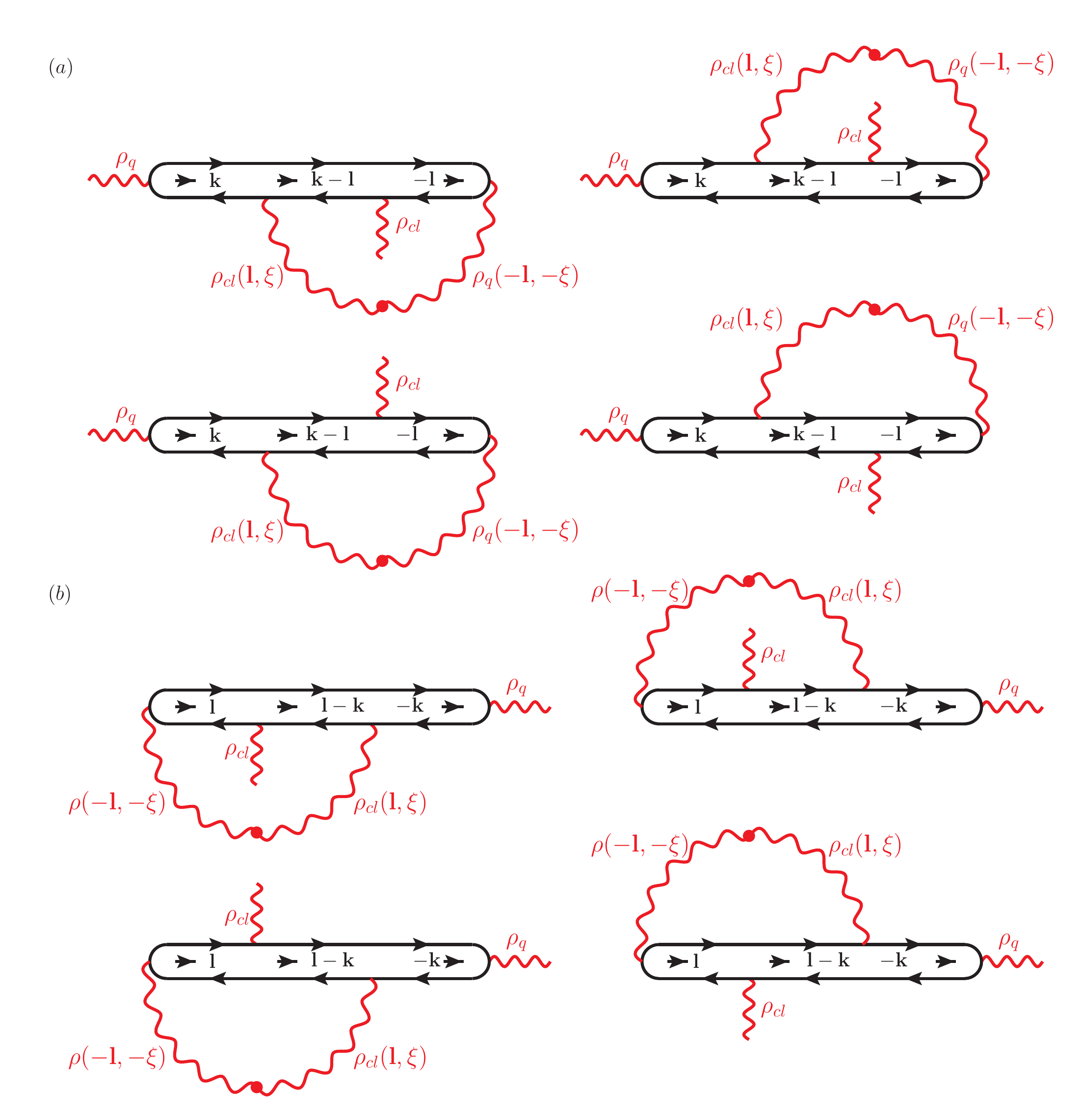}
	\caption{Class AII noncontributing diagrams (1/3).}
	\label{fig:App_a}
\end{figure}
\begin{figure}
	\centering
	\includegraphics[width=0.7\linewidth]{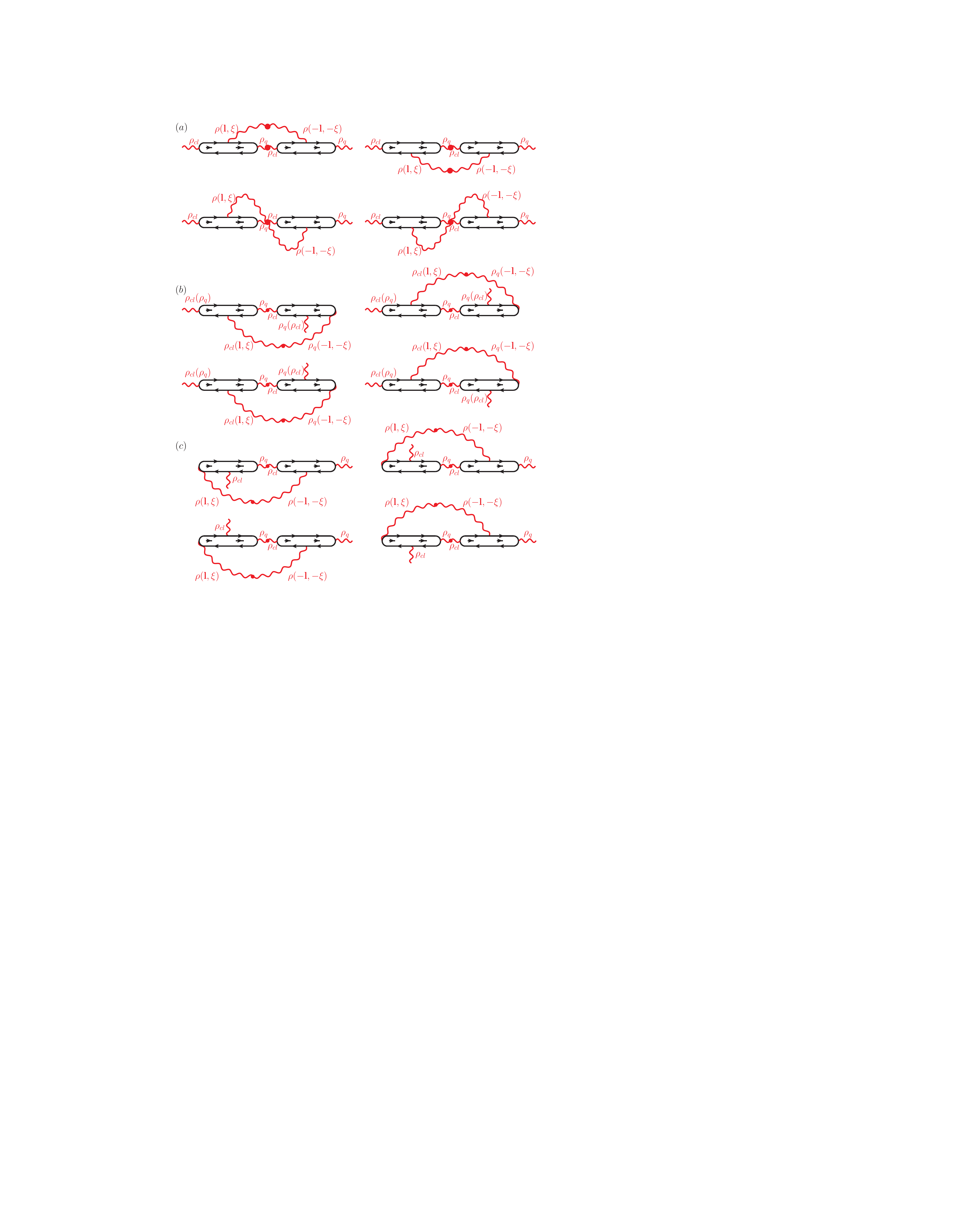}
	\caption{Class AII noncontributing diagrams (2/3).}
	\label{fig:App_b}
\end{figure}
\begin{figure}
	\centering
	\includegraphics[width=0.7\linewidth]{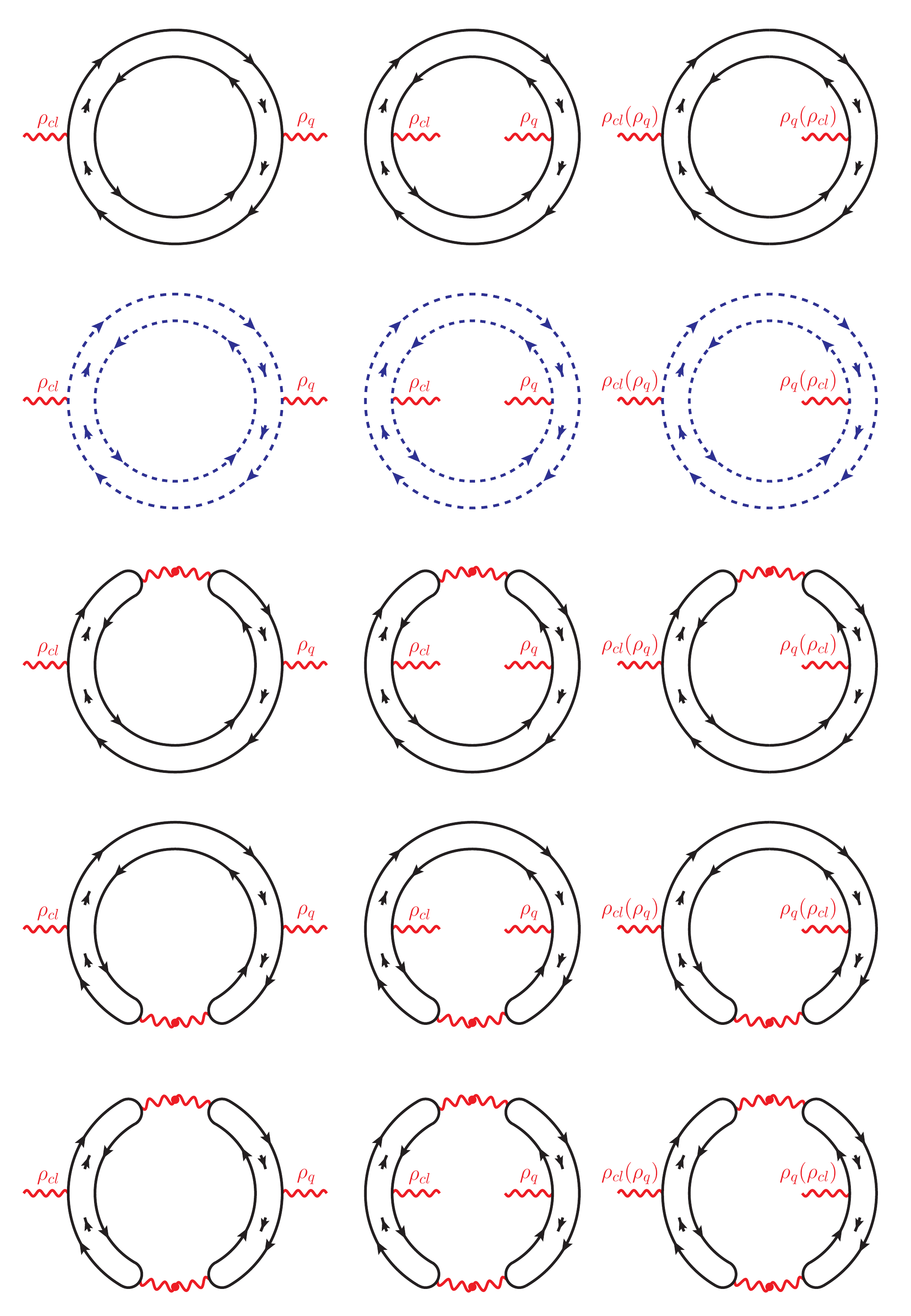}
	\caption{Class AII noncontributing diagrams (3/3).}
	\label{fig:App_c}
\end{figure}

\section{Class AII vanishing diagrams \label{Sec:App2}}

Additional diagrams that give vanishing net contribution 
to the renormalized charge density polarization function 
in the symplectic class AII are shown in Figs.~\ref{fig:App_a}--\ref{fig:App_c}.

\end{document}